\documentclass[aps,pre,reprint,onecolumn,superscriptaddress]{revtex4-2}
\pdfoutput=1
\usepackage{epsfig}
\usepackage{natbib}
\usepackage[normalem]{ulem}
\usepackage{xspace}
\usepackage{xcolor}
\usepackage{graphicx}
\usepackage{amssymb}
\usepackage{cancel}
\usepackage{lineno}
\usepackage{amsmath}
\usepackage{stackrel}
\usepackage{appendix}
\usepackage{url}
\usepackage{enumitem}
\usepackage{braket}
\usepackage{bbm}
\newcommand{\aref}[1]{Appendix \ref{#1}}
\usepackage{mathrsfs}
\usepackage{empheq}

\usepackage{array}
\usepackage{booktabs}
\let\omegaa\omega
\renewcommand{\omega}{k}

\usepackage{graphicx}% Include figure files
\usepackage{dcolumn}% Align table columns on decimal point
\usepackage{bm}% bold math
\usepackage[colorlinks= true, linkcolor=red, citecolor=blue, urlcolor=blue]{hyperref}% add hypertext capabilities
\usepackage{lipsum}
\usepackage{lipsum}
\usepackage{braket}
\usepackage{graphicx}
\usepackage{pgfplots}

\expandafter\let\csname equation*\endcsname\relax
\expandafter\let\csname endequation*\endcsname\relax
\usepackage{amsmath,amssymb}

\usepackage{listings}
\usepackage{xcolor}
\usepackage{tikz}
\usepackage{bm}
\usepackage{randomwalk}
\usepackage{bm}
\pgfplotsset{width=10cm, height=6cm}
\usetikzlibrary{graphs,graphs.standard}
\usetikzlibrary{decorations.pathreplacing}
\usetikzlibrary{patterns}

\definecolor{dgreen}{rgb}{0,0.7,0}

\usepackage{xparse}

\makeatletter
\NewDocumentCommand{\sump}{e{_}}
 {%
  \DOTSB
  \mathop{\IfNoValueTF{#1}{\sump@{}}{\sump@{#1}}}%
  \nolimits
 }
\newcommand{\sump@}[1]{\mathpalette\sump@@{#1}}
\newcommand{\sump@@}[2]{%
  \ifx#1\displaystyle
    {\sump@display{#2}}%
  \else
    \sum@\nolimits'_{#2}%
  \fi
}
\newcommand{\sump@display}[1]{%
  \sbox\z@{$\m@th\displaystyle\sum@\nolimits'$}%
  \sbox\tw@{$\m@th\displaystyle\sum@\limits_{#1}$}%
  \sbox\@tempboxa{$\m@th\displaystyle'$}
  \mathop{\sum@\nolimits' \kern-\wd\@tempboxa}\limits_{#1}%
  \ifdim\wd\z@>\wd\tw@
    \kern\dimexpr\wd\z@-\wd\tw@\relax
  \fi
}

\usepackage[many]{tcolorbox}

\usepackage{bbold}
\usepackage{mathtools}

\usepackage{xcolor}

\usepackage{amsfonts}
\usepackage{graphicx}
\usepackage{hyperref}

\usepackage{xcolor}
\usepackage{textcomp}
\usepackage{float}
\usepackage{soul}
\usepackage{stackrel}
\pdfoutput=1
\usepackage{esint}
\usepackage{wrapfig}
\usepackage{lipsum}

\usepackage{tikz}
\usepackage{graphicx}% Include figure files
\usepackage{bm}% bold math
\usepackage{hyperref}
\usepackage{multirow}
\usepackage{array}
\usepackage{booktabs}
\usepackage{ctable}
\usepackage{upgreek}
\usepackage{epsfig,psfrag,subfigure,amsopn}
\usepackage{mathrsfs}
\usepackage{amssymb}
\usepackage{amsbsy}
\usepackage{color}
\usepackage{cancel}
\usepackage{}
\usepackage{xr}
\usepackage{color}
\definecolor{dgreen}{rgb}{0,0.7,0}

\usepackage{ragged2e}
\begin{document}

\title{Splitting Probability in Monitored Quantum Dynamics}

\author{Prashant Singh}
\email{prashantsinghramitay@gmail.com}
\affiliation{Department of Physics, Bar-Ilan University, Ramat Gan 52900, Israel}
\author{Sabine Tornow}
		\affiliation{Research Institute CODE, University of the Bundeswehr Munich, D-81739 Munich, Germany}
\author{David A Kessler}
\affiliation{Department of Physics, Bar-Ilan University, Ramat Gan 52900, Israel}
\author{Eli Barkai}
\affiliation{Department of Physics, Bar-Ilan University, Ramat Gan 52900, Israel}
\affiliation{Institute of Nanotechnology and Advanced Materials, Bar-Ilan University, Ramat Gan 52900, Israel}
%\date{\today}

\begin{abstract}
We consider a continuous-time quantum walk stroboscopically subjected to projective measurements on two distinct targets with sampling time period $\tau$. We construct a general renewal framework for the first-detection amplitude at each target, from which we obtain the associated splitting probabilities. For Hamiltonians with reflection symmetry, we further show that each dual-target first-detection amplitude can be viewed as an interference effect between a pair of single-target first-detection amplitudes. This formulation yields exact expressions for the splitting probabilities and gives special sampling times at which they undergo pointwise discontinuous jumps due to a form of ergodicity breaking associated with dark states. As an illustrative example, we apply the framework to a one-dimensional tight-binding lattice chain with nearest-neighbor hopping, and study the splitting probabilities as functions of the initial position and sampling time. Below a critical sampling time, $0<\tau < \tau_c$ where $\tau_c = 2 \pi / \Delta E$ with $\Delta E$ being the energy bandwidth, the splitting probabilities for large system sizes become equal to $1/2$, independent of the initial position, as long as this position does not coincide with target sites. Above the critical sampling time, $\tau \geq \tau_c$, the splitting probabilities deviate from $1/2$ and develop oscillations as a function of initial position. The wavenumber associated with these oscillations is $Q=2\sin^{-1}(r\tau_c/\tau)$, where $r$ is a positive integer. We also explore the effects of relative phase and spatial separation between different components of a coherent initial state on the splitting probabilities. Finally, we study how a finite number of measurements modifies the splitting probabilities and broadens the resonances at special sampling times.

\end{abstract}

\pacs{Valid PACS appear here}% PACS, the Physics and Astronomy
\maketitle
%\tableofcontents
\begingroup
\hypersetup{
    linkcolor=red,
    linktoc=page
}
\tableofcontents
\endgroup

%\newpage
\section{Introduction}
The splitting probability is a fundamental concept in the theory of stochastic processes and first-passage phenomena \cite{Fellerbook, Rednerbook, Metzlerbook}. It measures the probability that the stochastic trajectory reaches one designated target before any of the other available targets, thus providing a natural way to characterize competition between multiple possible outcomes. A famous example arises in financial risk management, where a gambler tries to reach a target fortune before going bankrupt \cite{shoesmith1986huygens}. 
%Hence, the name \textit{gambler's ruin formula}. 
In this context, the splitting probability gives the likelihood of eventual success versus zero-fortune failure. Also, in transport problems, the transmission and reflection probabilities of particles through a medium can be modeled within the splitting probability framework \cite{ST-2022, nt-ctt-1}. 
%Recent works have shown that splitting probability can also be used to derive fundamental bounds on thermodynamic quantities, such as the entropy production rate \cite{Neri2015}. 
Further applications can be found in chemical kinetics \cite{BENICHOU2014225, Vezzani2024}, single-molecule experiments \cite{A1_2011, A1_2015} and in studies of polymers \cite{A1_2009, Rosso2026}. These diverse applications have led to systematic investigations of the splitting problem across a broad spectrum of stochastic processes, such as random walk and diffusion \cite{Fellerbook, Rednerbook, Metzlerbook, shoesmith1986huygens, STFP-2005, ST-2014, ST2023}, random acceleration \cite{STRA-2000, STRA-2005}, active models \cite{STAC-2018, STAC-2020, Park2022, SinghSS_2022, Malgaretti2023, Frydel2024}, resetting systems \cite{nt-ctt-2, ST-2019, nt-ctt-3}, fractional Brownian motion \cite{STSM-2010, Wiese2019}, and disordered systems \cite{ST-2024, disorder1}. We refer to standard textbooks \cite{Fellerbook, Rednerbook} for more details on the splitting problem. In this paper, we investigate a quantum analogue of this process.

As a concrete example of the classical problem, consider a discrete-space, continuous-time random walker hopping along a one-dimensional lattice chain of size $N$, with lattice sites labeled as $x \in \{1,2, \ldots,N \}$. Starting from an initial position $x_0$ (with $1 \leq x_0 \leq N$), the walker can jump to one of its neighboring sites with a rate $\gamma ~(> 0)$. We further have absorbing conditions at the left and the right boundaries, $x_L = 1$ and $x_R = N$, so that the walker is instantly absorbed and the process terminates upon reaching either boundary of the chain. In this set-up, the splitting probability characterizes the likelihood that the walker, starting from $x_0$, will be absorbed by one boundary before it reaches the other one.

%corresponds to the probability that the walker, starting from $x_0$, is absorbed at a specific boundary site, before visiting the other one. 

For this discrete-space, continuous-time random walker, the splitting probabilities, $P_L^{\rm (cl)}(x_0)$ and $P_R^{\rm (cl)}(x_0)$, are given by the well-known \textit{gambler's ruin formula} \cite{Rednerbook}
\begin{align}
P_L^{\rm (cl)}(x_0) &= \frac{ \left( N - x_0 \right)}{\left( N - 1 \right)}, \qquad
P_R^{\rm (cl)}(x_0) = 1 -P_L^{\rm (cl)}(x_0).
\label{unitary-eqn-classical}
\end{align}
The superscript `$\mathrm{cl}$' in our notation explicitly indicates the classical nature of the process, which will later be contrasted with the quantum case. These splitting probabilities are linear in $x_0$, and are independent of the hopping rate $\gamma$. The fact that they satisfy $P_L^{\rm (cl)}(x_0)+P_R^{\rm (cl)}(x_0)=1$ implies that the walker will be always absorbed by one of the boundaries. Also, Eq.~\eqref{unitary-eqn-classical} implies the proximity effect whereby $P_L^{(\text{cl})}(x_0) > P_R^{(\text{cl})}(x_0)$ if $x_0$ is closer to the left boundary and vice versa, so that starting closer to a boundary increases the probability of absorption there. Finally, we emphasize that the same expressions in Eq.~\eqref{unitary-eqn-classical} hold for the continuum diffusion process, as well as for the discrete-space, discrete-time random walk \cite{Rednerbook, ST-2022}.

Quantum walks are the analogues of classical random walks with applications in quantum information science and quantum computation \cite{Kempe2001, Vishwanath2001,SP3, SP2, Kempe2003,SP1, Childs2009, Bach2009, MULKEN2011,  SP4, SP5, SP6}. In a recent Letter \cite{myletter}, some of us introduced the splitting problem for monitored quantum walks and showed that it differs qualitatively from its classical counterpart in Eq.~\eqref{unitary-eqn-classical}. More specifically, the study examined a continuous-time, discrete-space quantum walker evolving on a finite graph under a general time-independent, Hermitian Hamiltonian. In addition, projective measurements are performed at two distinct target states at regular time intervals $\tau$ to monitor the walker in either target state. In contrast to classical processes, where the splitting problem, as explained above, is formulated in terms of absorbing boundaries, here it is defined through repeated projective measurements at the target states. Also $\tau$ will be referred to as the sampling time throughout this work. For the one-dimensional lattice chain with $N$ sites, with a Hamiltonian possessing reflection symmetry about the centre of the chain and localized measurements performed at the left and the right endpoints of the chain, the dual-target splitting problem can be mapped on to a pair of single-target first-detection problems. This showed that the splitting problem can be viewed as an interference effect between two single-target processes. The single-target detection problem has been quite extensively studied both theoretically \cite{QW1,  QW4, QW5, QW25, QW21, QW22, QW7, QW8, QW11,QW6, QW12, QW13, QW14, QW15, QW16, QW17, QW18,Das_2022, Das_20221, QW19, QW20,  QW26,  Gambassi2025,Godec2025, Vecchio2026} as well as in experimental platforms involving quantum computers, optical waveguides and trapped ions \cite{ExptWG-3, ExptWG-2, Chen2026, ExptQC-1, ExptQC-2, ExptQC-3, Ryan2025}. The mapping therefore allowed us to use established single-target methods to analyze the dual-target splitting problem.

\begin{figure}[t]
	\centering
	\includegraphics[scale=0.3]{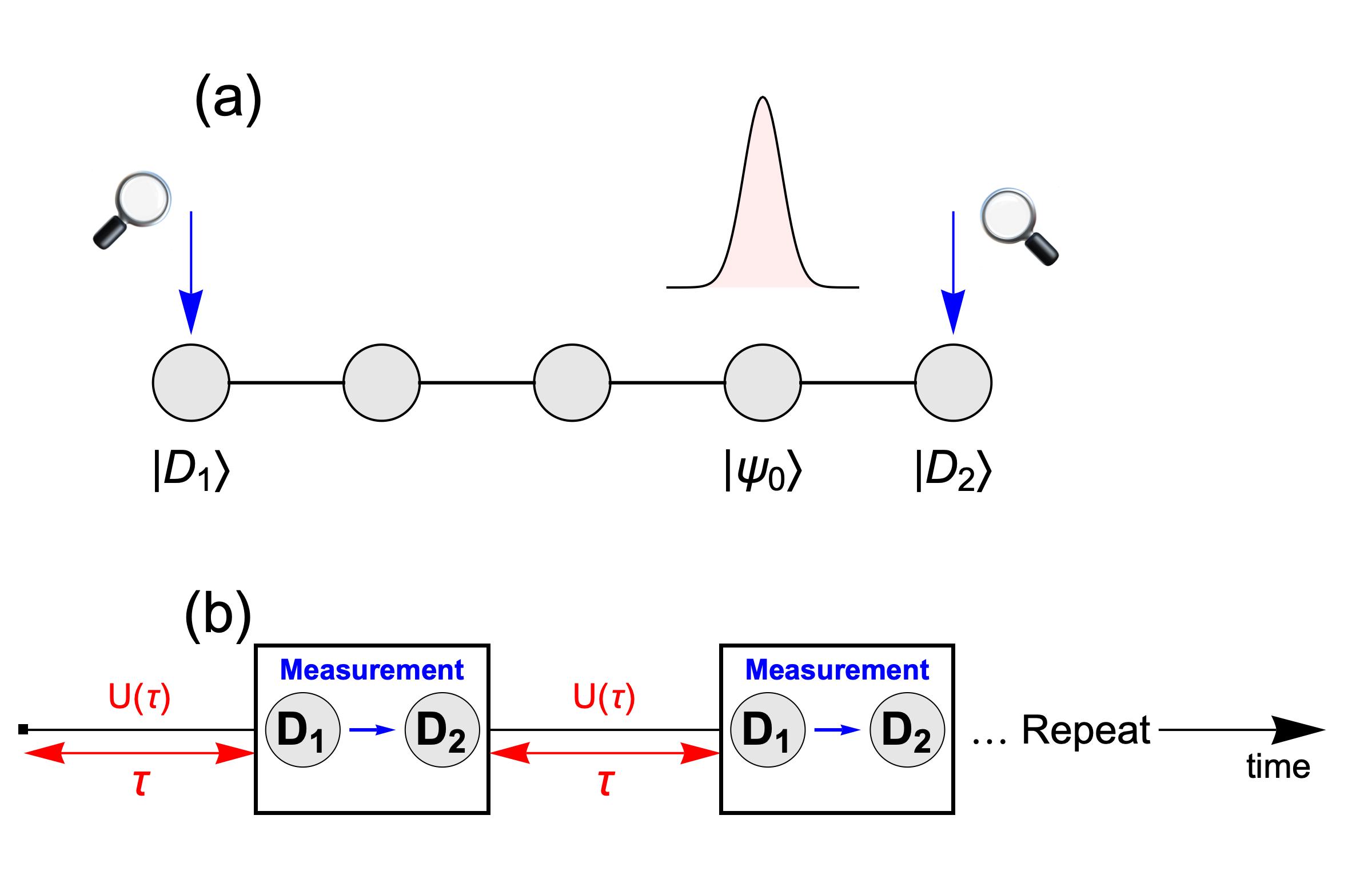} 
	\caption{Panel (a) shows a chain of size $N=5$ in which a system with $\ket{\psi_0}=\ket{4}$ (see the pink shaded region) evolves with the nearest-neighbor hopping Hamiltonian in Eq.~\eqref{main-eqq-1}. After every time interval $\tau$, projective measurements are performed at $\ket{D_1}=\ket{1}$ and $\ket{D_2}=\ket{N}$ to check the system at these states. Panel (b) illustrates the splitting probability detection protocol. During the first $\tau$ interval, the system evolves unitarily with the operator $U(\tau)$ in Eq.~\eqref{unitary-eqn}. After this evolution, a projective measurement $\ket{D_1}\bra{D_1}$ is performed at the first target state. If this detection is successful, the process terminates. If not, then the state of the system is projected onto the subspace complementary to $\ket{D_1}$, and a second projective measurement $\ket{D_2}\bra{D_2}$ is immediately made at the second target state. If the system is detected, the process again terminates. If this detection also fails, the system's state is projected onto the subspace complementary to $\ket{D_2}$. With this projected state, the system evolves unitarily for another interval $\tau$, and the procedure is repeated, in principle, for an infinite number of times or until a successful detection is made at either of the target states. We implement this detection protocol to study the splitting probabilities at the two targets.}
	\label{fig-scheme}
\end{figure}

For the nearest-neighbor hopping Hamiltonian with localized initial condition, $\ket{\psi_0} = \ket{x_0}$, as shown in Fig.~\ref{fig-scheme}(a), it was demonstrated that, in the large-$N$ limit, the splitting probabilities undergo a nonanalytic flat-to-fluctuating transition as a function of the sampling time. For $0< \tau < \tau_c$, both splitting probabilities are equal to $1/2$, independent of $x_0$ and $\tau$, while for $\tau \geq \tau_c$, they differ from $1/2$ and develop a fluctuating pattern of peaks and dips that depend sensitively on both $x_0$ and $\tau$. The critical sampling time is $ \tau _c = 2 \pi/ \Delta E $ in units of $\hbar$, where $\Delta E$ is the difference between the maximum and minimum energy eigenvalues. For finite system, one gets resonances in the splitting probabilities at special values of $\tau$, related to a form of ergodicity breaking associated with dark states.

In this paper, we extend the analysis on the splitting problem in monitored quantum systems and report several new results. We develop a renewal-theory formulation for the first-detection amplitudes at two arbitrary targets for a general Hamiltonian. This formulation connects the associated two detection amplitudes to quantities involving only the measurement-free unitary dynamics of the system. As a result, by combining renewal theory with the generating-function technique, one can treat the dual-target splitting problem entirely in terms of the measurement-free propagators. Renewal ideas have a long history in classical first-passage problems \cite{MontrollWeiss1965}, going back to Schr\"{o}dinger \cite{Schrodinger1915}, and have more recently proved useful in quantum first-detection problems \cite{QW6,QW11}. Here, we construct them for the quantum splitting problem.

We further investigate the dependence of the splitting probabilities on the initial position $x_0$ for the one-dimensional chain with nearest-neighbor hopping Hamiltonian. Apart from the flat-to-fluctuating transition at $\tau_c$, the splitting probabilities exhibit a notable spatial structure: for $\tau \geq \tau_c$, exhibiting pronounced oscillations as functions of $x_0$. Question is: what physical mechanism determines the characteristic spatial scales of these oscillations, and how do these scales depend on the sampling time $\tau$? To address this question, we analyze the oscillations in the Fourier space of $x_0$ and show that they are governed by phase-matching events between different momentum modes of the quantum walk. Their wavenumbers depend systematically on the sampling time $\tau$, as shown in Eq.~\eqref{skpssrs} below. Beyond localized initial conditions, we also consider $\ket{\psi_0}$ involving coherent superpositions over multiple sites. Does the flat-to-fluctuating transition at $\tau_c$ persist for such coherent initial states, and if so, how is it modified by the spatial separation and relative phase between different components of the initial state? As we show in the paper, the transition remains present, although some of its features differ from those found for localized initial conditions. Finally, we study how a finite number of measurements, as commonly encountered in experiments \cite{ExptQC-1, ExptQC-2, ExptQC-3}, affect the splitting probabilities.

The rest of our paper is structured as follows: In Sec.~\ref{sec-model}, we introduce our model and the splitting probability detection protocol. We outline the main results of our paper in Sec.~\ref{summary-sec}. Sec.~\ref{sec-general} derives the quantum renewal equations. The renewal framework is then combined with the generating function techniques in Sec.~\ref{sec-GF} to derive the splitting probabilities for arbitrary Hamiltonian, sampling time, initial condition and target states. For one-dimensional chain with Hamiltonians satisfying reflection symmetry and target states coinciding with the endpoints, we show in Sec.~\ref{sec-saubcb} that these general expressions reduce to the mapping mentioned above. Sec.~\ref{sec-tight} focuses on the nearest-neighbor hopping case. 
%We present the calculations for generic $\tau$ in Sec.~\ref{sec-smooth}. 
The Zeno limit is analyzed in Sec.~\ref{sec-Zeno}, followed by a discussion on the oscillations in the splitting probabilities and the flat-to-fluctuating transition in Sec.~\ref{sec-transition}. Sec.~\ref{sec-resonance} then derives the conditions for observing the resonances at special sampling times. We present detailed results for our case study, $N=5$, in Sec.~\ref{sec-example}. We discuss the role of coherent initial states in Sec.~\ref{sec-OIC}, and the effect of finite number of measurements in Sec.~\ref{sec-finite}. Finally, we present our conclusions and indicate future directions in Sec.~\ref{sec-conclusion}.

\section{Model and Preliminaries}
\label{sec-model}
Consider a discrete-state, continuous-time quantum system characterized by a time-independent, Hermitian Hamiltonian $H$, defined on a Hilbert space $\mathcal{H}$ with a finite $N$ states. The system is governed by the Schr\"odinger equation $i\hbar \ket{\dot{\psi}} = H \ket{\psi}$ with the initial condition $\ket{\psi_0}$ and overhead dot denoting the time derivative. For example, $H$ could represent the nearest-neighbor hopping Hamiltonian on a finite one-dimensional chain of size $N$
\begin{align}
H = - \gamma \sum _{x=1}^{N-1} \Big[  \ket{x} \bra{x+1}+\ket{x+1} \bra{x} \Big],  \label{main-eqq-1}
\end{align}
where $\gamma$ is the hopping amplitude; see Fig.~\ref{fig-scheme}(a).
%This assumption is justified for a spinless system with time-reversal symmetry, for example in the absence of magnetic flux or other sources of complex hopping phases.  
Then $H$ takes the form of a discrete Laplacian, which can be conveniently represented as the adjacency matrix of a graph \cite{MULKEN2011}. We later return to this model. For now, we keep our analysis general.
%We also choose a general initial condition $\ket{\psi_0}$.
%The initial condition $\ket{x_0}$ is chosen to be localized at a lattice site $x_0 \in \{ 1,2., \ldots,N \}$.

In the absence of any measurement, the system evolves unitarily with the operator
\begin{align}
U(t) = \exp(-i H t), \label{unitary-eqn}
\end{align}
where we set $\hbar=1$. When measurements are performed, this purely unitary evolution is however interrupted by wave function collapses at the measurement times.  
%Consequently, the measured-dynamics, as shown later, is formulated in terms of a non-Hermitian operator in one time period describing both the system's unitary dynamics and the detection protocol. This naturally relates the framework to non-Hermitian Floquet dynamics \cite{floquet1, floquet2}. 
We now explain our detection protocol at two distinct target states.
\subsection{Splitting probability detection protocol}
\label{sec-protocol}
Projective measurements are made on two distinct target states, $\ket{D_1}$ and $\ket{D_2}$, satisfying the orthogonality condition, $\braket{D_1|D_2} = 0$. 
%These states may, for example, correspond to the two endpoints of the one-dimensional chain, as shown in Fig.~\ref{fig-scheme}. However, we keep the formulation general here and specialize to that case only later. 
The measurements are performed stroboscopically at times $t=\tau,~2\tau,~3 \tau$ and so on, such that the time interval $\tau$ between successive measurements is kept fixed. As mentioned before, $\tau$ is called the sampling time. Our detection protocol is the following:
\begin{enumerate}
\item Starting from $\ket{\psi_0}$, we evolve the system for a time duration $\tau$ as $\ket{ \psi(\tau)} = U(\tau) \ket{\psi_0}$, where   $U(\tau) $ is the unitary operator in Eq.~\eqref{unitary-eqn}.

\item At the end of this evolution, we perform a projective measurement $\ket{D_1}\bra{D_1}$ to check the system at the first target state. If detected, the process terminates with detection at this target. If no detection occurs, then the state $\ket{ \psi(\tau)}$ is projected onto the subspace complementary to $\ket{D_1}$ as $\ket{\psi(\tau^+)} = \mathcal{N} \left( \mathbb{1}-\ket{D_1}\bra{D_1} \right) \ket{\psi(\tau)}$. Here $\mathbb{1}$ is the identity matrix, $\tau^+$ is the instant immediately after the null measurement and $\mathcal{N}$ is the normalization factor chosen such that $\braket{\psi(\tau^+)|\psi(\tau^+)}=1$.

%using the operator $\left( \mathbb{1}-\ket{D_1}\bra{D_1} \right)$, where $\mathbb{1}$ is the identity matrix.

\item  With $\ket{\psi \left( \tau^+  \right)}$, a second projective measurement $\ket{D_2}\bra{D_2}$ is immediately performed at the second target state. If the system is detected there, the process terminates with detection at this target. If this detection also fails, then $\ket{\psi \left( \tau^+  \right)}$ is projected onto the complementary subspace  using $\left(\mathbb{1} - \ket{D_2}\bra{D_2}\right)$.

\item The new projected state is then evolved for another duration $\tau$, followed by the repetition of steps~2-4, in principle, for an infinite number of times or until a successful detection is made.

\end{enumerate}

%During each interval, the detection proceeds in two stages. First, after the unitary evolution described by Eq.~\eqref{unitary-eqn} for a duration $\tau$, we perform a projective measurement $\ket{D_1}\bra{D_1}$ to check the walker at the first target state. If the walker is detected, the process terminates with detection at this target. If no detection occurs, then the walker's state is projected onto the complementary subspace using the operator $\left( \mathbb{1}-\ket{D_1}\bra{D_1} \right)$, where $\mathbb{1}$ is the identity matrix. With this projected state, a second projective measurement $\ket{D_2}\bra{D_2}$ is immediately performed at the second target site. If the walker is detected there, the process terminates with detection at this site. If this detection also fails, the walker's state is projected onto the complementary subspace  using $\left(\mathbb{1} - \ket{D_2}\bra{D_2}\right)$. After this second projection, the walker is then allowed to evolve for another duration $\tau$, followed by the next measurement. Notice that at the end of each $\tau$-interval, the walker may be detected at either $D_1$ or $D_2$ sites, or remain undetected in both. We repeat our detection procedure, in principle, for an infinite number of times or until a successful detection is made. 

With this protocol, we investigate the quantum splitting probabilities corresponding to detection at the target states $\ket{D_1}$ and $\ket{D_2}$. Fig.~\ref{fig-scheme} shows a schematic illustration of this protocol for the one-dimensional chain with the nearest neighbor hopping Hamiltonian in Eq.~\eqref{main-eqq-1}. Fig.~\ref{fig-scheme}(a) shows the chain of size $N=5$, with initial condition $\ket{\psi_0}= \ket{4}$, and measurements at the endpoints of the chain, $\ket{D_1}=\ket{1}$ and $\ket{D_2} = \ket{5}$. Fig.~\ref{fig-scheme}(b) illustrates the detection protocol: after each interval of unitary evolution $U(\tau)$, the measurement is first performed at the $\ket{D_1}$ state and, if unsuccessful, immediately at $\ket{D_2}$, before the next evolution cycle begins. Observe that in our setup, a projective measurement is first performed at $\ket{D_1}$, and subsequently at $\ket{D_2}$. One may equally consider the opposite ordering, where measurement is first made at the $\ket{D_2}$ state, and then at $\ket{D_1}$. We will later show that the two orderings are completely equivalent due to the orthogonality of $\ket{D_1}$ and $\ket{D_2}$, and the fact that the two measurements are performed immediately one after the other.

More generally, let $F_{n}^{(D_{1})}\left(  \psi _0 \right) $ and $F_{n}^{(D_{2})}\left(  \psi _0 \right) $ be the first-detection probabilities corresponding to the $\ket{D_1}$ and $\ket{D_2}$ target states, respectively, on the $n$-th measurement attempt.  The splitting probabilities are then given by
\begin{align}
P_{D_1} \left(  \psi _0 \right) = \sum _{n=1}^{\infty} F_{n}^{(D_{1})}\left(  \psi _0 \right),~~P_{D_2} \left(  \psi _0 \right) = \sum _{n=1}^{\infty} F_{n}^{(D_{2})}\left(  \psi _0 \right) . \label{anj8y14}
\end{align}
The sums run over all possible first-detection events, namely detection at the first measurement attempt, the second, the third, and so forth. We next summarize the main results of our paper.
%Our approach is to employ the renewal equation \eqref{main-eqq-00} or the related generating function techniques to find the detection amplitudes. We then use Eq.~\eqref{anj8y14} to compute the splitting probabilities.

%The main results of our paper are summarized next.

\section{Summary of the results}
\label{summary-sec}
\begin{enumerate}[label=(\arabic*), leftmargin=*, itemsep=0pt, topsep=0pt]
\item \textit{Quantum renewal equation:} We show that the probability $F_{n}^{(D_{\alpha})}\left(  \psi _0 \right) $ with $\alpha \in \{1,2\}$ is given by
\begin{align}
F_{n}^{(D_{\alpha})}\left(  \psi _0 \right)  = \big|\phi_{n}^{(D_\alpha)}(\psi _0) \big|^2,~~\text{with }\phi_{n}^{(D_\alpha)}(\psi _0)= \bra{D_{\alpha}} U(\tau) \Big[ \Big(\mathbb{1}-\hat{D} \Big) U(\tau) \Big]^{n-1}\ket{ \psi _0}, \label{anjiuq1}
\end{align}
where $\hat{D} =  \left[ \ket{D_1} \bra{D_1}+\ket{D_2} \bra{D_2} \right] $ is the projection operator in the detector subspace and $U(\tau)$ recall is the unitary operator in Eq.~\eqref{unitary-eqn}. The quantity $\phi_{n}^{(D_\alpha)}(\psi_0)$ is the first-detection amplitude at the $\ket{D_{\alpha}}$ target state at the $n$-th stroboscopic attempt. We prove that $\phi_{n}^{(D_\alpha)}(\psi_0)$ satisfies the relation
\begin{align}
\phi _{n}^{\left(D_{\alpha} \right)} \left( \psi _0 \right) = \bra{D_{\alpha}}   U(n \tau) \ket{\psi _0} - \sum _{\beta=1}^{2} \sum _{l=1}^{n-1} \phi _{l}^{\left(D_{\beta} \right)} \left(  \psi_0 \right) \bra{D_{\alpha}} U\left( (n-l)\tau \right) \ket{ D_{\beta} }. \label{main-eqq-00}
\end{align}	
For example, when $n=1$ and $n=2$, Eq.~\eqref{main-eqq-00} reduces to
\begin{align}
n=1:\qquad
\phi_{1}^{(D_{\alpha})}(\psi_0)
&=
\bra{D_{\alpha}}U(\tau)\ket{\psi_0},
\nonumber\\
n=2:\qquad
\phi_{2}^{(D_{\alpha})}(\psi_0)
&=
\bra{D_{\alpha}}U(2\tau)\ket{\psi_0}
-
\phi_{1}^{(D_1)}(\psi_0)
\bra{D_{\alpha}}U(\tau)\ket{D_1}
-
\phi_{1}^{(D_2)}(\psi_0)
\bra{D_{\alpha}}U(\tau)\ket{D_2}.
\label{main-eqq-n1-n2}
\end{align}
One can clearly observe the emerging renewal structure. For $n=2$, we see that the first-detection amplitude consists of three parts: (i) the direct propagation amplitude $\bra{D_{\alpha}} U(2\tau)\ket{\psi_0}$ from the initial state $\ket{\psi _0  }$ to $\ket{D_{\alpha}}$ in time $t=2 \tau$, (ii) a correction term subtracting histories in which the system is detected in the target state $\ket{D_1}$ at the earlier time duration $\tau$ and subsequently evolves from $\ket{D_1}$ to $\ket{D_{\alpha}}$ during the remaining time duration $\tau$, and (iii) an analogous correction term associated with detection in the second target state $\ket{D_2}$ at time $\tau$, followed by evolution from $\ket{D_2}$ to $\ket{D_{\alpha}}$.

\quad \quad Eq.~\eqref{main-eqq-00} is the quantum renewal equation mentioned in the Introduction. As discussed there, the renewal formalism has proved useful both in classical first-passage problems \cite{MontrollWeiss1965, Schrodinger1915} and in quantum single-target detection problems \cite{QW6,QW11}. Here, we use this approach and combine it with the generating function techniques to find the detection amplitudes $\phi _{n}^{\left(D_{\alpha} \right)} \left( \psi _0 \right)$. We then use Eq.~\eqref{anj8y14} to compute the quantum splitting probabilities.

\item \textit{Quantum mapping:} For a one-dimensional lattice chain of size $N$, with a general Hamiltonian symmetric under the reflection transformation $x \to (N+1-x)$, and with target states localized at the left and the right endpoints, $\ket{D_1}=\ket{x_L}=\ket{1}$ and $ \ket{D_2} = \ket{x_R}=\ket{N}$, the first-detection amplitudes in Eq.~\eqref{anjiuq1} can be expressed as a linear combination of two terms $\chi _{n}^{(\pm)} \left( \psi _0 \right)$
\begin{align}
\begin{cases}
& \phi _{n}^{\left(x_L \right)} \left( \psi _0 \right) = \frac{1}{\sqrt{2}} \left[ \chi _{n}^{\left(+\right)} \left( \psi _0 \right) + \chi _{n}^{(-)} \left( \psi _0 \right) \right], \\
& \phi _{n}^{\left(x_R \right)} \left( \psi _0 \right) = \frac{1}{\sqrt{2}}\left[ \chi _{n}^{\left(+\right)} \left( \psi _0 \right) - \chi _{n}^{(-)} \left( \psi _0 \right) \right].
\end{cases}
\label{main-eqq-101}
\end{align}
The auxiliary functions 
\begin{align}
\chi ^{(\pm)}_n(\psi_0) = \braket{d_{\pm} |U(\tau) \left[ \Big(\mathbb{1}-\ket{d_{\pm}} \bra{d_{\pm}} \Big) U(\tau)\right]^{n-1}|\psi_0}, \label{shjay781}
\end{align}
represent two distinct single-target first-detection amplitudes, with projective measurements performed on the states $\ket{d_{\pm}} = \left( \ket{1} \pm \ket{N}  \right)/\sqrt{2}$ \cite{QW6, QW7}. Eq.~\eqref{main-eqq-101} is the mapping mentioned before. The dual-target splitting problem in Eq.~\eqref{anjiuq1} is mapped to two auxiliary single-target detection problems: one involving the auxiliary target state $\ket{d_+}$, and the other involving a second auxiliary target state $\ket{d_-}$. The mapping is intrinsically quantum, relying on auxiliary target states that arise from superposition and therefore have no classical counterpart. Moreover, Eq.~\eqref{main-eqq-101} makes it possible to analyze the splitting dynamics using methods already developed for single-target detection problems \cite{QW6, QW16, QW7, QW8}. In particular, we employ this mapping to obtain exact expressions for the splitting probabilities for arbitrary Hamiltonians possessing parity symmetry, arbitrary values of $\tau$ and $N$, and an arbitrary initial state $\ket{\psi_0}$.

\item \textit{Nearest-neighbor hopping Hamiltonian:} We apply Eq.~\eqref{main-eqq-101} to a quantum walk with nearest-neighbor hopping Hamiltonian in Eq.~\eqref{main-eqq-1} such that $H\ket{E_{\ell}}=E_{\ell}\ket{E_{\ell}}$, with $E_{\ell}$ and $\ket{E_{\ell}}$ given in Eq.~\eqref{neww-eg-eq-1} $(\ell \in \{ 1,2, \ldots,N\})$. Whenever measurements are made at the left and the right endpoints of the chain, we denote the splitting probabilities by $P_L(\psi_0)$ and $P_R(\psi_0)$ corresponding to $\ket{D_1}=\ket{x_L}=\ket{1}$ and $\ket{D_2}= \ket{x_R}= \ket{N}$, respectively. 
For generic values of $\tau$, both $P_L(\psi _0)$ and $P_R(\psi _0)$ vary smoothly with $\tau$, satisfying the normalization condition $P_L(\psi_0)+P_R(\psi_0) = 1$, just as the classical case in Eq.~\eqref{unitary-eqn-classical}.
However, at certain special values of $\tau$, this smooth behavior breaks down and the splitting probabilities develop point-wise discontinuous jumps. At these resonant points, the normalization condition is violated, $P_L(\psi_0)+P_R(\psi_0) < 1$, due to the formation of a dark state in the system. We prove that resonances occur when
\begin{align}
& \left(E_{\ell} - E_{\ell'}\right)\tau = 0 \text{ mod } 2\pi ~~\text{and }\braket{\psi_0|\mathbb{D}} \neq 0, \label{main-eqq-3}
\end{align}
where $\ket{E_{\ell}}$ and $\ket{E_{\ell'}}$ are two distinct energy eigenstates that possess the same parity under reflection, and $\ket{\mathbb{D}} \sim  \left[  \sin \left( \pi \ell /(N+1)\right) \ket{E_{\ell'}}-\sin \left(\pi \ell ' /(N+1) \right) \ket{E_{\ell}} \right]$ is the dark state mentioned above.

\item \textit{Flat profile in the Zeno limit:} In the regime of small but finite $\tau$, the splitting probabilities for $H$ in Eq.~\eqref{main-eqq-1} take a simple form determined solely by the projections of the initial state onto the two target states
\begin{align}
P_L(\psi_0) \simeq \frac{1}{2} + \frac{1}{2} \left[ \left| \braket{x_L| \psi_0}  \right|^2-\left| \braket{x_R| \psi_0}  \right|^2 \right],~~P_R(\psi_0) \simeq \frac{1}{2} - \frac{1}{2} \left[ \left| \braket{x_L| \psi_0}  \right|^2-\left| \braket{x_R| \psi_0}  \right|^2 \right]. \label{main-eqq-6}
\end{align}
For example, if the initial condition is localized at a lattice site, $\ket{\psi_0} = \ket{x_0}$, $x_0 \in \{ 1,2, \ldots,N \}$, Eq.~\eqref{main-eqq-6} implies that $P_L(x_0) \simeq P_R(x_0) \simeq 1/2$ for $1<x_0<N$. Remarkably, for all initial positions distinct from the boundaries, both $P_L(x _0)$ and $P_R(x _0)$ are equal to $1/2$, independent of $x_0$ and $N$ (see Fig.~\ref{fig-tau-N15}). This flat profile is in sharp contrast to the classical random walk, where the splitting probabilities vary linearly with $x_0$, as shown in Eq.~\eqref{unitary-eqn-classical}. 

\item \textit{Flat-to-fluctuating transition at $\tau _c $:} In the limit of large system size, a phase-transition-like behavior emerges, in which both $P_L(x _0)$ and $P_R(x _0)$ undergo a sharp qualitative change as functions of $\tau$ (see Fig.~\ref{fig-transition}). This nonanalytic transition, as was also pointed out in the Introduction, occurs at a critical sampling time
\begin{align}
\Delta E~\tau_c = 2 \pi, \label{ajbidhiguqok}
\end{align}
with $\Delta E \approx 4 \gamma$ is the energy bandwidth for $N \gg 1$ for $H$ in Eq.~\eqref{main-eqq-1}. In the regime $0 < \tau < \tau_c$, both splitting probabilities are essentially equal to $1/2$, independent of $\tau$. Above this critical sampling time, $\tau \geq \tau_c$, both $P_L(x _0)$ and $P_R(x _0)$ deviate from $1/2$ and have a fluctuating dependence on $\tau$.

\item \textit{Emergence of oscillations for $\tau \geq \tau _c $:} We also study the dependence of $P_L(x _0)$ and $P_R(x _0)$ on the initial position $x_0$. For $\tau \ge \tau_c$, they exhibit pronounced oscillations as functions of $x_0$ (see Fig.~\ref{fig-tau-N15}). In the large-$N$ limit, the dominant wavenumber characterizing these oscillations is given by
\begin{align}
Q_{\rm max}(\tau)
=
2\sin^{-1}\left(
\frac{r \tau_c}{\tau}
\right),
\qquad r=1,2,\ldots . \label{skpssrs}
\end{align}
This equation gives real $Q_{\rm max}(\tau)$ only for $\tau \geq r\tau_c$, indicating that the minimum sampling time at which oscillations appear is $ \tau_c$ in Eq.~\eqref{ajbidhiguqok}. Close to $\tau_c$, the wavenumber for $r=1$ scales as $Q_{\rm max}(\tau) \simeq \pi -2 \sqrt{2(\tau-\tau_c)/\tau_c}$. We show that the oscillations are governed by phase-matching events between different momentum modes of the quantum walker, which can occur only for $\tau\geq\tau_c$.

\item \textit{Coherent initial state:} We consider the initial state
$\ket{\psi_0}=(\ket{x_0}+e^{i\Phi}\ket{x_0+d_0})/\sqrt{2}$, and study the effect of the phase difference $\Phi$ and the separation $d_0$ on the splitting probability. For the nearest-neighbor hopping
Hamiltonian, we show
\begin{align}
P_L(\psi_0)
=
\begin{cases}
1/2+\mathcal{K}_{d_0}(x_0,\tau)\sin\Phi,
& d_0~\text{odd},\\[6pt]
1/2+\mathcal{K}_{d_0}(x_0,\tau)\cos\Phi,
& d_0~\text{even},
\end{cases} \label{OIC-eq-7-sum}
\end{align}
where $\mathcal{K}_{d_0}(x_0,\tau)$ is given in
Eq.~\eqref{OIC-eq-8} [similarly for $P_R(\psi_0)$]. Thus, the splitting probability has $\sin\Phi$ or $\cos\Phi$ form, depending on whether the separation $d_0$ is odd or even. Interestingly, the separation $d_0$ also plays an important role in determining whether the flat-to-fluctuating transition persists for coherent initial states. As shown in Fig.~\ref{fig-OIC}, for even $d_0$, the transition at $\tau_c$ persists, whereas for odd $d_0$, it becomes less pronounced, particularly for small separations. However, as $d_0$ increases, the flat-to-fluctuating transition is recovered also for odd $d_0$.

\end{enumerate}

In what follows, we provide a rigorous derivation of all the results outlined above. We first formulate the monitored splitting problem in a general setting and then derive the corresponding renewal equations. We subsequently use this framework to obtain the conditions for the special sampling times and to analyze the nonanalytic transition.

\section{General Formalism}
\label{sec-general}
We begin by deriving the quantum renewal relation in Eq.~\eqref{main-eqq-00} for a general discrete-state, continuous-time quantum system with Hamiltonian $H$, general initial condition $\ket{\psi_0}$ and two arbitrary orthogonal detection states $\ket{D_1}$ and $\ket{D_2}$.
 
\subsection{First dual-target measurement at $t=\tau$}
Starting from $\ket{\psi_0}$, the system evolves with the unitary operator in Eq.~\eqref{unitary-eqn} for the duration $\tau$. Before we perform the first measurement, the state of the system at time $t = \left( \tau -\epsilon \right)$ with $\epsilon \to 0^+$ is
\begin{align}
\ket{\psi(\tau^-) }= U(\tau)  \ket{\psi _0}. \label{unitary-eqn-2}
\end{align}
Now the first measurement is performed at the $\ket{D_1}$ target state. The probability of detecting the system in this state is
\begin{align}
\mathcal{G}^{(D_1)}_1 \left(  \psi _0 \right) = \big| \langle D_1 \ket{\psi(\tau^-) } \big|^2 = \big| \bra{D_1}U(\tau) \ket{\psi _0} \big|^2.  \label{unitary-eqn-3}
\end{align}
We use $\mathcal{G}_{n}^{(D{\alpha})}(\psi_0)$ with $\alpha \in \{1,2 \}$ to denote the conditional probability that the first successful detection occurs at target $\ket{D_{\alpha}}$ on the $n$th measurement attempt, given that no detection occurred at either target in the earlier measurements.

%In Eq.~\eqref{unitary-eqn-3}, if the walker is detected at $D_1$, the process terminates. If it is not detected, then according to the standard projection postulate of quantum mechanics \cite{Cohenbook}, the state vector collapses onto the subspace orthogonal to $\ket{D_1}$, i.e., the component along $\ket{D_1}$ is projected out. Therefore, right after a null measurement at $D_1$, the state vector at time $t = \left( \tau +\epsilon \right)$ collapses to
In Eq.~\eqref{unitary-eqn-3}, a successful detection of $\ket{D_1}$ ends the process. However, if no detection occurs, then the projection postulate of quantum mechanics implies that the state $\ket{\psi(\tau^-)}$ is projected onto the subspace orthogonal to $\ket{D_1}$ \cite{Cohenbook}. The component of $\ket{\psi(\tau^-)}$ along $\ket{D_1}$ is removed. Thus, immediately after a failed detection of $\ket{D_1}$, the state of the system at time $t=(\tau+\epsilon)$ becomes
\begin{align}
\ket{ \psi(\tau^+)} = \mathcal{N} \Big( \mathbb{1} - \ket{D_1} \bra{D_1}\Big) \ket{\psi(\tau^-) } =  \mathcal{N} \Big( \mathbb{1} - \ket{D_1} \bra{D_1}\Big) U(\tau)  \ket{\psi _0}, \label{unitary-eqn-4}
\end{align}
where $\mathcal{N} = 1 / \sqrt{1-\mathcal{G}^{(D_1)}_1\left(  \psi _0 \right)} $ gives the correct normalization $\langle \psi(\tau^+ )\ket{ \psi(\tau ^+)} = 1$. With $\ket{ \psi(\tau^+ )} $, we next perform an immediate projective measurement of the second target state $\ket{D_2}$. The probability of detection of this state is 
\begin{align}
\mathcal{G}^{(D_2)}_1\left(  \psi _0 \right) = \big| \langle D_2 \ket{\psi(\tau^+) } \big|^2 = \frac{ \big| \bra{D_2} \Big( \mathbb{1} - \ket{D_1} \bra{D_1}\Big)   U(\tau) \ket{\psi _0} \big|^2}{\left( 1-\mathcal{G}^{(D_1)}_1  \left(  \psi _0 \right) \right)  }.  \label{unitary-eqn-5}
\end{align} 
Remember that this expression represents a conditional probability, conditioned on the failure of the preceding detection attempt of $\ket{D_1}$. Moreover, as the two target states are orthogonal, $\langle D_2 \ket{D_1} = 0$, Eq.~\eqref{unitary-eqn-5} then becomes
\begin{align}
\mathcal{G}^{(D_2)}_1 \left(  \psi _0 \right) = \frac{ \big| \bra{D_2}  U(\tau) \ket{\psi _0} \big|^2}{\left(  1-\mathcal{G}^{(D_1)}_1 \left(  \psi _0 \right) \right) }.  \label{unitary-eqn-6}
\end{align}
%This represents the probability that the walker is detected at $D_2$ at $t=\tau$, given that it was not detected at $D_1$ in the earlier measurement. 
Once again if the detection of $\ket{D_2}$ is successful, we terminate the process. Otherwise the state $\ket{\psi(\tau^+)}$ is projected onto the subspace complementary to  $\ket{D_2}$. Thus, after null detections of both target states during the first measurement cycle, the state becomes
\begin{align}
\ket{\psi(\tau^{++})} = \frac{\Big( \mathbb{1} - \ket{D_2} \bra{D_2} \Big) \ket{\psi \left( \tau^+\right)  }}{\sqrt{1-\mathcal{G}^{(D_2)}_1 \left(  \psi _0 \right)   }} = \frac{\Big( \mathbb{1} - \hat{D} \Big) U(\tau) \ket{\psi_0 }}{\sqrt{\left( 1-\mathcal{G}^{(D_1)}_1  \left(  \psi _0 \right) \right)  \left( 1-\mathcal{G}^{(D_2)}_1  \left( \psi _0 \right) \right)} }, \label{unitary-eqn-7}
\end{align}
where $`++$' in $\ket{\psi(\tau^{++})}$ denotes that two measurements were performed at time $\tau$.  Also, the operator $\hat{D}$ stands for the projection in the detector subspace, and is defined as
\begin{align}
\hat{D} = \ket{D_1} \bra{D_1}+\ket{D_2} \bra{D_2}. \label{unitary-eqn-8}
\end{align}
Eqs.~\eqref{unitary-eqn-4} and \eqref{unitary-eqn-7} describe the state of the system. We will soon turn to the first-detection probabilities.

\subsection{Second dual-target measurement at $t=2 \tau$}
The measurement at the end of the second time interval $t=2\tau$ also proceeds in the same way. Just before this measurement, the state of the system is $\ket{\psi(2 \tau^{-})} = U(\tau) \ket{ \psi \left( \tau^{++} \right) }$. The conditional probability that the system is detected at the $\ket{D_1}$ target, given that it was not detected in either of the target states at the end of the first measurement time interval, is
\begin{align}
\mathcal{G}^{(D_1)}_2 \left( \psi _0 \right) =  \frac{\big| \bra{D_1} U(\tau) \Big(\mathbb{1}-\hat{D} \Big) U(\tau) \ket{ \psi _0}   \big|^2}{\left( 1-\mathcal{G}^{(D_1)}_1 \left(  \psi _0 \right) \right)  \left( 1-\mathcal{G}^{(D_2)}_1 \left(  \psi _0 \right) \right)}. \label{unitary-eqn-9}
\end{align} 
Similarly, the probability of detection of $\ket{D_2}$ at time $t = 2 \tau$, conditioned on the system not being detected in any of the target states before, is
\begin{align}
\mathcal{G}^{(D_2)}_2  \left(  \psi _0 \right)   =  \frac{\big| \bra{D_2} U(\tau) \Big( \mathbb{1}-\hat{D} \Big) U(\tau) \ket{ \psi _0}   \big|^2}{\left( 1-\mathcal{G}^{(D_1)}_2  \left(  \psi _0 \right) \right) \left( 1-\mathcal{G}^{(D_1)}_1  \left(  \psi _0 \right) \right)  \left( 1-\mathcal{G}^{(D_2)}_1  \left(  \psi _0 \right) \right)}. \label{unitary-eqn-10}
\end{align} 
Continuing this iteration for general $n$, we obtain the conditional probability $\mathcal{G}_{n}^{(D_{\alpha})}(\psi _0)~\left( \text{recall }\alpha \in \{1,2 \} \right)$ that the system is first detected at $\ket{D_{\alpha}}$ at the $n$-th measurement attempt as
\begin{align}
& \mathcal{G}^{(D_1)}_n \left(  \psi _0 \right) =  \frac{\big| \bra{D_1} U(\tau) \Big[ \Big( \mathbb{1}-\hat{D} \Big) U(\tau) \Big]^{n-1}\ket{ \psi _0}   \big|^2}{\prod_{j=1}^{n-1}\Big[ \left( 1-\mathcal{G}^{(D_1)}_{j}  \left(  \psi _0 \right) \right)  \left( 1-\mathcal{G}^{(D_2)}_{j} \left(  \psi _0 \right) \right) \Big] }, \label{unitary-eqn-11} \\
& \mathcal{G}^{(D_2)}_n\left( \psi _0 \right)  =  \frac{\big| \bra{D_2} U(\tau) \Big[ \Big( \mathbb{1}-\hat{D} \Big) U(\tau) \Big]^{n-1}\ket{ \psi _0}   \big|^2}{\left( 1-\mathcal{G}^{(D_1)}_{n} (\psi_0)\right)~\prod_{j=1}^{n-1}\Big[  \left( 1-\mathcal{G}^{(D_1)}_{j} \left(  \psi _0 \right) \right)  \left( 1-\mathcal{G}^{(D_2)}_{j}  \left(  \psi _0 \right) \right) \Big]}. \label{unitary-eqn-12} 
\end{align} 
We next use these conditional probabilities to obtain the splitting probabilities of the target states.
\subsection{Quantum splitting probability}
Remember that $F_{n}^{(D_{1})}\left(  \psi _0 \right) $ and $F_{n}^{(D_{2})}\left(  \psi _0 \right) $ are the first-detection probabilities of the $\ket{D_1}$ and $\ket{D_2}$ target states, respectively, on the $n$-th measurement attempt. To obtain them, we recall that both $\mathcal{G}^{(D_1)}_n \left(  \psi _0 \right) $ and $\mathcal{G}^{(D_2)}_n \left(  \psi _0 \right)$ in Eqs.~\eqref{unitary-eqn-11} and \eqref{unitary-eqn-12} are the conditional detection probabilities, conditioned on the fact that no successful detection has occurred in any of the preceding $(n-1)$ measurement attempts. The unconditional first-detection probabilities,  $F_{n}^{(D_{1})}\left(  \psi _0 \right) $ and $F_{n}^{(D_{2})}\left(  \psi _0 \right) $, can be obtained from the conditional probabilities by using the relation
 \begin{align}
\resizebox{0.8\textwidth}{!}{$
\begin{aligned}
& F_{n}^{(D_1)}  \left(  \psi _0 \right) =   \mathcal{G}^{(D_1)}_n \left(  \psi _0 \right) ~\prod_{j=1}^{n-1}\Big[ \left( 1-\mathcal{G}^{(D_1)}_{j} \left(  \psi _0 \right)  \right)  \left( 1-\mathcal{G}^{(D_2)}_{j}  \left( \psi _0 \right)\right) \Big], \\[6pt]
& F_{n}^{(D_2)}  \left(  \psi _0 \right) =   \mathcal{G}^{(D_2)}_n  \left(  \psi _0 \right) ~\left( 1-\mathcal{G}^{(D_1)}_{n} \left( \psi _0 \right) \right)~\prod_{j=1}^{n-1}\Big[  \left( 1-\mathcal{G}^{(D_1)}_{j}  \left(  \psi _0 \right) \right)  \left( 1-\mathcal{G}^{(D_2)}_{j}  \left( \psi _0 \right) \right) \Big].
\end{aligned}
$} \label{unitary-eqn-13hau18}
\end{align}
The product term  $\prod_{j=1}^{n-1} \big[\cdots\big]$ in the first line gives the probability that both $\ket{D_1}$ and $\ket{D_2}$ were not detected in any of the earlier $(n-1)$ measurements. To understand it, let us analyze our measurement process. Starting from $\ket{\psi_0}$, we evolve the system for a time interval $\tau$. At the first attempt, $n=1$, the probability of first detecting the system at $\ket{D_1}$ is $\mathcal{G}^{(D_1)}_1(\psi_0)$ given in Eq.~\eqref{unitary-eqn-9}. In computer simulations, we generate a uniform random number. If this random number is smaller than $\mathcal{G}^{(D_1)}_1(\psi_0)$, the first detection occurs at $\ket{D_1}$. Otherwise, the measurement at $\ket{D_1}$ fails with probability $
\left[ 1-\mathcal{G}^{(D_1)}_1(\psi_0) \right]$. Then, a new random number is generated to test detection of $\ket{D_2}$. This detection is successful with probability $\mathcal{G}^{(D_2)}_1(\psi_0)$. Otherwise both measurements at $n=1$ fail, the probability of which is $\left[1-\mathcal{G}^{(D_1)}_1(\psi_0)\right] \left[1-\mathcal{G}^{(D_2)}_1(\psi_0)\right]$. We then evolve the system for another $\tau$ duration and proceed to the $n=2$ measurement at $\ket{D_1}$ target. In this way, we continue the process until a successful measurement is recorded, which then defines the random first-detection event for this realization. A second realization is obtained by returning to the same initial state $\ket{\psi_0}$ and repeating the measurement protocol with a new set of random numbers. By performing the procedure over many independent realizations, we obtain the first-detection probabilities $F_{n}^{(D_1)}  \left(  \psi _0 \right)$ and $F_{n}^{(D_2)}  \left(  \psi _0 \right)$.

%beginning again with the measurement at $D_1$. Repeating this procedure shows that reaching the $n$th attempt requires failure at both targets during every preceding attempt, which gives the product in Eq.~\eqref{unitary-eqn-13hau18}.

%Multiplying this probability by the conditional probability $\mathcal{G}_n^{(D_\alpha)}(x _0)$ yields the corresponding unconditional probability $F_n^{(D_\alpha)}(x _0)$. In the second line, there is an additional factor of $(1 - \mathcal{G}_n^{(D_1)} (x _0))$. This factor accounts for the failed detection at $D_1$ at the $n$-th attempt, so that the walker gets detected at $D_2$.

%represents the probability that all previous $(n-1)$ measurement attempts were unsuccessful, meaning that the walker was not detected at $D_1$ or $D_2$. This factor is required when converting from the conditional probability $Q_n^{(D_\alpha)}(x _0)$, which assumes no earlier detections, to the unconditional probability $F_n^{(D_\alpha)}(x _0)$, which explicitly accounts for the likelihood of this condition being satisfied. In the first relation, this product alone is sufficient, as we only need the walker to be detected at the $D_1$ site at the $n$-th attempt. In contrast, there is an additional factor of $(1 - Q_n^{(D_1)} (x _0))$ in the second relation. This factor accounts for the failed detection at $D_1$ at the $n$-th attempt, so that the walker gets detected at $D_2$.

Combining Eq.~\eqref{unitary-eqn-13hau18} with Eqs.~\eqref{unitary-eqn-11}-\eqref{unitary-eqn-12} gives us the first-detection probability at the $n$-th measurement in the $\ket{D_{\alpha}}$ target as
\begin{align}
\boxed{\phi_n^{(D_{\alpha})}(\psi_0)= \bra{D_{\alpha}} U(\tau) \Big[ \Big( \mathbb{1}-\hat{D} \Big) U(\tau) \Big]^{n-1}\ket{\psi _0},~~\text{and }   F_{n}^{(D_{\alpha})}\left(  \psi _0 \right) = \big| \phi_n^{(D_{\alpha})}(\psi_0)  \big|^2  } \label{unitary-eqn-13} 
\end{align}
This result was outlined before in Eq.~\eqref{anjiuq1}. One can now plug Eq.~\eqref{unitary-eqn-13} in Eq.~\eqref{anj8y14} to get the splitting probabilities of detecting $\ket{D_1}$ or $\ket{D_2}$ target states.

%Equipped with the first detection probabilities, we sum them over all values of $n$
%\begin{align}
%P_{D_{\alpha}} \left( \psi _0\right) =  \sum _{n=1}^{\infty} F_{n}^{(D_{\alpha})} \left(  \psi _0 \right), \label{unitary-eqn-131} 
%\end{align}

\subsection{Quantum renewal equation}
Starting from Eq.~\eqref{anjiuq1}, we now derive the quantum renewal relation for the first-detection amplitude $\phi _n^{(D_{\alpha})}(\psi_0)$ in Eq.~\eqref{main-eqq-00}. This renewal approach is not only instructive from a theoretical perspective, but also serves to simplify the solution, as we will demonstrate in the next section.

 %We now derive a renewal equation for $\phi _{n}^{\left(D_{\alpha} \right)}\left( x _0 \right)$. 

Let us look at $\phi _{n}^{\left(D_{\alpha} \right)}\left( \psi _0 \right)$ for the first few values of $n$. Using Eq.~\eqref{anjiuq1} for $n=1$, we find 
\begin{align}
\phi _{1}^{\left(D_{\alpha} \right)} \left( \psi _0 \right) =  \bra{D_{\alpha}}   U(\tau) \ket{\psi _0}, \nonumber
\end{align}
and for $n=2$
\begin{align}
\phi _{2}^{\left(D_{\alpha} \right)} \left( \psi _0 \right) = \bra{D_{\alpha}}   U(2  \tau) \ket{\psi _0} -  \phi _{1}^{\left(D_{1} \right)} \left( \psi _0 \right) \bra{D_{\alpha}} U(\tau) \ket{D_{1}} -  \phi _{1}^{\left(D_{2} \right)} \left( \psi _0 \right) \bra{D_{\alpha}} U(\tau) \ket{D_{2}}.  \nonumber
\end{align}
One can clearly see an emerging renewal structure. In \aref{appen-sec-renewal}, we use induction to show that the renewal structure is true even for general $n$
\begin{align}
\boxed{\phi _{n}^{\left(D_{\alpha} \right)}  \left( \psi _0 \right)  = \bra{D_{\alpha}}   U(n \tau) \ket{\psi _0} - \sum _{\beta=1}^{2} \sum _{l=1}^{n-1} \phi _{l}^{\left(D_{\beta} \right)}  \left( \psi_0 \right)  \bra{D_{\alpha}} U\left( (n-l)\tau \right) \ket{ D_{\beta} }.}  \label{unitary-eqn-17}  
\end{align}	
This relation was quoted above in Eq.~\eqref{main-eqq-00}.
The amplitude $\phi_{n}^{(D_{\alpha})} \left( \psi _0 \right) $ of detecting the system in $ \ket{D_{\alpha}}$ target at the $n$-th attempt consists of two contributions: the direct propagation amplitude $\bra{D_{\alpha}} U(n\tau)\ket{\psi_0}$ from the initial state $\ket{\psi _0  }$ to $ \ket{D_{\alpha}}$ in time $n\tau$, and a set of correction terms subtracting all histories where the system was already detected earlier at $ \ket{ D_{\beta}}$ (with $\beta \in \{ 1,2 \}$) at some earlier step $l<n$, and then propagated from $ \ket{ D_{\beta} }$ to $\ket{D_{\alpha}}$ in the remaining time $(n-l)\tau$. This recursive relation captures the essential renewal nature of the quantum splitting problem. Later, we will exploit its convolution form using generating-function techniques to calculate the splitting probabilities.

\subsection{Splitting probability and the order of measurements}
Recall that in our detection protocol explained in Sec.~\ref{sec-protocol}, after each time duration $\tau$, a projective measurement is first performed at $\ket{D_1}$ and, if no detection occurs, subsequently at $\ket{D_2}$. One may equally consider the opposite ordering, where $\ket{D_2}$ is measured first and $\ket{D_1}$ second. The renewal equation \eqref{unitary-eqn-17} is, however, manifestly invariant under the interchange 
$\ket{D_1} \leftrightarrow \ket{D_2}$, indicating that it is independent of the measurement order. As mentioned before, this invariance is due to the orthogonality of $\ket{D_1}$ and $\ket{D_2}$, and the fact that 
the two measurements are performed immediately one after the other. 

\section{Generating function}
\label{sec-GF}
Given the convolution structure of $\phi _{n}^{\left(D_{\alpha} \right)} \left(  \psi _0 \right) $ in Eq.~\eqref{unitary-eqn-17}, one may solve it using the generating function approach. For this, we perform the Z-transform (discrete Laplace transform) as
\begin{align}
\widetilde{\phi} _{D_{\alpha}}(z) = \sum _{n=1}^{\infty} z^{n} \phi_n^{\left(  {D_{\alpha}}  \right) }\left(  \psi _0 \right). \label{unitary-eqn-18} 
\end{align}
We have not explicitly shown the initial state dependence in the generating function $\widetilde{\phi} _{D_{\alpha}}(z)$ to keep the notation compact. Z-transforming Eq.~\eqref{unitary-eqn-17} gives
\begin{align}
\begin{pmatrix}
1+\widetilde{U}_{11}(z) & \widetilde{U}_{12}(z) \\
\widetilde{U}_{21}(z)& 1+\widetilde{U}_{22}(z) 
\end{pmatrix}
\begin{pmatrix}
\widetilde{\phi} _{D_{1}}(z) \\
\widetilde{\phi} _{D_{2}}(z)
\end{pmatrix}
 = \begin{pmatrix}
 \bra{D_1} \widetilde{U}(z) \ket{ \psi _0} \\
 \bra{D_2} \widetilde{U}(z) \ket{ \psi _0}
\end{pmatrix}, \label{unitary-eqn-19} 
\end{align}
where we have used the notation
\begin{align}
& \widetilde{U}(z) = \sum _{n=1}^{\infty} z^n U(n \tau) = \frac{z e^{-iH \tau}}{1-z e^{-iH \tau}}, \label{unitary-eqn-20}  \\
& \widetilde{U}_{\alpha \beta}(z) = \bra{D_{\alpha}} \widetilde{U}(z) \ket{ D_{\beta}},~~\text{with } \alpha, \beta \in \{ 1,2\}. \label{unitary-eqn-21} 
\end{align}
The idea is to solve Eq.~\eqref{unitary-eqn-19} to obtain generating functions $\widetilde{\phi} _{D_{1}}(z)$ and $\widetilde{\phi} _{D_{2}}(z)$. The splitting probability $P_{ D_{\alpha}  }(\psi_0)$ in Eq.~\eqref{anj8y14} can be expressed in terms of these generating functions. To see this, we first write $P_{ D_{\alpha}  }(\psi_0)$ as
\begin{align}
P_{D_{\alpha}}\left(\psi _0\right) & =\sum _{n=1}^{\infty} \big| \phi _n ^{\left( D_{\alpha} \right) }  \big|^2  = \sum _{n=1}^{\infty} \sum _{l=1}^{\infty} \phi _n ^{ \left(  D_{\alpha} \right)} \left[ \phi _l ^{\left(  D_{\alpha} \right) } \right]^* ~\delta _{n,l}, \label{unitary-eqn-2i2} 
\end{align}
where $*$ stands for the complex conjugate and $\delta_{n,l}$ is the Kronecker delta that takes value one if $n=l$, and zero otherwise. Using
\begin{align}
\delta _{n,l} = \frac{1}{2 \pi} \int _{0}^{2 \pi} d \omegaa ~e^{i \omegaa(n-l)},
\end{align}
along with the definition of the generating function in Eq.~\eqref{unitary-eqn-18}, one can simplify Eq.~\eqref{unitary-eqn-2i2} to
\begin{align}
\boxed{
\begin{aligned}
& P_{D_{\alpha}}\left(\psi_0\right)
=
\frac{1}{2\pi}
\int_0^{2\pi}
d\omegaa\,
\left|
\widetilde{\phi}_{D_{\alpha}}
\left(e^{i\omegaa}\right)
\right|^2,
\\[1ex]
\text{with } &\widetilde{\phi}_{D_{\alpha}}(z)
=
\frac{
\left[1+\widetilde{U}_{\bar{\alpha}\bar{\alpha}}(z)\right]
\bra{D_{\alpha}}\widetilde{U}(z)\ket{\psi_0}
-
\widetilde{U}_{\alpha\bar{\alpha}}(z)
\bra{D_{\bar{\alpha}}}\widetilde{U}(z)\ket{\psi_0}
}{
\left[1+\widetilde{U}_{11}(z)\right]
\left[1+\widetilde{U}_{22}(z)\right]
-
\widetilde{U}_{12}(z)\widetilde{U}_{21}(z)
},
\end{aligned}
}
\label{unitary-eqn-22}
\end{align}
where recall $\bar{\alpha}=2$ if $\alpha=1$ and vice versa.  Eq.~\eqref{unitary-eqn-22} provides us the exact expression of the splitting probability for arbitrary Hamiltonian $H$, orthogonal target states $\ket{D_{1}}$, $\ket{D_{2}}$, the initial condition $\ket{\psi_0}$ and the sampling time $\tau$.

%Once we have $ \widetilde{\phi} _{D_{\alpha}} \left( z \right)$ from Eq.~\eqref{unitary-eqn-19}, we plug it in Eq.~\eqref{unitary-eqn-22} with $z=e^{i \omega} $, and compute the splitting probabilities for the quantum walker.
% We now employ this framework to analyze a walker on a tight-binding chain with nearest-neighbor hopping.

\subsection{Mapping of the splitting problem to two separate single-target detection problems}
\label{sec-saubcb}
In the summary section~\ref{summary-sec}, we noted that for a finite one-dimensional lattice chain whose Hamiltonian possesses reflection symmetry about the center of the chain, and with target states localized at the left and the right endpoints of the chain, the dual-target splitting problem can be mapped onto a pair of single-target first detection problems, each with its own auxiliary target. In the Letter \cite{myletter}, we derived this mapping using an induction method. Here we provide an alternate derivation based on our renewal theory.

The reflection/parity operator is defined as
\begin{align}
\mathcal{P} \ket{x} = \ket{N+1-x}, \label{new-map-eqq1}
\end{align}
corresponding to reflection about the center of the chain. For Hamiltonians satisfying the parity symmetry, the commutation between $H$ and $\mathcal{P}$ vanishes
\begin{align}
[H ,\mathcal{P}] = 0, \implies \mathcal{P}H \mathcal{P} =H. \label{new-map-eqq2}
\end{align}
To obtain the second relation, we have used the identity $\mathcal{P}^{-1} = \mathcal{P}$. Eq.~\eqref{new-map-eqq2} also implies $\mathcal{P} \widetilde{U}(z) \mathcal{P} = \widetilde{U}(z)$. We also fix the detection states localized at the endpoints of the chain, $\ket{D_1}=\ket{x_L}=\ket{1}$ and $\ket{D_2}=\ket{x_R}=\ket{N}$.  Eq.~\eqref{unitary-eqn-21} now gives
\begin{align}
& \widetilde{U}_{11}(z)=\braket{1|\widetilde{U}(z)|1}=\braket{N|\mathcal{P}\widetilde{U}(z) \mathcal{P}|N} = \widetilde{U}_{22}(z), \\
& \widetilde{U}_{12}(z)=\braket{1|\widetilde{U}(z)|N}=\braket{N|\mathcal{P}\widetilde{U}(z) \mathcal{P}|1} = \widetilde{U}_{21}(z).
\end{align}
The generating functions in Eq.~\eqref{unitary-eqn-22} then simplify to
\begin{align}
& \widetilde{\phi}_{x_L}(z) = \frac{1}{\sqrt{2}} \left[ \frac{\langle d_+ \big| \widetilde{U}(z) \big| \psi_0 \rangle }{1+\widetilde{U}_{11}(z)+ \widetilde{U}_{12}(z)} ~+~ \frac{\langle d_- \big| \widetilde{U}(z) \big| \psi_0 \rangle }{1+\widetilde{U}_{11}(z)- \widetilde{U}_{12}(z)} \right], \label{neww-eg-eq-6} \\
& \widetilde{\phi}_{x_R}(z) = \frac{1}{\sqrt{2}} \left[ \frac{\langle d_+ \big| \widetilde{U}(z) \big| \psi_0 \rangle }{1+\widetilde{U}_{11}(z)+ \widetilde{U}_{12}(z)} ~-~ \frac{\langle d_- \big| \widetilde{U}(z) \big| \psi_0 \rangle }{1+\widetilde{U}_{11}(z)- \widetilde{U}_{12}(z)} \right], \label{neww-eg-eq-7}
\end{align}
where
\begin{align}
& \ket{d_{\pm}} = \frac{\left( \ket{1} \pm \ket{N} \right)}{\sqrt{2}}.
\end{align}
Notice that $\mathcal{P} \ket{d_{\pm}} = \pm  \ket{d_{\pm}}$, which gives $\braket{d_+|\widetilde{U}(z)|d_-} = \braket{d_-|\widetilde{U}(z)|d_+}=0$. With this result, we obtain
\begin{align}
 \widetilde{U}_{11}(z)+ \widetilde{U}_{12}(z)=\braket{d_+ | \widetilde{U}(z) | d_+},~~\text{and }  \widetilde{U}_{11}(z)- \widetilde{U}_{12}(z)= \braket{d_-| \widetilde{U}(z) | d_-} .
\end{align}
These relations allow us to recast Eqs.~\eqref{neww-eg-eq-6}-\eqref{neww-eg-eq-7} as
\begin{align}
& \widetilde{\phi}_{x_L}(z) = \frac{1}{\sqrt{2}} \left[ \frac{\langle d_+ \big| \widetilde{U}(z) \big| \psi_0 \rangle }{1+\braket{d_+ | \widetilde{U}(z) | d_+}} ~+~ \frac{\langle d_- \big| \widetilde{U}(z) \big| \psi_0 \rangle }{1+ \braket{d_- | \widetilde{U}(z) | d_-}  } \right], \label{neww-eg-eq-6n1} \\
& \widetilde{\phi}_{x_R}(z) = \frac{1}{\sqrt{2}} \left[ \frac{\langle d_+ \big| \widetilde{U}(z) \big| \psi_0 \rangle }{1+\braket{d_+ | \widetilde{U}(z) | d_+}} ~-~ \frac{\langle d_- \big| \widetilde{U}(z) \big| \psi_0 \rangle }{1+ \braket{d_- | \widetilde{U}(z) | d_-}  } \right]. \label{neww-eg-eq-7n1}
\end{align}
Following earlier studies \cite{QW6, QW16}, we identify the terms
\begin{align}
\widetilde{\chi} _{\pm}(z) = \frac{\langle d_{\pm} \big| \widetilde{U}(z) \big| \psi_0 \rangle }{1+\braket{d_{\pm} | \widetilde{U}(z) | d_{\pm}}} , \label{aux-eq-1}
\end{align}
on the right hand side of Eqs.~\eqref{neww-eg-eq-6n1} and \eqref{neww-eg-eq-7n1} as the generating functions associated with two different single-target detection problems, that of the first detection of the quantum walker on the target state $\ket{d_{\pm}}$. These single-target detection amplitudes in the $n$-domain are denoted by $\chi ^{(\pm)}_n(\psi_0)$ and are given by Eq.~\eqref{shjay781} \cite{QW6, QW16, QW7, QW8}. Recasting Eqs.~\eqref{neww-eg-eq-6n1} and \eqref{neww-eg-eq-7n1}
in the $n$-domain, we obtain 
\begin{align}
\boxed{
\begin{aligned}
\phi _{n}^{\left(x_L \right)} \left( \psi _0 \right)
&=
\frac{1}{\sqrt{2}}
\left[
\chi _{n}^{\left(+\right)} \left( \psi _0 \right)
+
\chi _{n}^{(-)} \left( \psi _0 \right)
\right],\\
\phi _{n}^{\left(x_R \right)} \left( \psi _0 \right)
&=
\frac{1}{\sqrt{2}}
\left[
\chi _{n}^{\left(+\right)} \left( \psi _0 \right)
-
\chi _{n}^{(-)} \left( \psi _0 \right)
\right].
\end{aligned}
}
\label{main-eqq-101-po}
\end{align}
Eq.~\eqref{main-eqq-101-po} is the mapping announced previously in Eq.~\eqref{main-eqq-101}. Thus, the detection amplitudes for the splitting problem with two boundary targets can be expressed as linear combinations of the amplitudes $\chi ^{(\pm)}_n(\psi_0)$, 
associated with two separate single-target problems: one involving detection at the auxiliary target state $\ket{d_+}$, and the other involving detection at another target state $\ket{d_-}$. A useful consequence of this mapping is that methods developed for the single-target first-detection problem can be applied directly to the present dual-target splitting problem. Moreover, this mapping is purely quantum and has no analogue in classical systems, since $\ket{d_{\pm}}$ cannot be interpreted as physical states in the classical probabilistic sense. Indeed, the detection of a particle in a coherent superposition of the two detector states, with a relative phase between them, as in $\ket{d_\pm}$, has no classical analogue.

The mapping in Eq.~\eqref{main-eqq-101-po} also provides a useful physical interpretation of the splitting process. The detection amplitudes $\phi _{n}^{\left(x_L \right)} \left( \psi _0 \right)$ and $\phi _{n}^{\left(x_R \right)} \left( \psi _0 \right)$ can be viewed as an interference effect between a pair of single-target detection processes. As we show later, this interference is responsible for several characteristic features of the splitting problem, including the flat-to-fluctuating transition and the emergence of spatial oscillations in the splitting probabilities above a critical sampling time.

\subsection{Single-target detection probability}
For later use, we evaluate the total detection probabilities associated with the auxiliary single-target detection problems. It can be expressed in terms of the generating functions as \cite{QW11}
\begin{align}
\rho _{\pm} \left( \psi _0 \right) = \frac{1}{2 \pi} \int_0^{2 \pi} d \omegaa ~ \big|  \widetilde{\chi}_{\pm} \left(e^{i \omegaa} \right)   \big|^2. \label{new-pseb-eq-1}
\end{align}
Unlike in the classical random walk, these probabilities can be less than one in the monitored quantum setting even for finite systems \cite{QW16,QW11}. The integration in Eq.~\eqref{new-pseb-eq-1} was carried out in terms of the eigenstates of the unitary operator $U(\tau)$ in Eq.~\eqref{unitary-eqn} \cite{QW16}. The eigenvalues of $U(\tau)$ take the form 
$e^{-i\lambda_{\ell}}$, where 
\begin{align}
\lambda_{\ell} = E_{\ell} \tau \text{ mod }2 \pi, \label{neww-eg-eq-1q}
\end{align}
are the so-called quasienergies. Recall that $H\ket{E_\ell}=E_\ell\ket{E_\ell}$, with the index $\ell$ labeling all energy eigenstates.

For generic sampling times, each distinct energy eigenvalue $E_{\ell}$ produces a distinct phase factor, and therefore a distinct quasienergy $\lambda_{\ell}$. Therefore, the number of distinct quasienergies is equal to the number of distinct energy eigenvalues for most sampling times. At special values of $\tau$, however, two or more different energies $\{E_{\ell}, E_{\ell'} \}$ can lead to the same phase factor, $e^{-iE_{\ell}\tau} = e^{-iE_{\ell'}\tau}$, and hence the same quasienergy. Consequently, the number of distinct quasienergies is reduced relative to the number of distinct energies.
%As an example, Fig.~\ref{fig-phase-merge}(a) shows that for an $N=5$ chain with $\tau=1$ and $\gamma=1$, all five phase factors are distinct, yielding five distinct quasienergies. At special values of $\tau$, however, two or more different energies $\{E_k, E_{\ell} \}$ can lead to the same phase factor, $e^{-iE_k\tau} = e^{-iE_{\ell}\tau}$, and hence the same quasienergy. This is illustrated again in Fig.~\ref{fig-phase-merge}(b) for $\gamma=1$ and $\tau=2\pi/\sqrt{3}$, where three phase factors merge, leaving only three distinct quasienergies. For special sampling times, we therefore obtain a degeneracy in the quasienergy spectrum. 
Let us assume that $g_{\ell}$ is the degeneracy associated with the $\lambda_{\ell}$ quasienergy and we denote these degenerate states as $\left\{ \ket{E_{{\ell},m}} \right\}$, with $m=1,2,\ldots,g_{\ell}$. These states satisfy $U(\tau) \ket{E_{{\ell},m}}=e^{-iE_{\ell} \tau} \ket{E_{{\ell},m}}$. Then, the unitary operator can be expressed as
\begin{align}
U(\tau) =  \sum _{{\ell}} e^{-i \lambda _{\ell}}~ L_{\ell}, ~~\text{with } L_{\ell} = \sum _{m=1}^{g_{\ell}} \ket{E_{{\ell}, m}} \bra{E_{\ell, m}}. \label{neww-eg-eq-3}
\end{align}
Here, the summation runs over all distinct values of $\lambda _{\ell}$. Also, note that both the quasienergies $\lambda_{\ell}$, and the corresponding degeneracies $g_{\ell}$ are, in general, functions of $\tau$, although we do not make this dependence explicit in the notation.

In the quasienergy basis, $\rho _{\pm} \left(  \psi _0 \right) $ take the form \cite{QW16}
\begin{align}
\rho _{\pm}( \psi _0) = \sump_{{\ell}} 
      \frac{\big| \sum_{{\ell}=1}^{g_{\ell}} 
         \braket{d_{\pm} \big| E_{{\ell},m}} \braket{E_{{\ell},m} \big|  \psi _0} \big|^2}
           { \sum_{m=1}^{g_{\ell}} \big| \braket{d_{\pm} \big| E_{{\ell},m}} \big|^2}. \label{new-pseb-eq-2}
\end{align}
For $\rho _{+}( \psi _0) $, the primed summation is restricted only to those distinct quasienergy states that have a non-vanishing overlap with $\ket{d_{+}}$ [and similarly for $\rho _{-}( \psi _0) $].

\subsection{Splitting probability}
We now turn to the splitting probability for the quantum walker on a one-dimensional chain, with parity symmetric Hamiltonian and target states localized at the left and the right endpoints, $\ket{D_1}=\ket{x_L}=\ket{1}$ and $\ket{D_2}=\ket{x_R}=\ket{N}$. As above, we denote these probabilities by $P_L( \psi _0)$ and $P_R( \psi _0)$, corresponding to the target states being the left and the right endpoints. Using our mapping~\eqref{main-eqq-101} in Eq.~\eqref{unitary-eqn-22}, we get
\begin{align}
\boxed{
\begin{aligned}
P_L(\psi_0)
&=
\frac{1}{2}\rho_+(\psi_0)
+
\frac{1}{2}\rho_-(\psi_0)
+
\zeta_{+-}(\psi_0),\\
P_R(\psi_0)
&=
\frac{1}{2}\rho_+(\psi_0)
+
\frac{1}{2}\rho_-(\psi_0)
-
\zeta_{+-}(\psi_0),
\end{aligned}
}
\label{neww-eg-eq-17}
\end{align}
where $\rho_{\pm}(\psi_0)$ are given in Eq.~\eqref{new-pseb-eq-2} and $\zeta_{+-}( \psi_0)$ represents the interference effect between the two detection amplitudes associated with 
$\ket{d_+}$ and $\ket{d_-}$. Its expression reads
\begin{align}
\zeta _{+  -}( \psi _0) = \text{Re} \left[  \frac{1}{2\pi} \int _0 ^{2 \pi} d \omegaa ~\widetilde{\chi}_+\left( e^{i \omegaa} \right)   ~\left(  \widetilde{\chi}_-\left( e^{i \omegaa} \right) \right)^*   \right], \label{aux-eq-2}
\end{align}
where $\widetilde{\chi} _{\pm}(z)$ are given in Eq.~\eqref{aux-eq-1}. Also, $\text{Re}[\ldots]$ denotes the real part of $[\ldots]$, while $(\ldots)^*$ indicates the complex conjugate of $(\ldots)$.

Eqs.~\eqref{neww-eg-eq-17}, together with Eqs.~\eqref{new-pseb-eq-2} and \eqref{aux-eq-2}, determine the splitting probabilities for any one-dimensional tight-binding Hamiltonian possessing reflection symmetry with initial condition $\ket{\psi_0}$, and projective measurements performed at the boundary sites at regular stroboscopic time intervals $\tau$. We next examine these expressions for the nearest-neighbor hopping Hamiltonian.

%Two distinct scenarios emerge depending on whether the quasi-energy spectrum is non-degenerate or contains degeneracies, leading respectively to smooth or discontinuous behavior in the splitting probabilities. We first analyze the non-degenerate case in detail, followed by the degenerate case.
%\bluew{\textbf{Check the generality of the mapping}}

\section{Nearest-neighbor hopping Hamiltonian}
\label{sec-tight}
We now apply our general results to the nearest-neighbor hopping Hamiltonian in Eq.~\eqref{main-eqq-1}, which we rewrite here for continuity
\begin{align}
H = - \gamma \sum _{x=1}^{N-1} \Big[  \ket{x} \bra{x+1}+\ket{x+1} \bra{x} \Big], \quad x \in \{ 1,2, \ldots,N\}. \label{unitary-eqn-23} 
\end{align}
We briefly recall basic properties of this Hamiltonian. Observe that this Hamiltonian satisfies the parity symmetry, $[H, \mathcal{P}] = 0$ with $\mathcal{P}$ in Eq.~\eqref{new-map-eqq1}. The eigenvalues and the eigenstates, denoted by $E_{\ell}$ and $\ket{E_{\ell}}$ for $\ell \in \{1,2,\ldots,N\}$, are
\begin{align}
E_{\ell}=-2 \gamma \cos \left( k_{\ell} \right),~~\ket{E_{\ell}} = \sqrt{\frac{2}{N+1}}~ \sum _{x=1}^{N} ~\sin \left( k_{\ell} x \right)~\ket{x},~~k_{\ell} = \frac{\pi \ell}{N+1}. \label{neww-eg-eq-1}
\end{align}
Equivalently, the site basis state can be expanded in the energy eigenstates as
\begin{align}
\ket{x} = \sqrt{\frac{2}{N+1}}~ \sum _{\ell=1}^{N} ~\sin \left( k_{\ell} x \right)~\ket{E_\ell}, \label{neww-eg-eq-2}
\end{align}
while the auxiliary single-target states are
\begin{align}
& \ket{d_{+}}  =  \frac{2}{\sqrt{N+1}} \sum_{\substack{\ell=1 \\ \ell \in  \,\text{odd}}}^{N} \sin \left( k_{\ell} x \right)~\ket{E_\ell}, \label{neww-eg-eq-8}  \\
& \ket{d_{-}} =  \frac{2}{\sqrt{N+1}} \sum_{\substack{\ell=1 \\ \ell \in  \,\text{even}}}^{N} \sin \left( k_{\ell} x \right)~\ket{E_\ell}.  \label{neww-eg-eq-82}
\end{align}
The summation in the first line involves only odd values of $\ell$, while in the second line, it only involves even values. We next use our result in Eqs.~\eqref{neww-eg-eq-17} to compute the splitting probabilities of this model.
\begin{figure}[t]
	\centering
	\includegraphics[scale=0.25]{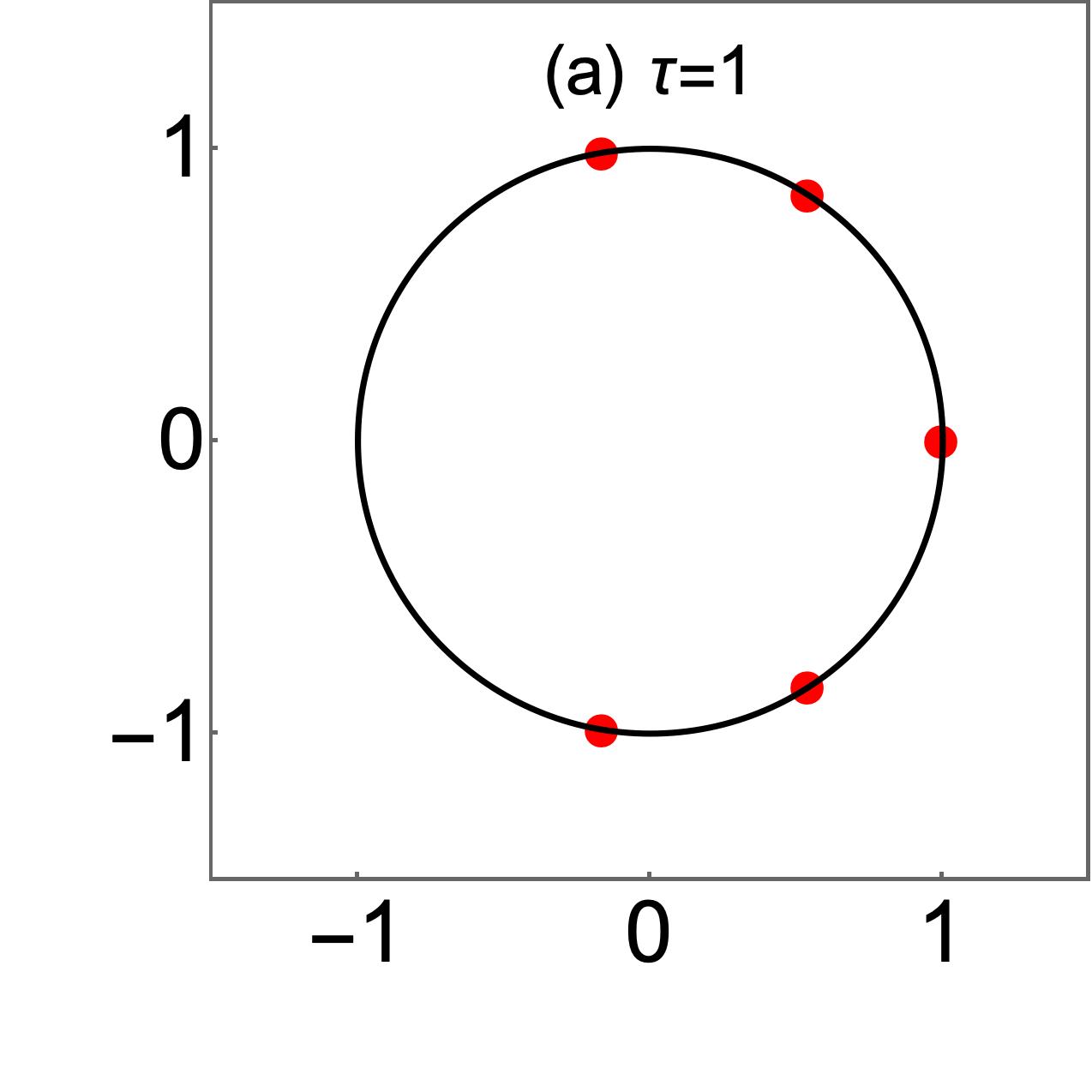} \hspace{1.0cm}
	\includegraphics[scale=0.25]{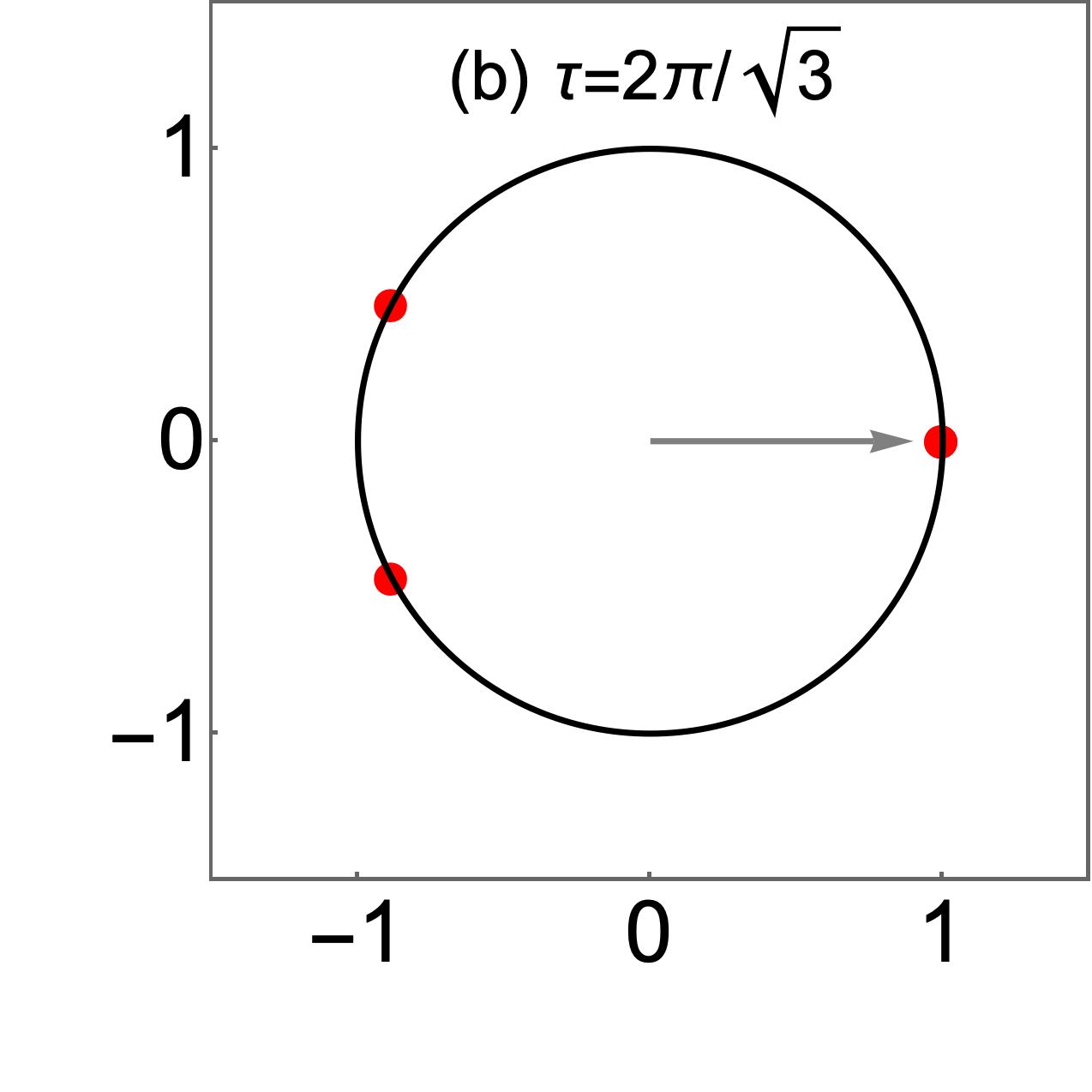} 
	\caption{The nearest-neighbor hopping Hamiltonian in Eq.~\eqref{unitary-eqn-23}, with $N=5$, has five eigenvlaues given in Eq.~\eqref{neww-eg-eq-1}, namely $\{E_1=-\sqrt{3} \gamma, E_2=-\gamma, E_3=0, E_4=\gamma, E_5=\sqrt{3} \gamma \}$. These eigenvalues generate five distinct phases $\{ e^{-iE_\ell\tau} \}$ shown in red in panel~(a) for $\gamma =1$ and $\tau=1$. Consequently, we get five distinct quasienergies for this case. By contrast, for  $\gamma =1$ and $\tau = 2 \pi / \sqrt{3}$, we have merging of three phases, $e^{-i E_1 \tau} =e^{-i E_3 \tau} =e^{-i E_5 \tau} =1$, indicated by the arrow in panel~(b). As a result, there are only three distinct quasienergies for $\tau = 2 \pi / \sqrt{3}$. As explained in the paper, the merging of phases is responsible for the discontinuous-jumps in the splitting probabilities (see Fig.~\ref{fig-Ptau}) and gives rise to the resonance condition in Eq.~\eqref{main-eqq-3}. }
	\label{fig-phase-merge}
\end{figure}

\subsection{Generic sampling times and non-degenerate $\lambda_\ell$}
\label{sec-smooth}
For generic sampling times, we explained before that the number of distinct quasienergies is equal to the number of distinct energies.  For $H$ in Eq.~\eqref{unitary-eqn-23}, all energies in Eq.~\eqref{neww-eg-eq-1} are non-degenerate, and consequently, all $N$ quasienergies are distinct. For example, in Fig.~\ref{fig-phase-merge}(a), we have shown that for $N=5$, $\tau=1$ and $\gamma=1$, all five phase factors are distinct, yielding five distinct quasienergies. At special values of $\tau$, however, two or more different energies $\{ E_{\ell}, E_{\ell'} \}$ can lead to the same phase factor, $e^{-iE_\ell\tau} = e^{-iE_{\ell'}\tau}$, and hence the same quasienergy.
This is illustrated again in Fig.~\ref{fig-phase-merge}(b) for $\gamma=1$ and $\tau=2\pi/\sqrt{3}$, where three phase factors merge, leaving only three distinct quasienergies. For such special sampling times, we therefore obtain a degeneracy in the quasienergy spectrum. We first consider the nondegenerate case and return to the degenerate case later.

%For generic $\tau$, the number of distinct quasi-energies $\lambda_k$, as explained before, is the same as the number of distinct energies $E_k$; see Fig.~\ref{fig-phase-merge}(a). 
%Only for certain special values of $\tau$, two or more quasi-energies coincide, resulting in fewer distinct quasi-energy levels, a feature that will be rigorously analyzed later. For this case, the splitting probabilities can be simplified compared to their general expressions in Eq.~\eqref{neww-eg-eq-17}.

%In this section, we investigate the generic-$\tau$ case, for which the splitting probabilities in Eq.~\eqref{neww-eg-eq-17} can be further simplified.

%As mentioned before, all $N$ energy levels are non-degenerate, implying that for general $\tau$, there are $N$ distinct quasi-energies, and so 

For generic $\tau$, the quasienergy degeneracy factor appearing in the expressions of $\rho _{\pm}(x _0)$ in Eq.~\eqref{new-pseb-eq-2} is unity, \textit{i.e.,} $g_\ell = 1$ for all $\ell$. So the quasienergy state $\ket{E_{\ell,1}}$ can be simply written as $\ket{E_\ell}$. This yields
\begin{align}
\rho _{\pm}( \psi_0) = \sum_{\ell=1}^{N}{}^{\prime}  \frac{\big| 
         \braket{d_{\pm} \big| E_{\ell}} \braket{E_{\ell} \big|  \psi _0} \big|^2}
           {  \big| \braket{d_{\pm} \big| E_{\ell}} \big|^2}.   \label{aux-eq-3}
\end{align}
Recall that the primed summation is restricted to only those values of $\ell$ for which the overlap 
$\braket{d_{\pm} \big| E_{\ell}}$ is non-zero. 
Using Eqs.~\eqref{neww-eg-eq-8} and \eqref{neww-eg-eq-82}, we note that 
$\braket{d_{+} \big| E_{\ell}} = 0$ when $\ell$ is even, 
while $\braket{d_{-} \big| E_{\ell}} = 0$ when $\ell$ is odd. Eq.~\eqref{aux-eq-3} then becomes
\begin{align}
\rho _+(  \psi _0) = \sum_{\substack{\ell=1 \\ \ell \in  \,\text{odd}}}^{N} \big| \braket{E_\ell |  \psi _0} \big|^2,~~\rho _-( \psi _0) = \sum_{\substack{\ell=1 \\ \ell \in  \,\text{even}}}^{N} \big| \braket{E_\ell |  \psi _0} \big|^2. \label{aux-eq-4}
\end{align}
%Similarly, the auxiliary generating functions $\widetilde{\chi} _{\pm}(z) $, which determine the interference term $\zeta _{+  -}(x _0)$ in Eq.~\eqref{aux-eq-2}, are
Plugging them in Eq.~\eqref{neww-eg-eq-17}
\begin{align}
\boxed{
\begin{aligned}
P_L(\psi_0)
&=
\frac{1}{2}
+
\zeta_{+-}(\psi_0),\\
P_R(\psi_0)
&=
\frac{1}{2}
-
\zeta_{+-}(\psi_0),
\end{aligned}
}
\label{aux-eq-5}
\end{align}
where recall that $\zeta_{+-}( \psi_0)$ represents the interference effect between the two auxiliary single-target detection processes. Using Eq.~\eqref{aux-eq-2}, it can be written as
\begin{align}
\zeta _{+-}( \psi_0) = \text{Re} \left[ \oint _{|z|=1} \mathcal{I}( \psi_0, z)~dz \right], \label{pole-eq-1}
\end{align}
involving the real part of a contour integral of $\mathcal{I}(\psi_0,z)$ around the unit circle in the complex plane. The integrand $\mathcal{I}(\psi_0,z) = \widetilde{\chi}_+(z)   \widetilde{\chi}_- \left( 1/z \right)^* /  2 \pi i z$ can be explicitly written using Eq.~\eqref{aux-eq-1} as
\begin{align}
\mathcal{I}(\psi_0, z) = \frac{N+1}{8 \pi  i z} &  \left[ \frac{\sum _{\ell \in ~\rm{odd}} \sin (\omega _\ell)  \braket{E_\ell| \psi_0} \left( \frac{e^{2 i \gamma \tau \cos (\omega _\ell)}}{1-z ~e^{2 i \gamma \tau \cos (\omega _\ell)}}  \right)  }{   \sum _{\ell \in ~\rm{odd}} \sin ^2 (\omega _\ell)  \left( \frac{1}{1-z ~e^{2 i \gamma \tau \cos (\omega _\ell)}}  \right)      } \right]  \nonumber \\
& ~~~~~~~\times \left[ \frac{\sum _{\ell \in ~\rm{even}} \sin (\omega _\ell) \braket{\psi_0|E_\ell} \left( \frac{e^{-2 i \gamma \tau \cos (\omega _\ell)}}{z- ~e^{-2 i \gamma \tau \cos (\omega _\ell)}}  \right)  }{   \sum _{\ell \in ~\rm{even}} \sin^2 (\omega _\ell)  \left( \frac{1}{z- ~e^{-2 i \gamma \tau \cos (\omega _\ell)}}  \right)      } \right], \label{pole-eq-2}
\end{align}
where remember $\omega_\ell = \pi \ell/(N+1)$. Once again, we emphasize that Eqs.~\eqref{aux-eq-5}-\eqref{pole-eq-2} are valid for the nearest-neighbor hopping Hamiltonian at generic sampling times, and the special sampling times at which quasienergy degeneracies occur will be treated separately later. Since $\zeta_{+-}(\psi_0)$ depends on both the sampling time and the initial condition, it gives rise to the rich dependence of $P_L(\psi_0)$ and $P_R(\psi_0)$ on $\tau$ and $\ket{\psi_0}$, which we discuss next.

\section{Universal Zeno features}
\label{sec-Zeno}
For the Hamiltonian in Eq.~\eqref{unitary-eqn-23}, we discussed in Eq.~\eqref{main-eqq-6} that, in the Zeno limit, the splitting probability depends only on the projection of the initial state onto the two targets. To prove this result, let us consider the $0<\gamma \tau \ll 1$ limit of Eq.~\eqref{aux-eq-5}. 
%takes the constant value $1/2$ for all initial positions $x_0$ in the interior of the chain, with deviations occurring only at $x_0=1$ and $x_0=N$. To prove this result, let us consider the $0<\gamma \tau \ll 1$ limit of Eq.~\eqref{aux-eq-5}. 
As evident, one then needs to evaluate the interference term $\zeta _{+-}(\psi_0)$ in this expression, for which we turn to Eqs.~\eqref{pole-eq-1}-\eqref{pole-eq-2}, and simplify them for $\gamma \tau \ll 1$. In particular, the function $\mathcal{I}(\psi_0, z) $ in Eq.~\eqref{pole-eq-2} becomes
\begin{align}
\mathcal{I}(\psi_0, z) \simeq  \frac{N+1}{8 \pi  i z} &  \left[ \frac{\sum _{\ell \in ~\rm{odd}} \sin (\omega _\ell)  \braket{E_\ell| \psi_0}  }{   \sum _{\ell \in ~\rm{odd}} \sin ^2 (\omega _\ell)     } \right]   \left[ \frac{\sum _{\ell \in ~\rm{even}} \sin (\omega _\ell) \braket{\psi_0|E_\ell}  }{   \sum _{\ell \in ~\rm{even}} \sin^2 (\omega _\ell)     } \right]. \label{aux-eq-ssq1}
\end{align}
We next use 
\begin{align}
\sum _{\ell \in ~\rm{odd}} \sin^2 (\omega _\ell) = \sum _{\ell \in ~\rm{even}} \sin^2 (\omega _\ell) = \frac{N+1}{4},
\end{align}
and recast Eq.~\eqref{aux-eq-ssq1} as
\begin{align}
\mathcal{I}(\psi_0, z) \simeq  \frac{ \braket{d_+|\psi_0} \braket{\psi_0|d_-}}{2 \pi i z} \simeq \frac{1}{4 \pi i z} \left[  |\braket{1| \psi_0}|^2-|\braket{N| \psi_0}|^2-2 i \text{Im} \left( \braket{1|\psi_0} \braket{\psi_0|N} \right) \right],
\end{align}
where $\ket{d_{\pm}}$ are given in Eqs.~\eqref{neww-eg-eq-8}-\eqref{neww-eg-eq-82}. Plugging the form of $\mathcal{I}(\psi_0, z)$ in Eq.~\eqref{pole-eq-1} gives
\begin{align}
\zeta _{+-}(\psi_0) \simeq \frac{1}{2} \left( |\braket{1| \psi_0}|^2-|\braket{N| \psi_0}|^2 \right).
\end{align}
Inserting this result in Eq.~\eqref{aux-eq-5}, we obtain the Zeno-limit expressions of the splitting probabilities outlined in Eq.~\eqref{main-eqq-6} for the one-dimensional tight-binding chain with nearest-neighbor hopping. Interestingly, in the Zeno limit, the splitting probabilities are completely independent of the detailed structure of the initial state $\ket{\psi_0}$ in the interior of the chain and depend only on its projections onto the two boundary target sites.

\begin{figure}[t]
	\centering
	\includegraphics[scale=0.2]{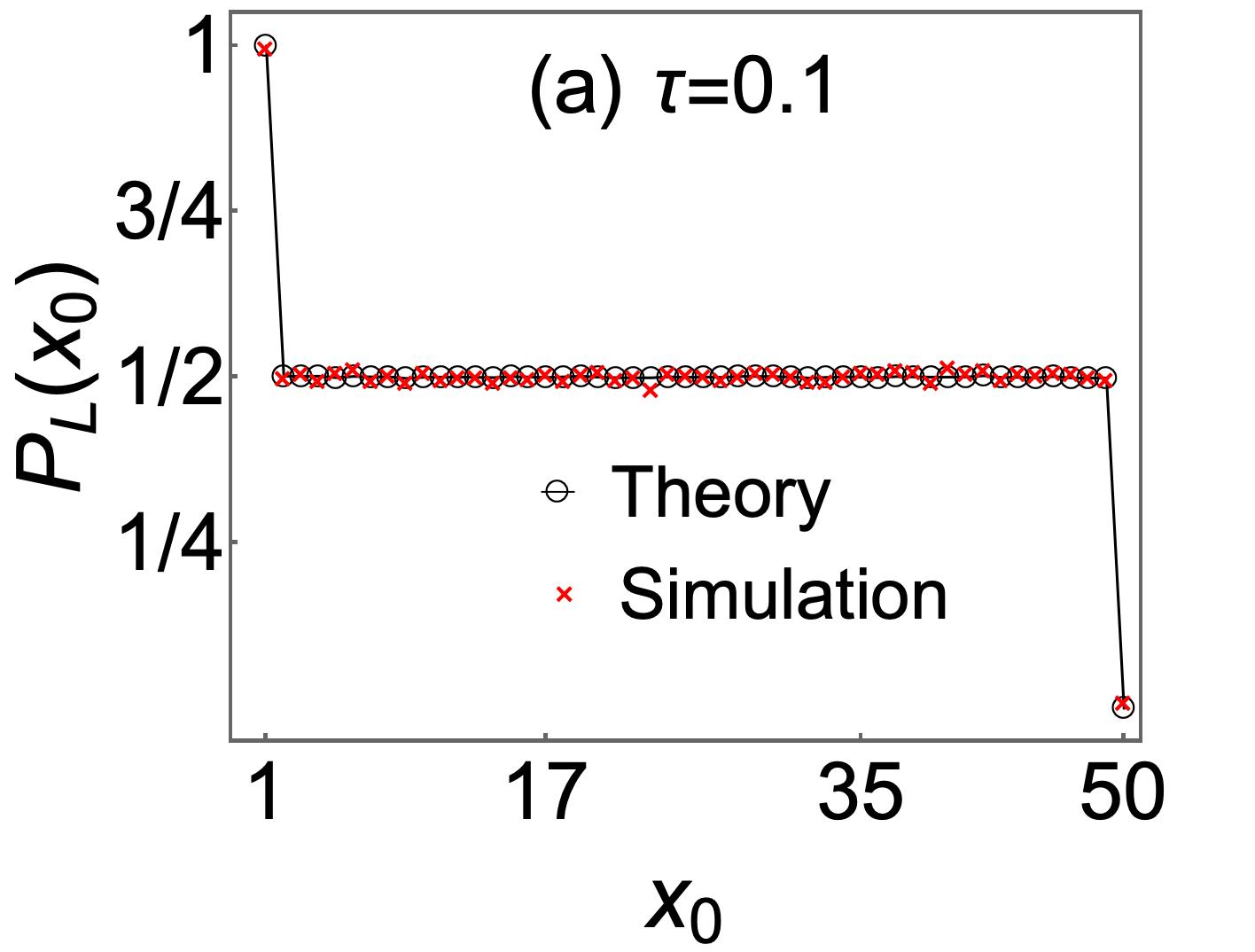} 
	\includegraphics[scale=0.2]{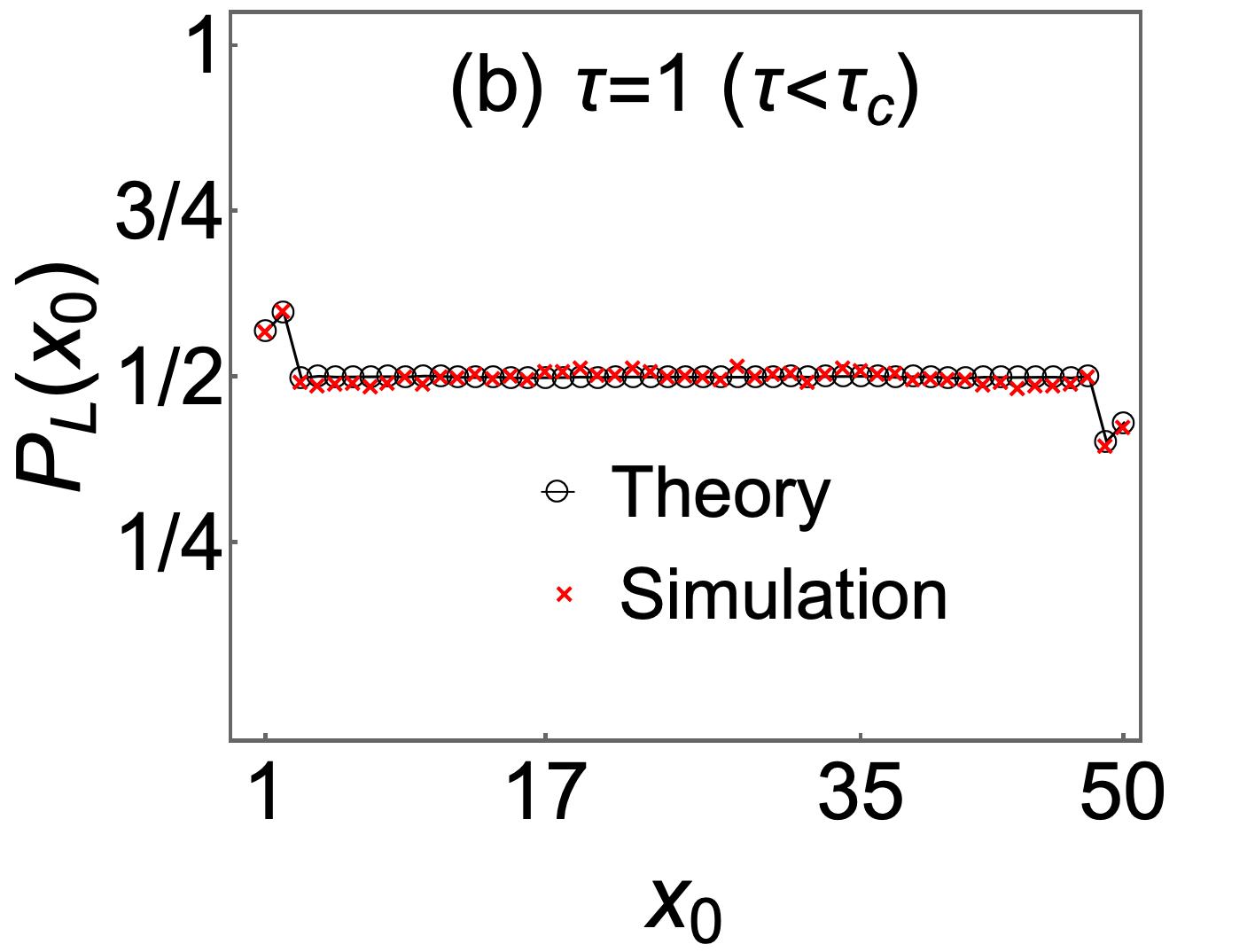} 
	\includegraphics[scale=0.2]{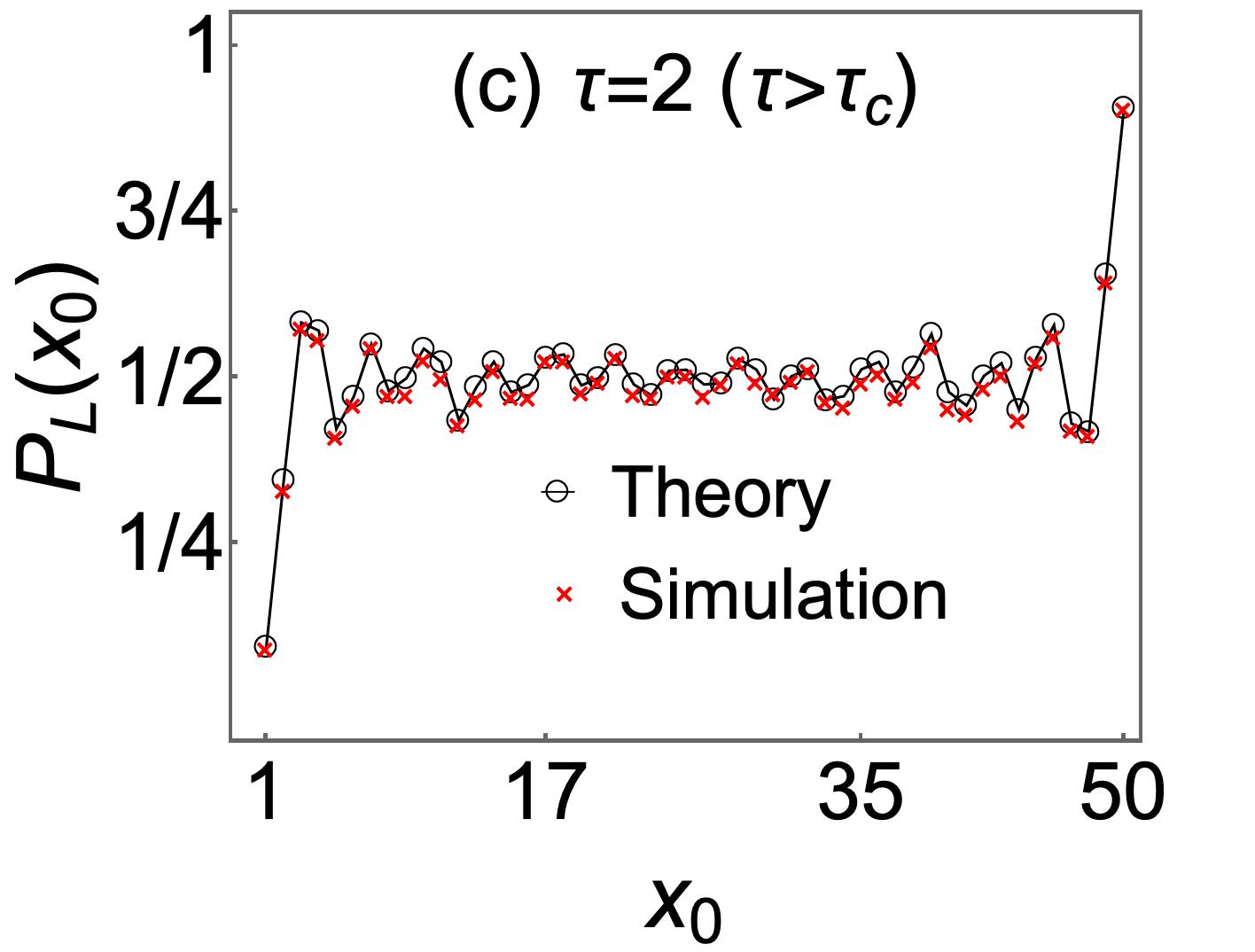} 
	\caption{Splitting probability $P_L(x_0)$ as a function of the initial position $x_0$ for the nearest-neighbor tight-binding chain of size $N=50$ in Eq.~\eqref{main-eqq-1}. 
Panel~(a) shows the theoretical prediction for the Zeno limit in Eq.~\eqref{neww-eg-eq-33} together with numerical simulations for $\gamma = 1$ and $\tau = 0.1$; see \aref{sec-appen-simulation} for details on the numerical simulations. 
Panels~(b,c) display simulation data of $P_L(x_0)$ for $\tau = 1$ and $\tau = 2$, respectively, compared with the theoretical expression in Eq.~\eqref{aux-eq-5}. 
These values of the sampling time are chosen such that they lie below and above the critical value $\gamma \tau_c = \pi/2~(\approx 1.57)$. As evident, $P_L(x_0)$ for $\tau=0.1$ and $\tau=1$ remains essentially flat and equal to $1/2$, except close to the boundary sites. In contrast, for $\tau=2$, $P_L(x_0)$ develops pronounced oscillations as a function of $x_0$. In the large-$N$ limit, these oscillations are characterized by the wavenumber in Eq.~\eqref{pos-eq-2}. The splitting probabilities are defined only for integer values of $x_0$; lines are included merely as a visual aid.
}
\label{fig-tau-N15}
\end{figure}

To illustrate this result further, let us consider a localized initial condition, $\ket{\psi_0}=\ket{x_0}$. In this case, Eq.~\eqref{main-eqq-6} immediately gives
\begin{align}
\boxed{P_ L(x_0) \simeq \begin{cases}
1,~~~~\text{for }x_0=1, \\
0, ~~~~\text{for }x_0=N, \\
\frac{1}{2}, ~~~~\text{otherwise },
\end{cases}
~~P_ R(x_0) \simeq \begin{cases}
0,~~~~\text{for }x_0=1, \\
1, ~~~~\text{for }x_0=N, \\
\frac{1}{2}, ~~~~\text{otherwise }.
\end{cases}}
\label{neww-eg-eq-33}
\end{align}
In Fig.~\ref{fig-tau-N15}(a), we compare our theoretical result with the numerical simulations. The simulation is performed for $N=50$, $\gamma=1$ and $\tau=0.1$. It is in excellent agreement with our theory. We refer to \aref{sec-appen-simulation} for details on the numerical simulations. Remarkably, for small sampling time, both splitting probabilities are equal to $1/2$, independent of the initial position $x_0$, as long as $x_0$ is in the interior of the tight-binding chain. At the boundary sites, the walker is detected by the corresponding detector with probability one and by the opposite detector with probability zero; see Fig.~\ref{fig-tau-N15}(a). Notably, this behavior is completely universal, and holds irrespective of the system size $N$. 

To understand the flat behavior physically, recall that for small values of $\tau$, the amplitudes of detection at the boundary sites, starting from anywhere in the interior of the chain, scale as $\sim \tau$. The corresponding detection probability between the successive measurements then scales as $\sim \tau ^2$ \cite{Misra1977Zeno}. Since $\tau$ is small, this probability is very small. Moreover, each unsuccessful detection at the boundaries collapses the wave function into the interior of the chain. Therefore, the walker is restricted mostly in the interior, and with only a small probability of order $\tau^2$ does it get detected at the boundary sites. As a  result, the walker typically takes a large number of measurement attempts to be detected unless its initial position already coincides with one of the detectors. Any initial asymmetry present in the system is effectively erased at such long detection times, leading to the flat splitting probability with a value equal to $1/2$.

%\redw{Such a measurement-induced uniformity has also been found in another context. The study by Yi et al., for example, examines a quantum walk monitored repeatedly at a single site, which eventually reaches a uniform state \cite{Yi2011}. However, unlike our work, their setup involves monitoring at a single site for a fixed number of measurements, without regard to when the first detection occurs. In our study, by contrast, the walker is monitored at two target sites and the process is terminated immediately upon a successful detection at the either target. Yet, the independence of $P_L(x_0)$ and $P_R(x_0)$ with $x_0$ for $1<x_0<N$ strongly indicates that the uniformity observed by Yi et al. may extend even when multiple sites are monitored. Exploring this presents an interesting avenue for future work.}

%in which for repeated measurements in a single site. Contrary to our setup, however, this work does not address the first-detection problem but instead monitors the system at a single site over a long yet fixed number of measurements.

%Another interesting observation is that for $N \to \infty$ with $x_0$ of order $1$, we do not recover the result of semi-infinite 

%\subsection{Similarity to disordered systems} 
%The quantum Zeno feature discussed above admits an illuminating analogy with classical disordered systems \cite{disorder1}. Consider a classical random walker on a finite lattice segment where each site is associated with a random energy well, and absorbing conditions at the two boundaries. Starting from an arbitrary initial position in the interior of the segment, the walker typically becomes trapped in a deep well at low temperatures and strong disorder, while thermal activation allows only rare jumps to neighboring sites. Once a jump occurs, the situation repeats -- strong localization followed by rare jumps. This mechanism ensures that the walker spends most of its time in the interior of the chain. This behavior is similar to the quantum case discussed above, with one crucial distinction: in the classical system, strong disorder effectively confines the walker to the interior, while in the quantum case, the confinement results purely from repeated failed measurements at the boundaries. Consequently, in both cases, the walker reaches the boundaries only after a long time, and over such long time scales, the system loses all memory of its initial position. As a result, the splitting probability becomes completely flat, taking the value $1/2$ for any initial position in the interior, with deviations from $1/2$ appearing only when $x_0$ coincides with the boundaries \cite{disorder1}.

\subsection{Singular case $\tau=0$}
In experiments, such as those conducted on waveguides \cite{ExptWG-1, ExptWG-2, ExptWG-3, Chen2026}, and quantum computers \cite{ExptQC-1, ExptQC-2, ExptQC-3}, the sampling time $\tau$ can be small, though it remains finite ($0 < \gamma \tau \ll 1$). Now, we turn to the $\tau = 0$ case to show that the Zeno limit is singular. In other words, we cannot recover the $\tau = 0$ result by using Eq.~\eqref{neww-eg-eq-33} for $0<\gamma \tau \ll 1$ and plugging $\tau = 0$ in this expression.

%The case $\tau = 0$ turns out to be  a singular point not captured by our $0 < \gamma \tau \ll 1$ analysis. This is because the starting point of our $0<\gamma \tau \ll 1$ derivation in Eq.~\eqref{neww-eg-eq-33} is the expression for the splitting probabilities in Eq.~\eqref{aux-eq-5}, which was itself derived under the assumption of non-degenerate quasi-energies. This assumption, however, breaks down for $\tau = 0$, where all quasi-energies $\lambda_k = 0$ for every $k$, see Eq.~\eqref{neww-eg-eq-1q}. Consequently, Eq.~\eqref{neww-eg-eq-33} does not hold for $\tau=0$. At this point, it becomes necessary to consider the general expressions of $P_L(x_0)$ and $P_R(x_0)$ in Eq.~\eqref{neww-eg-eq-17}.

For $\tau = 0$, we show in \aref{appen-sec-Zeno} that the splitting probabilities are given by
\begin{align}
P_L(\psi_0) =  |\braket{1| \psi_0}|^2,~~P_R(\psi_0) =  |\braket{N| \psi_0}|^2,\label{neww-eg-eq-28kou}
\end{align}
which, for $\ket{\psi_0} = \ket{x_0}$, becomes
\begin{align}
 P_L(x_0) =\delta _{x_0, 1},~~P_R(x_0) =\delta _{x_0, N}.
\label{neww-eg-eq-28}
\end{align}
This result is clearly different from Eq.~\eqref{neww-eg-eq-33}. For $\tau = 0$, the system is never detected if $x_0$ is not at the boundary. These results admit a simple physical interpretation based on quantum Zeno physics \cite{Misra1977Zeno, Teuscher2004Turing}. For wave functions initially localized in the interior of the chain, continuous measurements at the boundaries always yield a failed detection. Each failed measurement collapses the wave function in the interior of the chain, restricting the walker completely in this part. As a result, the walker is never detected at the boundary sites, and both splitting probabilities vanish. Combining Eq.~\eqref{neww-eg-eq-28} with Eq.~\eqref{neww-eg-eq-33}, we thus find that both $P_L(x_0)$ and $P_R(x_0)$ exhibit a universal discontinuous jump at $\tau = 0$, changing from $0$ to $1/2$ on slightly increasing $\tau$.

Importantly, the jump at $\tau=0$ is not merely a theoretical construct, but also leaves an experimental fingerprint in quantum computers, where a finite number of measurement attempts leads to the broadening of the otherwise discontinuous jumps \cite{ExptQC-1, ExptQC-2, ExptQC-3}. As explained above, the $1/2$ value of the splitting probabilities in Eq.~\eqref{neww-eg-eq-33} crucially depends on the walker typically taking a large number of measurements attempts for a successful detection. We later show that putting a cap on the number of measurements leads to deviations from this value. A finite number of measurements leads to a broadening effect of the otherwise pointwise discontinuous jump, an effect discussed later. In particular, the limit of an infinite number of measurements used to detect the quantum walker and the $\gamma \tau \ll 1$ limit do not commute. 
%This analysis is left for future work.

%Finally, we remark that such a discontinuous jump is a purely quantum feature that is not observed for a classical random walker. In the latter case, we later show that the splitting probabilities smoothly approach the gambler's ruin formula in Eq.~\eqref{unitary-eqn-classical} as the measurement time approaches zero.

\begin{figure}[t]
	\centering
	\includegraphics[scale=0.35]{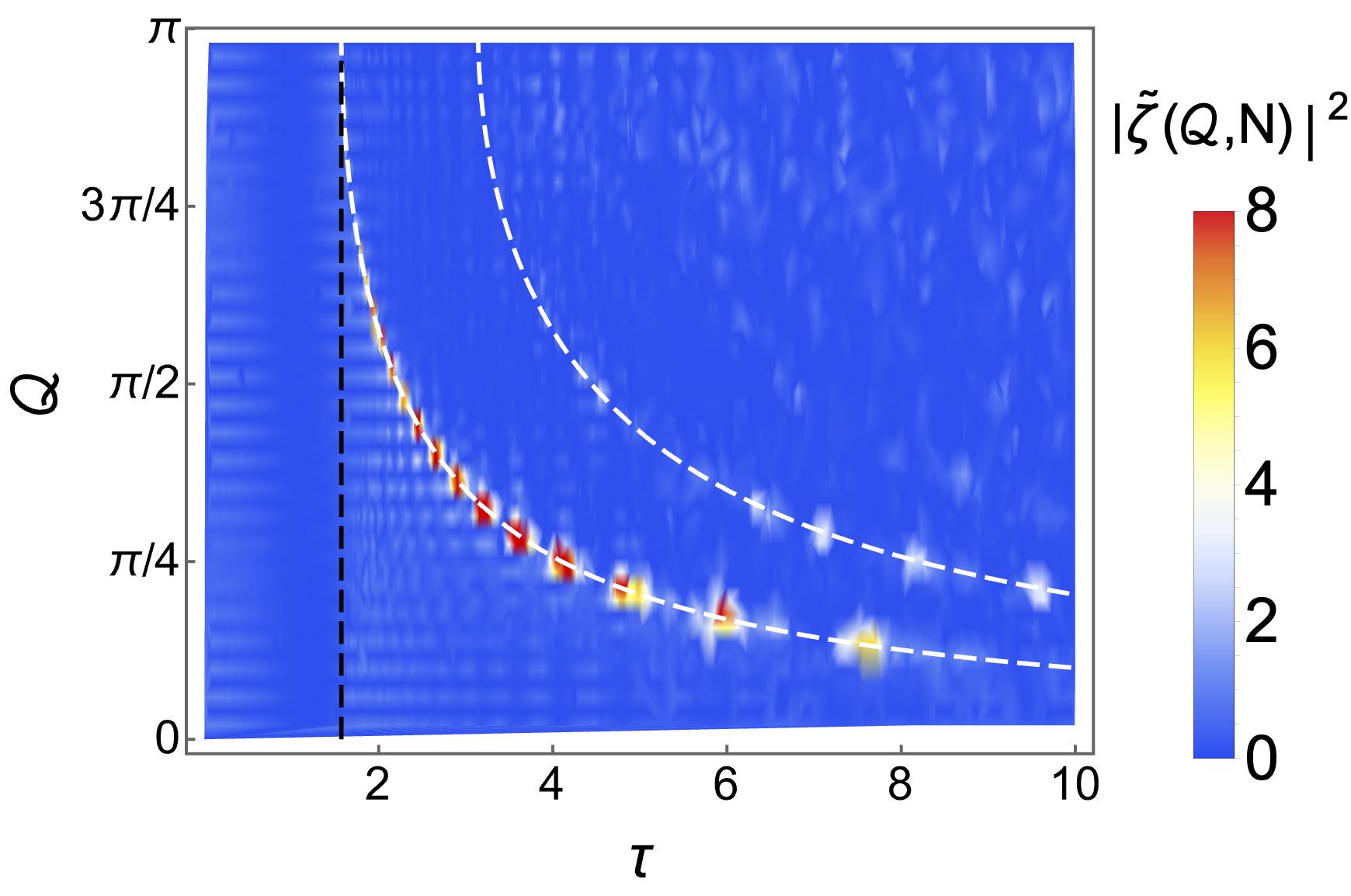} 
	\caption{Heat map of the power spectrum $\left|\widetilde{\zeta}(Q,N)\right|^2$ for the nearest-neighbor hopping Hamiltonian, shown as a function of the discrete wavenumber
$Q\in\left\{\pi/(N+1),\,2\pi/(N+1),\,\ldots,\,N\pi/(N+1)\right\}$
and the sampling time $\tau$, for $N=50$ and $\gamma=1$. To construct the plot, we randomly sample a pair $(Q,\tau)$, evaluate $\zeta_{+-}(x_0)$ numerically from Eqs.~\eqref{pole-eq-1}-\eqref{pole-eq-2}, and then use Eq.~\eqref{SNRPSS} to obtain the corresponding value of $\left|\widetilde{\zeta}(Q,N)\right|^2$. Repeating this procedure for $10^4$ randomly sampled $(Q,\tau)$ pairs yields the heat map. The dashed black line in the plot marks $\gamma \tau_c=\pi/2(\approx 1.57)$. For $\tau \geq \tau_c$, spectral branches emerge at which the value of the power spectrum is pronounced.  As shown in the paper, in the large-$N$ limit, these branches are given by
$Q_{\rm max}(\tau)=2\sin^{-1}\!\left(r \tau_c/\tau \right)$, with $r=1,2,\ldots$. The white dashed curves show the predictions for $r=1$ and $r=2$. These dominant wavenumbers govern the oscillations in $\zeta_{+-}(x_0)$ and, through Eq.~\eqref{aux-eq-5}, the oscillatory dependence of the splitting probability on the initial position $x_0$, as observed for $\tau \geq \tau_c$ in Fig.~\ref{fig-tau-N15}. Observe that, since $N=50$ is finite, the allowed wavenumbers $Q$ are discrete, and the spectral branches therefore appear as separated bright points rather than continuous lines.
}
\label{fig-osc}
\end{figure}

%\section{Beyond Zeno: Flat-to-fluctuating transition}
%\label{sec-transition}
\section{Flat-to-fluctuating transition and oscillations in the splitting probability}
\label{sec-transition}
For localized initial states, $\ket{\psi_0}=\ket{x_0}$, we further examine how the splitting probabilities depend on the initial position $x_0$ for general sampling times for the nearest-neighbor hopping Hamiltonian. More general initial states will be considered later. For the former case, we use  Eq.~\eqref{aux-eq-5} for $P_L(x_0)$, along with the interference term $\zeta_{+-}(x_0)$ in Eqs.~\eqref{pole-eq-1}-\eqref{pole-eq-2}. Following Eq.~\eqref{neww-eg-eq-1}, we also use $\braket{E_s|x_0} = \sqrt{2/(N+1)} \sin \left( k_s x_0 \right)$ ($s \in \{1,2, \ldots,N \}$) in this expression of $\zeta_{+-}(x_0)$. Fig.~\ref{fig-tau-N15} shows the plot of $P_L(x_0)$ as a function of $x_0$ for $\tau = 1$ and $\tau = 2$, along with the near-Zeno limit case ($\tau=0.1$). In all cases, the hopping amplitude is fixed at $\gamma = 1$ and $N=50$. In Fig.~\ref{fig-tau-N15}(b), we see that the profile for $\tau = 1$ remains essentially flat and equal to $1/2$, closely resembling the profile in the near-Zeno limit in Fig.~\ref{fig-tau-N15}(a). Deviations occur only when initial position is close to the endpoints of the chain. In contrast, for $\tau = 2$ in Fig.~\ref{fig-tau-N15}(c), this behavior changes qualitatively: $P_L(x_0)$ loses its flat behavior and develops oscillations with respect to $x_0$. Similar behavior is obtained for $P_R(x_0)$. 

\begin{figure}[t]
	\centering
	\includegraphics[scale=0.21]{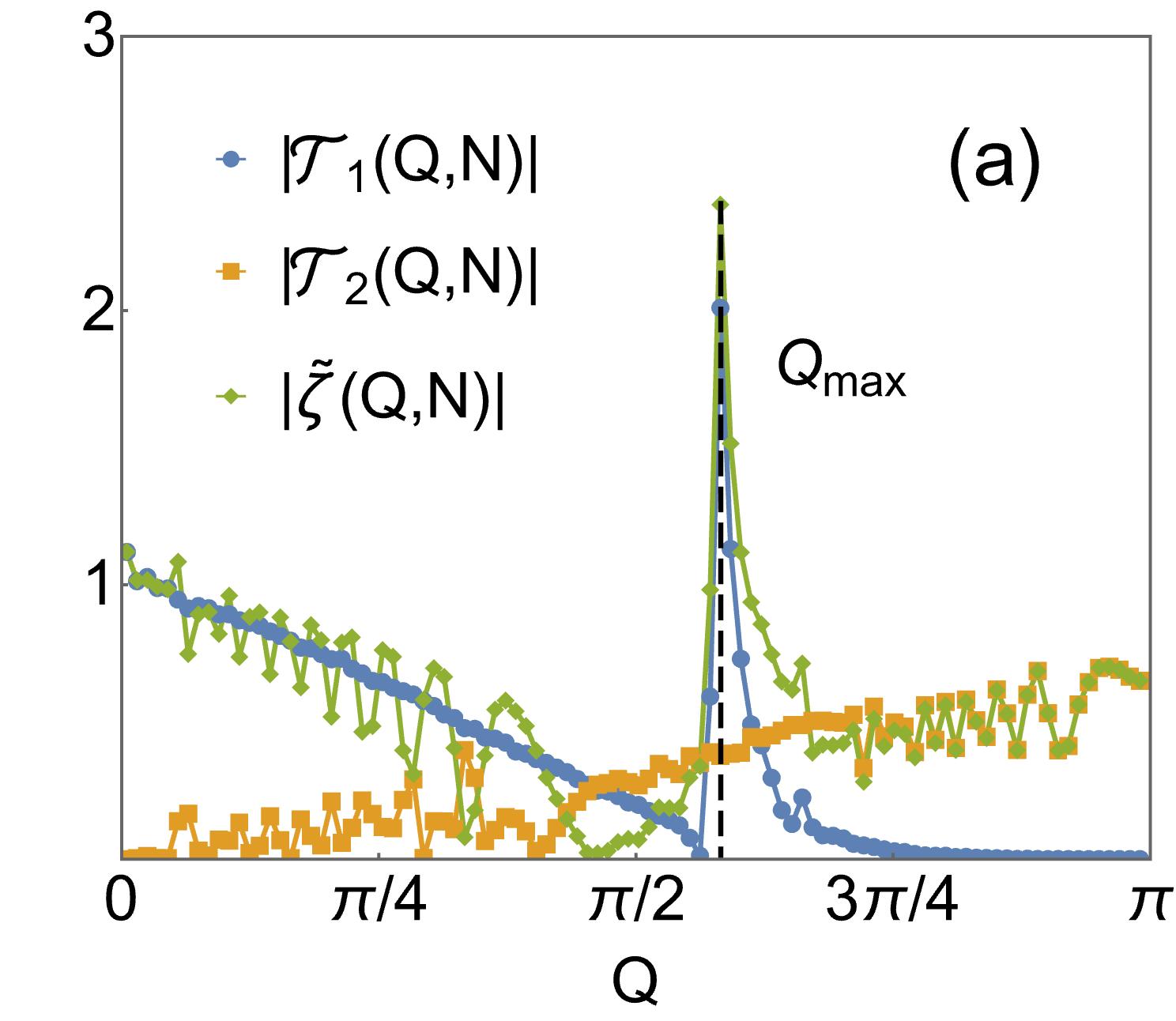}
	\includegraphics[scale=0.2]{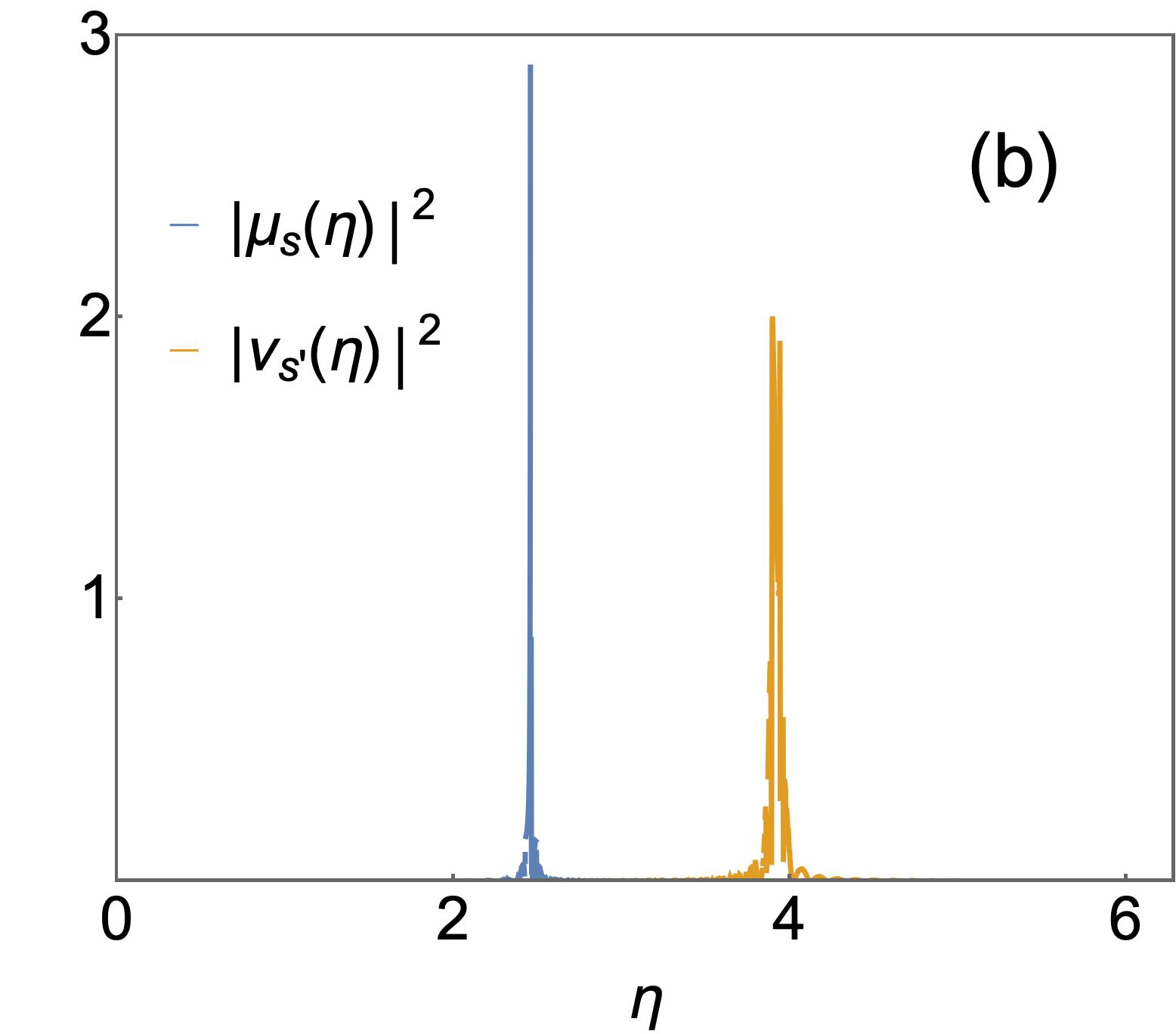}
	\includegraphics[scale=0.2]{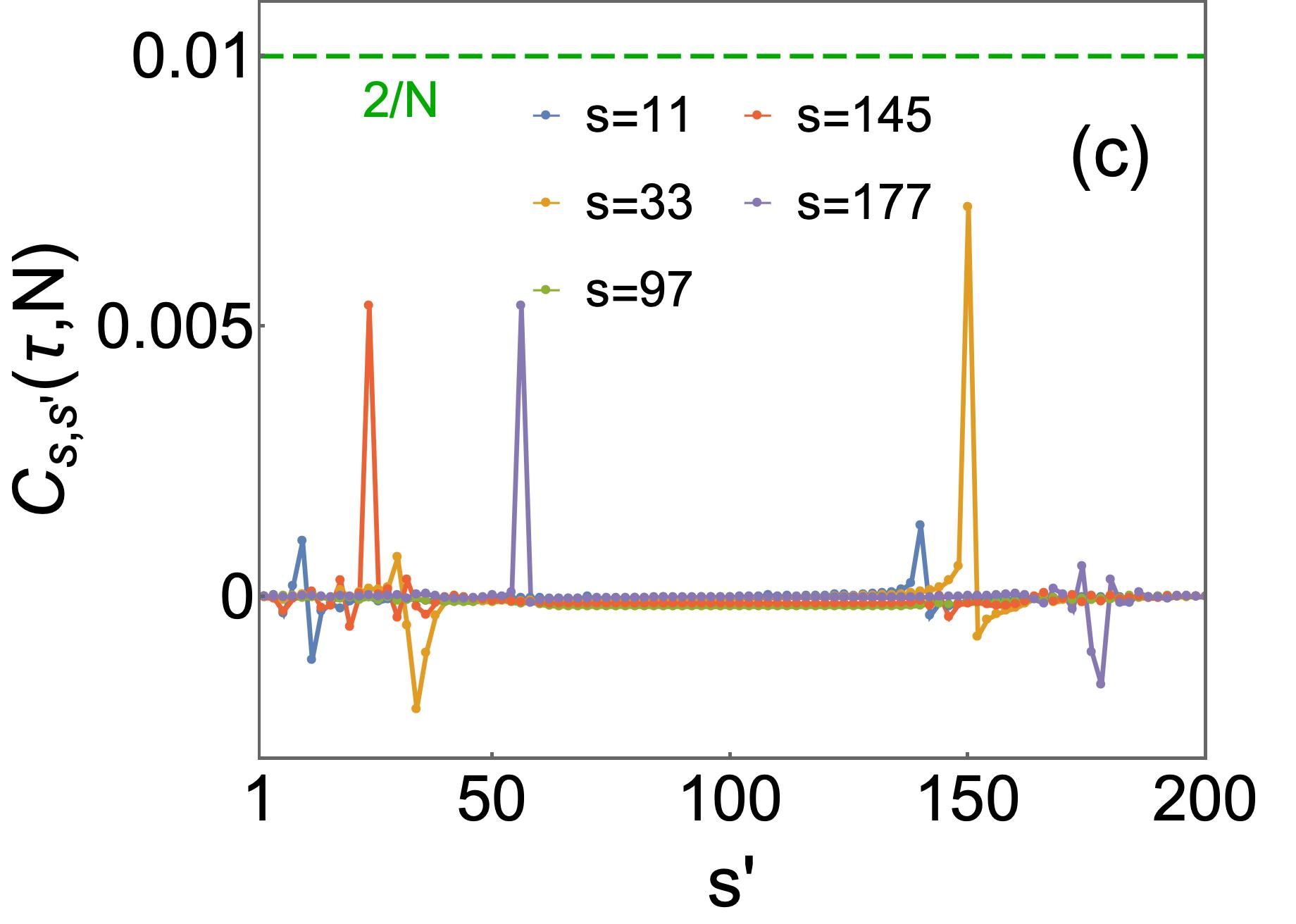}
	\caption{(a) The absolute value of the Fourier transform $\widetilde{\zeta}(Q,N)$ in Eq.~\eqref{SS-pos-eq-5} is plotted as a function of $Q = \pi j /(N+1)$ for odd $j$, $N=200$, $\gamma=1$, and $\tau=2$, along with the two terms   $|\mathcal{T}_1(Q,N)|$ and $|\mathcal{T}_2(Q,N)|$ in Eqs.~\eqref{SS-pos-eq-7} and \eqref{SS-pos-eq-8}. The maximum of $|\widetilde{\zeta}(Q,N)|$ occurs at the same $Q$-value as the maximum of $|\mathcal{T}_1(Q,N)|$; denoted by $Q_{\rm max}(\tau)(\approx 1.81)$ and depicted by the dashed line. We show in Eq.~\eqref{pos-eq-2} that for large $N$, $Q_{\rm max}(\tau) = 2 \sin^{-1} \left( \pi r /2 \gamma \tau \right)$, where $r$ is a positive integer. (b) The functions $|\mu_s(\eta)|^2$ and $|\nu_{s'}(\eta)|^2$ in Eqs.~\eqref{SS-pos-eq-3} and \eqref{SS-pos-eq-4}, respectively, are plotted as functions of $\eta$ for $s=19$ and $s'=60$, with the same parameters as in panel (a). These functions determine $\mathcal{C}_{s,s'}(\tau,N)$ through Eq.~\eqref{SS-pos-eq-2}. Both functions are sharply peaked at $\eta^{\mu_s \rm peak} \approx 2.4$ and $\eta^{\nu_{s'} \rm peak} \approx 3.9$, respectively, consistent with the peak locations given in Eqs.~\eqref{SS-pos-eq-9} and \eqref{SS-pos-eq-10}. (c) The function $\mathcal{C}_{s,s'}(\tau,N)$ in Eq.~\eqref{SS-pos-eq-2} is plotted with respect to the even index $s'$ for five representative fixed odd values of $s$, using the same parameters as in panel (a). In the large-\(N\) limit, we showed that $\mathcal{C}_{s,s'}(\tau,N)\leq 2/N$, where the value $2/N$ is indicated by the green dashed line. For the finite system shown here, $N=200$, the bound is not saturated. However, the maximal value of $\mathcal{C}_{s,s'}(\tau,N)$ approaches $2/N$ as $N$ increases.}
	\label{fig-osc-appen}
\end{figure}

From Eq.~\eqref{neww-eg-eq-33}, the oscillations in the splitting probabilities arise due to $\zeta_{+-}(x_0)$. It is therefore useful to analyze its Fourier transform 
\begin{align}
\widetilde{\zeta}\left( Q, N \right)  = \sum_{x_0=1}^{N} e^{-i Q x_0} \zeta_{+-}(x_0) \label{SNRPSS}, 
\end{align}
where $Q = \pi j /(N+1)$ with $j  \in \{1,2, \ldots,N \}$ being the wavenumber. Plugging the expression for $\zeta_{+-}(x_0)$ from Eqs.~\eqref{pole-eq-1}-\eqref{pole-eq-2} in Eq.~\eqref{SNRPSS}, we numerically obtain this Fourier transform. The resulting heat map of the power spectrum, $|\widetilde{\zeta}(Q,N)|^2$, as a function of $Q$ and $\tau$, for $N=50$ and $\gamma=1$, is shown in Fig.~\ref{fig-osc}. To construct the heat map, we randomly generate a $(Q,\tau)$ pair, evaluate $\zeta_{+-}(x_0)$ for this pair numerically from Eqs.~\eqref{pole-eq-1}-\eqref{pole-eq-2} for different $x_0$, and then substitute these values into Eq.~\eqref{SNRPSS} to obtain the corresponding value of $|\widetilde{\zeta}(Q,N)|^2$. We then repeat this procedure for $10^4$ randomly sampled $(Q,\tau)$ pairs yielding the heat map in Fig.~\ref{fig-osc}.

Interestingly, for $\tau \geq \tau_c$, where $\tau_c$ is given in Eq.~\eqref{ajbidhiguqok}, we observe pronounced branches in Fig.~\ref{fig-osc}, depicted by white dashed lines, along which the value of $|\widetilde{\zeta}(Q,N)|^2$ is significantly enhanced relative to the surrounding region. By contrast, for $\tau<\tau_c$, no such branches are observed. Physically, the enhanced Fourier modes imply that $\zeta_{+-}(x_0)$ develops oscillations as a function of the initial position $x_0$ only for $\tau > \tau_c$, which, according to Eq.~\eqref{neww-eg-eq-33}, directly leads to oscillations in the splitting probabilities. This provides a Fourier-space signature of the oscillatory behavior observed in Fig.~\ref{fig-tau-N15}, and identifies
$\tau_c$ as the threshold for the emergence of the oscillations. Furthermore, we see in Fig.~\ref{fig-osc} that the dominant wavenumber, $Q_{\rm max}(\tau)$, at which the power spectrum is maximal depends on the sampling time; see the white dashed lines. This dependence was stated earlier in Eq.~\eqref{skpssrs}. We now derive this result.

%Since the oscillations in $P_L(x_0)$ originate from the interference term $\zeta_{+-}(x_0)$ [see Eq.~\eqref{aux-eq-5}], it is useful to analyze its Fourier transform, $\widetilde{\zeta}\left( Q \right)  = \sum_{x_0=1}^{N} e^{-i Q x_0} \zeta_{+-}(x_0)$, where $Q \in \left\{ \frac{\pi}{N+1},   \frac{2\pi}{N+1}, \ldots,  \frac{N \pi}{N+1} \right\} $ is the wavenumber.
We first recast the expression of $\zeta_{+-}(x_0)$ in Eqs.~\eqref{pole-eq-1}-\eqref{pole-eq-2} by using $\braket{E_s|x_0} = \sqrt{2/(N+1)} \sin \left( \omega_s x_0 \right)$, as mentioned before. We then get
\begin{align}
& \zeta_{+-}(x_0) = \sum_{\substack{s=1\\ s\ \mathrm{odd}}}^{N}~\sum_{\substack{s'=1\\ s'  \ \mathrm{even}}}^{N} \mathcal{C}_{s, s'}(\tau, N) \sin \left( \omega_s x_0 \right) \sin \left( \omega_{s'} x_0 \right),\label{SS-pos-eq-1}
\end{align}
where $\omega_s$ is the momentum mode given in Eq.~\eqref{neww-eg-eq-1}, and $\mathcal{C}_{s, s'}(\tau, N)$ is defined as
\begin{align}
 & \mathcal{C}_{s, s'  }(\tau, N) = \text{Re} \left[ \frac{1}{4 \pi} \int_0^{2 \pi}  d \eta ~ \mu_{s} \left( \eta \right)  \nu_{s'} \left( \eta \right)   \right] , \label{SS-pos-eq-2}\\
 & \mu_s\left( \eta  \right) = \frac{\sin \left( \omega_s \right) e^{2i \gamma \tau \cos(\omega_s)}}{\sin^2 \left( \omega_s \right) + \left[1-e^{i \eta} e^{ 2i \gamma \tau \cos(\omega_s) } \right] \sum_{\substack{m\neq s\\ m\ \mathrm{odd}}}  \left[ \frac{\sin^2 \left( \omega_m \right)}{1-e^{i \eta} e^{2i \gamma \tau \cos(\omega_m)} } \right] }, \label{SS-pos-eq-3} \\
 &   \nu_{s'}\left( \eta \right) = \frac{\sin \left( \omega_{s'} \right) e^{-2i \gamma \tau \cos(\omega_{s'})}}{\sin^2 \left( \omega_{s'} \right) + \left[1-e^{-i \eta} e^{ -2i \gamma \tau \cos(\omega_{s'}) } \right] \sum_{\substack{m\neq {s'}\\ m\ \mathrm{even}}}  \left[ \frac{\sin^2 \left( \omega_m \right)}{1-e^{-i \eta} e^{-2i \gamma \tau \cos(\omega_{m})} } \right] }. \label{SS-pos-eq-4}
\end{align}
Performing the Fourier transform of Eq.~\eqref{SS-pos-eq-1}, we get
\begin{align}
\widetilde{\zeta}\left( Q,N \right) & = \sum_{\substack{s=1\\ s\ \mathrm{odd}}}^{N}\sum_{\substack{s'=1\\ s'  \ \mathrm{even}}}^{N}  \frac{\mathcal{C}_{s, s'}(\tau, N) }{4}\Bigg[
e^{-i\frac{(Q-\omega_s+\omega_{s'})(N+1)}{2}}
\frac{\sin\!\left[\frac{N}{2}(Q-\omega_s+\omega_{s'})\right]}
{\sin\!\left[\frac{1}{2}(Q-\omega_s+\omega_{s'})\right]}
+
e^{-i\frac{(Q+\omega_s-\omega_{s'})(N+1)}{2}}
\frac{\sin\!\left[\frac{N}{2}(Q+\omega_s-\omega_{s'})\right]}
{\sin\!\left[\frac{1}{2}(Q+\omega_s-\omega_{s'})\right]} \Bigg. \nonumber \\
&~~~~~~~~~~~~~~~ \Bigg. -
e^{-i\frac{(Q-\omega_s-\omega_{s'})(N+1)}{2}}
\frac{\sin\!\left[\frac{N}{2}(Q-\omega_s-\omega_{s'})\right]}
{\sin\!\left[\frac{1}{2}(Q-\omega_s-\omega_{s'})\right]}
-
e^{-i\frac{(Q+\omega_s+\omega_{s'})(N+1)}{2}}
\frac{\sin\!\left[\frac{N}{2}(Q+\omega_s+\omega_{s'})\right]}
{\sin\!\left[\frac{1}{2}(Q+\omega_s+\omega_{s'})\right]}
\Bigg]. \label{SS-pos-eq-5}
\end{align}
Eq.~\eqref{SS-pos-eq-5} contains terms of the form $\sim \sin(Ny)/\sin(y)$, which become highly oscillatory for large $N$ and develop sharp peaks of order $N$ as $y \to 0$. These sharply peaked contributions simplify $\widetilde{\zeta}(Q,N)$ in Eq.~\eqref{SS-pos-eq-5} considerably in the large-$N$ limit. In Appendix~\ref{appen-sec-FT}, we show that, for $N \gg 1$, $\widetilde{\zeta}(Q,N)$ with $Q=\pi j/(N+1)$ takes different asymptotic forms depending on whether $j$ is odd or even. Here we present the calculation for odd $j$ and relegate the even-$j$ case to Appendix~\ref{appen-sec-FT}. In the appendix, we show that for odd $j$ and $N \gg 1$, Eq.~\eqref{SS-pos-eq-5} reduces to
\begin{align}
&\widetilde{\zeta}\left( Q,N \right)  \simeq \mathcal{T}_1 \left( Q,N \right) + \mathcal{T}_2 \left( Q,N \right), \label{SS-pos-eq-6}\\
\text{with }& \mathcal{T}_1 \left( Q,N \right) =  \frac{N}{4} \sum_{\substack{s=1\\ s\ \mathrm{odd}}}^{N}\sum_{\substack{s'=1\\ s' \ \mathrm{even}}}^{N}  \mathcal{C}_{s, s'}(\tau, N)    \delta_{j, \left|s-s' \right|}, \label{SS-pos-eq-7} \\
&\mathcal{T}_2 \left( Q,N \right) =  -\frac{N}{4} \sum_{\substack{s=1\\ s\ \mathrm{odd}}}^{N}\sum_{\substack{s' =1\\ s' \ \mathrm{even}}}^{N}  \mathcal{C}_{s, s'}(\tau, N)  \delta_{j, N+1- \left|N+1- s-s' \right|}. \label{SS-pos-eq-8}
\end{align}
Remember our goal is to determine the dominant wavenumber $Q_{\rm max}(\tau)$ at which the power spectrum $|\widetilde{\zeta}(Q,N)|^2$, and equivalently $|\widetilde{\zeta}(Q,N)|$, attains its maximum. Since $\widetilde{\zeta}(Q,N)$ in
Eq.~\eqref{SS-pos-eq-6} is real and is composed of the two terms, $\mathcal{T}_1(Q,N)$ and $\mathcal{T}_2(Q,N)$, it is useful to examine their contributions separately. In Fig.~\ref{fig-osc-appen}(a), we plot $|\mathcal{T}_1(Q,N)|$, $|\mathcal{T}_2(Q,N)|$,
and $|\widetilde{\zeta}(Q,N)|$ as functions of $Q = \pi j /(N+1)$ with odd $j$ for $N=200$, $\gamma=1$ and $\tau=2$. We observe that the maximum of $|\widetilde{\zeta}(Q,N)|$ occurs at the same $Q_{\rm max}(\tau)$ as the maximum of $|\mathcal{T}_1(Q,N)|$.  This value is indicated by the dashed line in Fig.~\ref{fig-osc-appen}(a). We have verified for other system sizes and sampling times that the maxima of $|\widetilde{\zeta}(Q,N)|$ and $|\mathcal{T}_1(Q,N)|$ likewise occur at the same $Q_{\rm max}(\tau)$. We therefore proceed to determine $Q_{\rm max}(\tau)$ by analyzing $\mathcal{T}_1(Q,N)$.

From the expression for $\mathcal{T}_1(Q,N)$ in Eq.~\eqref{SS-pos-eq-7}, we see that each pair $(s,s')$, with odd $s$ and even $s'$ is weighted by $\mathcal{C}_{s, s'}(\tau,N)$. Hence, the dominant contributions to $\mathcal{T}_1(Q,N)$ arise from those $(s, s')$ pairs for which $\mathcal{C}_{s, s'}(\tau,N)$ is maximal or close to its maximum value. Let us next determine the maximal value of $\mathcal{C}_{s, s'}(\tau,N)$. 

\textit{Maximum of $\mathcal{C}_{s, s'}(\tau,N)$}: According to Eq.~\eqref{SS-pos-eq-2}, $\mathcal{C}_{s,s'}(\tau,N)$ involves integral of the product of two functions, $\mu_s(\eta)$ and $\nu_{s'}(\eta)$, defined in Eqs.~\eqref{SS-pos-eq-3} and \eqref{SS-pos-eq-4}, respectively. As functions of $\eta \in [0,2\pi]$ with fixed $s$ and $s'$, we show in Appendix~\ref{appen-sec-FT} that, for large $N$, these two functions are sharply peaked at 
\begin{align}
\eta^{{\mu_s} \rm peak}
&\simeq
-2\gamma\tau\cos(\omega_s)
\pmod{2\pi},
\label{SS-pos-eq-9}
\\
\eta^{{\nu_{s'}} \rm peak}
&\simeq
-2\gamma\tau\cos(\omega_{s'})
\pmod{2\pi}.
\label{SS-pos-eq-10}
\end{align}
Once again we emphasize $s$ is odd and $s'$ is even. We have illustrated Eqs.~\eqref{SS-pos-eq-9}-\eqref{SS-pos-eq-10} in Fig.~\ref{fig-osc-appen}(b) for $s=19$, $s' = 60$, $N=200$, $\gamma=1$ and $\tau=2$, where we use Eqs.~\eqref{SS-pos-eq-3} and \eqref{SS-pos-eq-4} to plot the absolute values of $\mu_s(\eta)$ and $\nu_{s'}(\eta)$ as functions of $\eta$. We see both of them are peaked at $\eta^{\mu_s \rm peak} \approx 2.4$ and $\eta^{\nu_{s'} \rm peak} \approx 3.9$, consistent with Eqs.~\eqref{SS-pos-eq-9}-\eqref{SS-pos-eq-10}. From Eq.~\eqref{SS-pos-eq-2}, it then follows that the integration over $\eta$ gives a pronounced contribution to $\mathcal{C}_{s,s'}(\tau,N)$ only when the locations of two peaks overlap, i.e., $\eta^{{\mu_s} \rm peak} \approx \eta^{{\nu_{s'}} \rm peak}$. When the two peaks are well separated, one function is negligible in the region where the other is large, so their product, and consequently its contribution to the integral, is negligible.
For example, for the $(s,s')$ pair mentioned above, we saw that the peak locations are well-separated and one finds $\mathcal{C}_{s,s'}(\tau,N)\sim10^{-5}$. By contrast, for $(s,s')=(33,150)$, one gets $\eta^{\mu_s {\rm peak}}\approx\eta^{\nu_{s'} {\rm peak}}\approx2.8$, and correspondingly $\mathcal{C}_{s,s'}(\tau,N)\sim10^{-3}$. Thus, pronounced values of $\mathcal{C}_{s,s'}(\tau,N)$ arise when the peak locations coincide. Using Eqs.~\eqref{SS-pos-eq-9}--\eqref{SS-pos-eq-10}, this condition can be written as
\begin{align}
\boxed{\left| E_{s}-E_{s'}\right| \tau = 2 \pi r, \qquad r=1,2, \ldots,} \label{SS-pos-eq-11}
\end{align}
such that $\ket{E_s}$ has even parity while $\ket{E_{s'}}$ has odd parity (corresponding to $s$ being odd and $s'$ being even, as proven later). The associated $\mathcal{C}_{s,s'}(\tau,N)$ is derived in Appendix~\ref{appen-sec-FT} to be
\begin{align}
\mathcal{C}_{s,s'}(\tau,N) \simeq \frac{4}{N}~ \frac{\sin \left( \omega_s \right) \sin \left( \omega_{s'} \right) \left[ \sin^2 \left( \omega_{s} \right)+ \sin ^2\left( \omega_{s'}  \right)\right] }{ \left[ \sin^2 \left( \omega_{s} \right)+ \sin ^2\left( \omega_{s'}  \right)\right]^2 + 16 \alpha_s(\tau)^2 \left[ \sin^2 \left( \omega_{s} \right) - \sin ^2\left( \omega_{s'}  \right)\right]^2 }, \label{SS-pos-eq-13}
\end{align}
with $\alpha_s(\tau)$ given in Eq.~\eqref{auiq71aa}. We further maximize $\mathcal{C}_{s,s'}(\tau,N) $ in Eq.~\eqref{SS-pos-eq-13} by noting that the second term in its denominator is nonnegative, and $\mathcal{C}_{s,s'}(\tau,N) $ is maximized when this term vanishes, i.e., $\sin^2 \left( \omega_{s} \right) =\sin^2 \left( \omega_{s'} \right) $. Since $\omega_{s} \neq \omega_{s'}$ (remember $s \neq s'$), we obtain the second condition
\begin{equation}
\boxed{\omega_s+\omega_{s'}=\pi.}
\label{SS-pos-eq-14}
\end{equation}
The corresponding maximum value is $\mathcal{C}_{s,s'}(\tau,N)\simeq 2/N$. Eqs.~\eqref{SS-pos-eq-11}-\eqref{SS-pos-eq-14} give the conditions for maximizing $\mathcal{C}_{s,s'}(\tau,N)$ with respect to $s$ and $s'$ in the large-$N$ limit. In Fig.~\ref{fig-osc-appen}(c), we have used the exact expression of $\mathcal{C}_{s,s'}(\tau,N) $ in Eq.~\eqref{SS-pos-eq-2} to plot it as a function of the even index $s'$ for several representative fixed odd values of $s$, for $N=200,~\gamma=1$ and $\tau=2$. For all pairs $(s,s')$, $\mathcal{C}_{s,s'}(\tau,N)$ remains below its large-$N$ maximum value of $2/N$, indicated by the green dashed line in Fig.~\ref{fig-osc-appen}(c). Notice that for finite system sizes, such as $N=200$ in Fig.~\ref{fig-osc-appen}(c), the values of $\mathcal{C}_{s,s'}(\tau,N)$ remain strictly below $2/N$. However, as the system size increases, the maximal value of $\mathcal{C}_{s,s'}(\tau,N)$ approaches $2/N$.

We also emphasize that although the maximal $\mathcal{C}_{s,s'}(\tau,N)$ decays as $2/N$ with the system size, the function $ \mathcal{T}_1 \left( Q,N \right) $ remains finite because of the overall $N$ prefactor in its expression in Eq.~\eqref{SS-pos-eq-7}.

\textit{Determination of $Q_{\rm max}(\tau)$}: Having obtained the conditions under which $\mathcal{C}_{s,s'}(\tau,N)$ is maximal, we now return to the determination of  $Q_{\rm max}(\tau)$. As discussed before and illustrated in Fig.~\ref{fig-osc-appen}(a), the maxima of $|\mathcal{T}_1(Q,N)|$ and $|\widetilde{\zeta}(Q,N)|$ occur at the same  $Q_{\rm max}(\tau)$. From Eq.~\eqref{SS-pos-eq-7}, $|\mathcal{T}_1(Q,N)|$ is maximum when $\mathcal{C}_{s, s'}(N, \tau)$ is maximum, and the corresponding wavenumber is $Q_{\rm max}(\tau) = |\omega_{s} - \omega_{s'}|$, where the momentum modes $\omega_s$ and $\omega_{s'}$ satisfy Eqs.~\eqref{SS-pos-eq-11}-\eqref{SS-pos-eq-14} for $N \gg 1$. In particular, Eq.~\eqref{SS-pos-eq-11} can be expressed as
\begin{align}
&2 \gamma \tau \left|
\cos(\omega_s)-\cos(\omega_{s'})
\right| = 2 \pi r, \nonumber \\
\implies & 4 \gamma \tau \sin \left( \frac{|\omega_s-\omega_{s'}|}{2} \right) \sin \left( \frac{\omega_s+\omega_{s'}}{2}\right) = 2 \pi r.
\end{align}
Using Eq.~\eqref{SS-pos-eq-14} alongwith $Q_{\rm max}(\tau) = |\omega_{s} - \omega_{s'}|$, we obtain
\begin{equation}
\boxed{Q_{\max}(\tau)
=
2\sin^{-1}\left(\frac{r \tau_c}{\tau}\right),\qquad r =1,2, \ldots.} \label{pos-eq-2}
\end{equation}
While this result was derived from Eqs.~\eqref{SS-pos-eq-6}-\eqref{SS-pos-eq-8} for $Q=\pi j/(N+1)$ with odd $j$, we show in Appendix~\ref{appen-sec-FT} that Eq.~\eqref{pos-eq-2} also holds for even $j$ in the large-$N$ limit. Thus for both odd and even $j$, the dominant wavenumber is given by Eq.~\eqref{pos-eq-2} in the large-$N$ limit. The result was quoted before in Eq.~\eqref{skpssrs}. Different values of $r$ correspond to different branches observed in Fig.~\ref{fig-osc}. The predictions for $r=1$ and $r=2$ in Eq.~\eqref{pos-eq-2} are shown by the white dashed curves in the figure, and they show excellent agreement with $Q_{\rm max}(\tau)$ in the heat map of $\left| \widetilde{\zeta}(Q,N) \right|^2$ in Fig.~\ref{fig-osc}. Since the system
size $N=50$ is finite, the allowed wavenumbers $Q$ are discrete, and the
spectral branches in Fig.~\ref{fig-osc} therefore appear as separated bright points rather than
continuous lines. Notice that $Q_{\rm max}(\tau)$ is real only for $\tau \ge r\tau_c$. Thus, the minimum sampling time for observing a dominant wavenumber is $\tau = \tau_c$. 

%For $\tau\ge\tau_c$, the oscillations in $\zeta_{+-}(x_0)$ are characterized by $Q_{\rm max}$ in Eq.~\eqref{pos-eq-2} and, through Eq.~\eqref{aux-eq-5}, lead to the oscillations of the splitting probabilities with $x_0$ observed in Fig.~\ref{fig-tau-N15}.

\subsection{Phase-matched momentum pairs}
It is also interesting to look at the interpretation of the condition in Eq.~\eqref{SS-pos-eq-11} in terms of phase-matched momentum pairs. In the large-$N$ limit, the allowed momenta become effectively
continuous over the interval $[0,\pi]$; see Eq.~\eqref{neww-eg-eq-1}. One can therefore drop the discrete indices $s$ and $s'$ and denote the corresponding momenta simply by $k$ and $k'$. Then, the dispersion relation in Eq.~\eqref{neww-eg-eq-1} for the nearest-neighbor Hamiltonian becomes $E_k = - 2 \gamma \cos k$. Combining this with the phase-matching condition in Eq.~\eqref{SS-pos-eq-11}, we get
\begin{equation}
    \left|\cos k-\cos k'\right|
    =
    \frac{\pi r}{\gamma\tau},
    \qquad r=1,2,\ldots .
    \label{eq:phase-match-continuous}
\end{equation}
Let us consider $k'<k$ (so $\cos k' > \cos k$) and $r=1$. For a given $k$, the allowed momenta that satisfy the condition~\eqref{SS-pos-eq-11} [equivalently Eq.~\eqref{eq:phase-match-continuous}] are
\begin{equation}
    k'(k, \tau)=
    \cos^{-1}\left(
    \cos k
    +
    \frac{\pi }{\gamma\tau}
    \right).
    \label{eq:phase-match-partner}
\end{equation}
From these possible values of $k'$, the one satisfying Eq.~\eqref{SS-pos-eq-14} gives the dominant contribution to the oscillations, as explained above in Eq.~\eqref{pos-eq-2}.

Let us now determine when such a phase-matched wave number $k'$ exists, which can then be chosen to satsify Eq.~\eqref{SS-pos-eq-14}. For $k'$ in Eq.~\eqref{eq:phase-match-partner} to be real, the argument of the inverse cosine in Eq.~\eqref{eq:phase-match-partner} must lie within
the interval $[-1,1]$, namely $  -1\leq
    \cos k+\frac{\pi}{\gamma\tau}
    \leq 1$. The lower bound is readily satisfied since $\cos k\geq-1$ and
$\pi/(\gamma\tau)>0$. On the other hand, the upper bound gives $ \cos k\leq \left( 1-\frac{\pi}{\gamma\tau} \right)$, which implies $ k \geq \cos^{-1} \left( 1-\frac{\pi}{\gamma\tau} \right)$. Therefore, the range of $k$ for which there exists a $k'$ in Eq.~\eqref{eq:phase-match-partner} satisfying
the phase-matching condition in Eq.~\eqref{SS-pos-eq-11} is
\begin{align}
k_{\rm min}(\tau) \leq k \leq \pi,~~\text{where }k_{\rm min}(\tau) = \cos^{-1} \left( 1-\frac{\pi}{\gamma\tau} \right). \label{eq:phase-match-partner2}
\end{align}
However $k_{\rm min}(\tau)$ is real only for $\tau \geq \tau_c$,  where $\tau_c = \pi / 2 \gamma$ is given in Eq.~\eqref{ajbidhiguqok}. This means that a physically meaningful range of \(k\) exists only for $\tau \geq \tau_c$. This can also be seen directly from Eq.~\eqref{eq:phase-match-partner}.
For $\tau<\tau_c$, the requirement that $k'$ be real implies $\cos k < -1$, which is not possible for real values of $k$.  Hence, no real
phase-matched pair $(k,k')$ exists for $\tau<\tau_c$. Only for
$\tau\geq\tau_c$ can a pair of momentum modes satisfying the
phase-matching condition in Eq.~\eqref{SS-pos-eq-11} exist. As explained above, these phase-matched pairs, when combined with Eq.~\eqref{SS-pos-eq-14}, give pronounced contributions to the interference term $\zeta_{+-}(x_0)$, which in turn gives rise to the oscillations of the splitting probabilities with $x_0$.

\begin{figure}[t]
	\centering
	\includegraphics[scale=0.25]{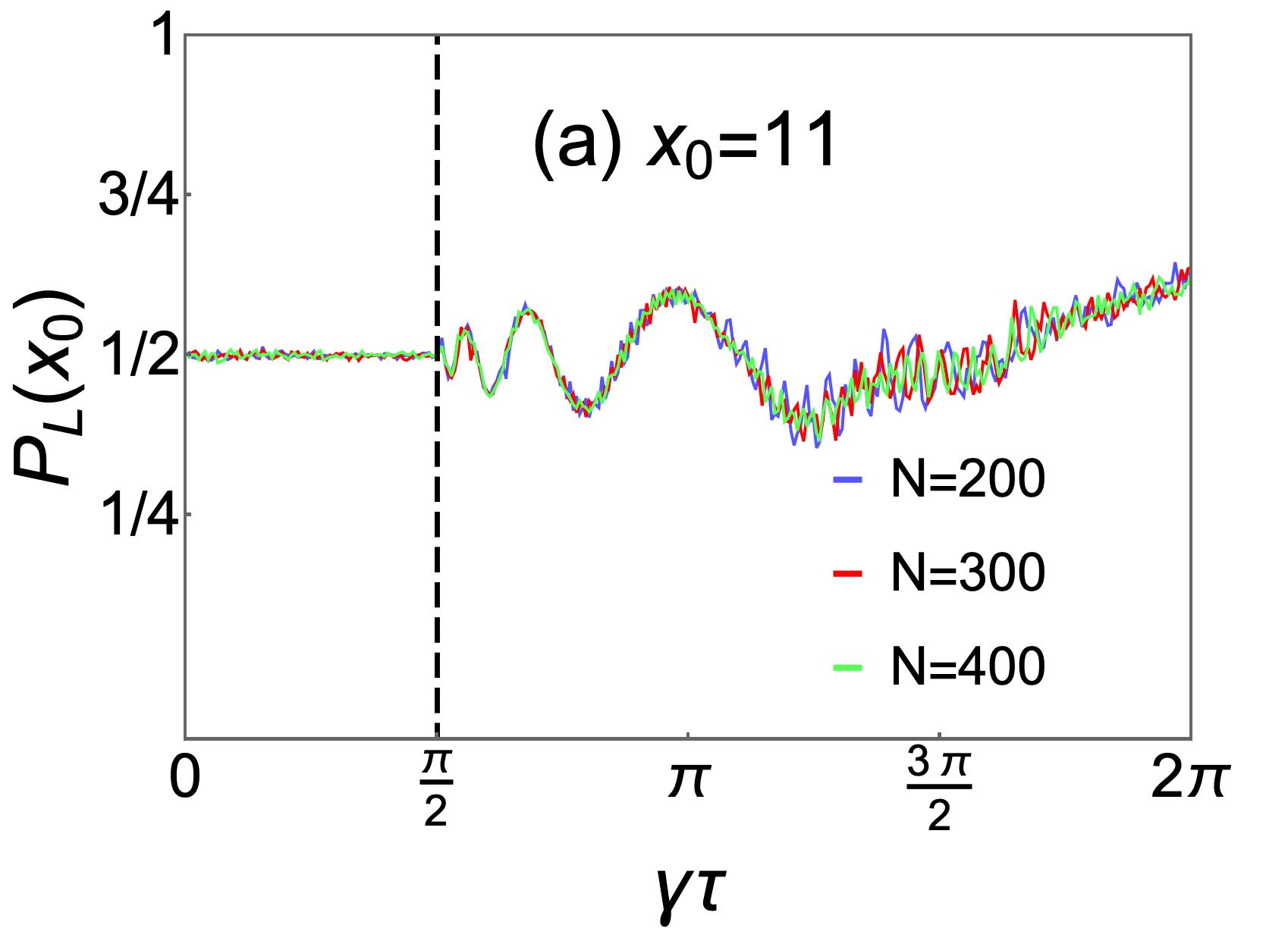} 
	\includegraphics[scale=0.25]{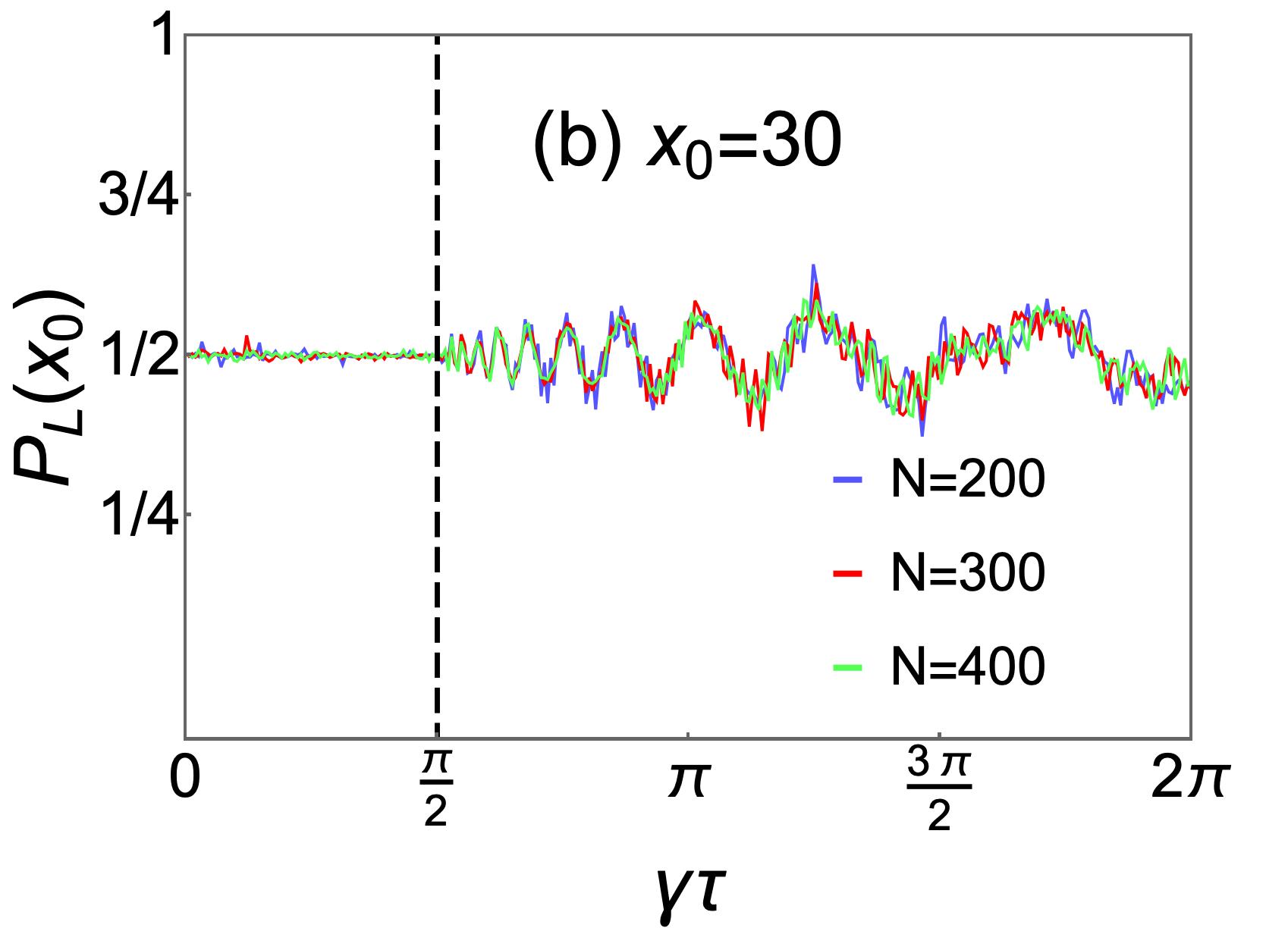} 
\caption{Illustration of the flat-to-fluctuating transition in  $P_L(x_0)$ at the critical sampling time $\gamma \tau _c = \pi/2$, using its expression in Eq.~\eqref{aux-eq-5} for $x_0=11$ and $x_0=30$. Each panel shows results for three different system sizes, illustrating that $P_L(x_0)$ becomes essentially independent of $N$ once the system is sufficiently large. In both panels, $P_L(x_0)$ is essentially $1/2$ for $\tau \leq \tau _c$, independent of the initial position and the sampling time. The critical point $\gamma \tau _c = \pi/2$ is indicated by the dashed vertical line. Beyond this point, for $\tau > \tau_c$, the flat behavior breaks down, and we find the emergence of a fluctuating regime, characterized by structures of pronounced peaks and dips. In this regime, $P_L(x_0)$ deviates from $1/2$ and the overall profile depends both on $\tau$ and $x_0$. For all plots, only the smooth part of the splitting probabilities is shown; the sharp resonant discontinuities found for special values of $\tau$ are omitted for clarity (see Fig.~\ref{fig-Ptau} for resonances). Since $P_L(x_0)+P_R(x_0)=1$, the same transition can also be found for $P_R(x_0)$.
}
\label{fig-transition}
\end{figure}

\subsection{Flat-to-fluctuating transition}
\begin{figure}[t]
	\centering
	\includegraphics[scale=0.25]{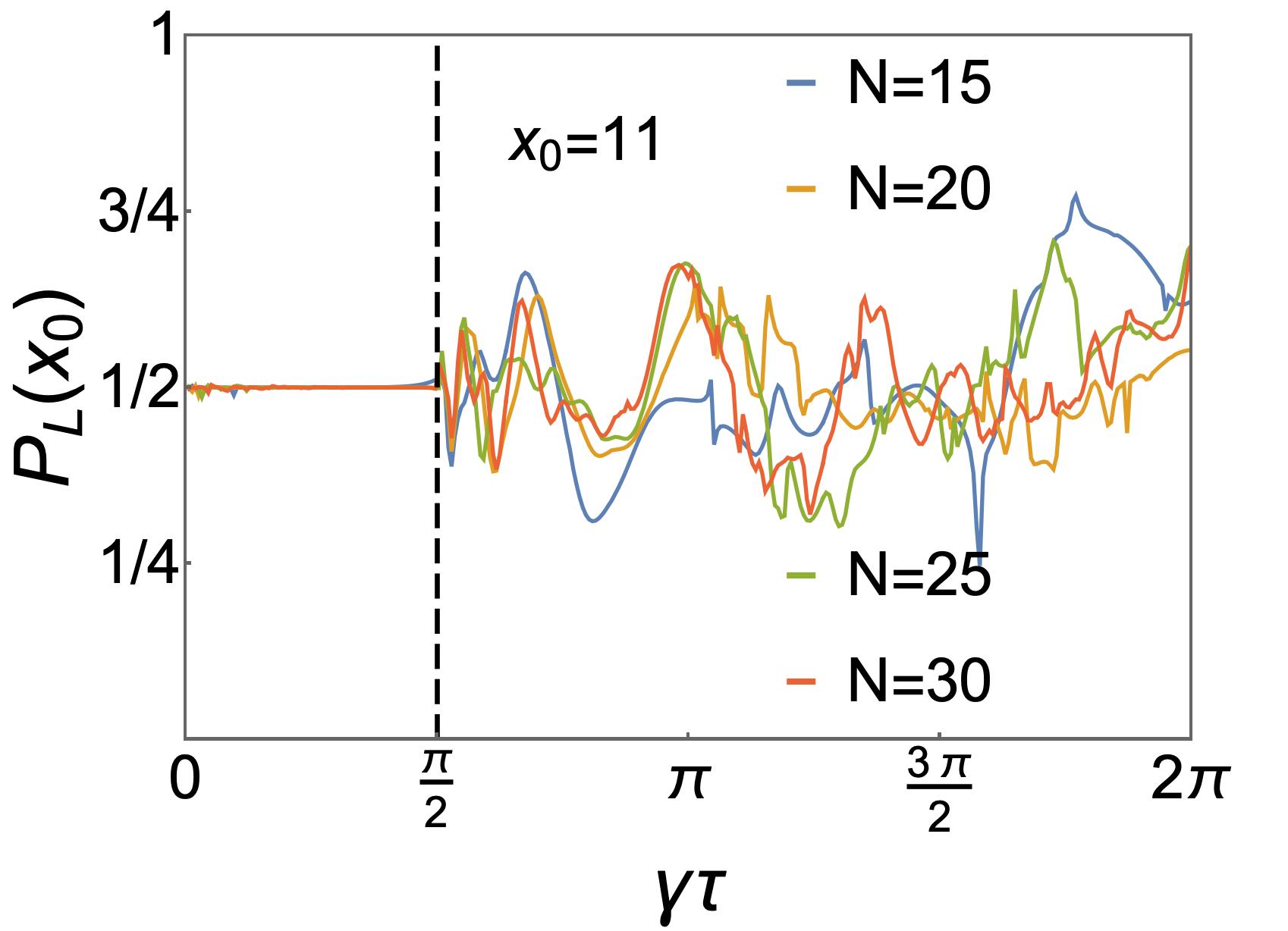} 
	\includegraphics[scale=0.25]{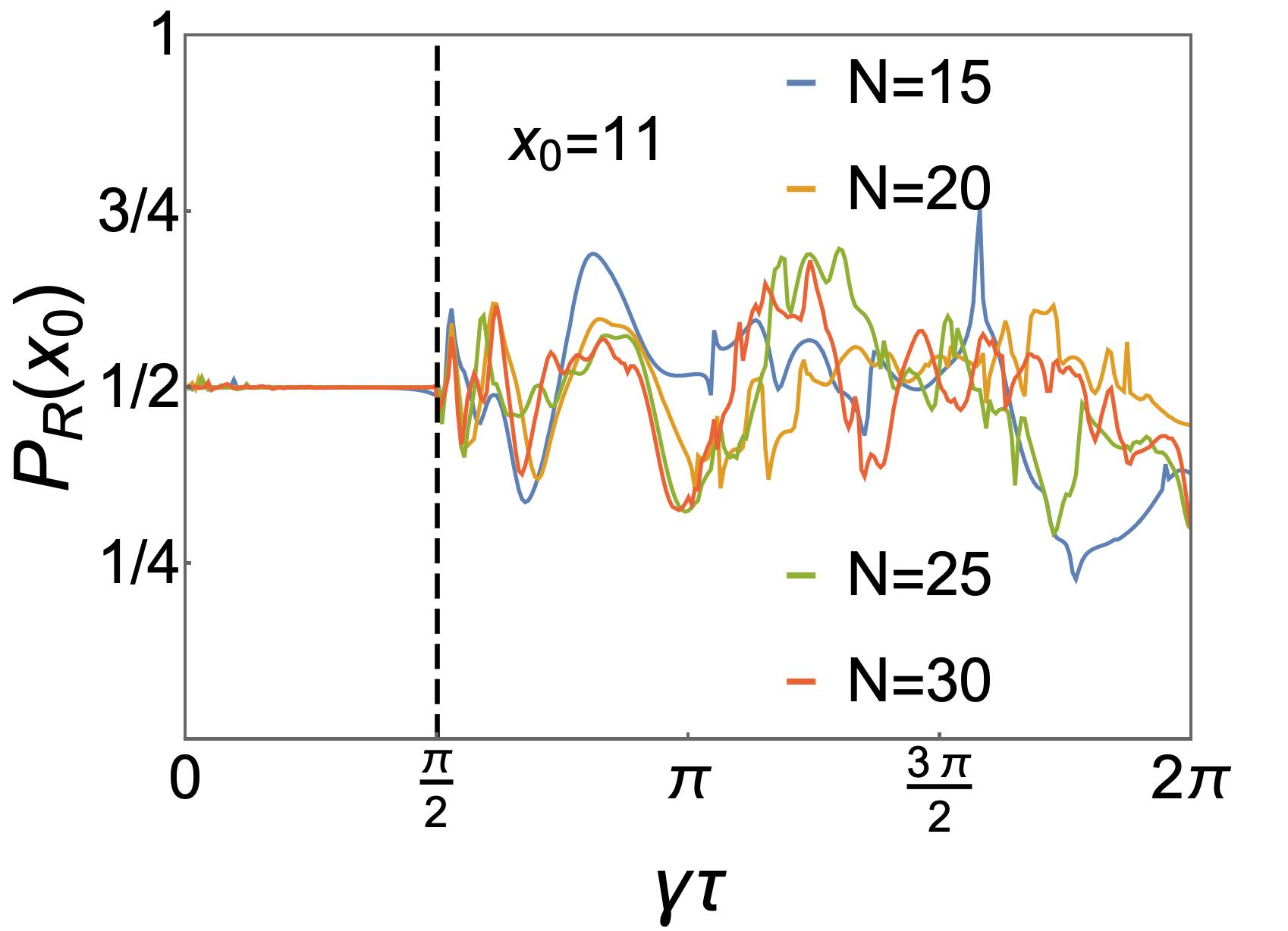} 
\caption{Emergence of the flat-to-fluctuating transition in the splitting probabilities $P_L(x_0)$ (left panel) and $P_R(x_0)$ (right panel) for initial position $x_0 = 11$ and system sizes $N = 15, 20, 25, 30$. Even at these moderate sizes, the transition is visible: the flat regime ($\gamma \tau \leq \pi/2$) shows probabilities near $1/2$, while beyond the critical point ($\gamma \tau > \pi/2$), fluctuations appear with dependence on the values of $N$.}
\label{fig-transition-2}
\end{figure}

So far, we studied the behavior of the splitting probability with respect to the initial position. In Sec.~\ref{summary-sec}, we pointed out that, for large system sizes, the splitting probability undergoes a flat-to-fluctuating transition as the sampling time is varied. To see this, we plot $P_L(x_0)$ as a function of $\gamma \tau$ in Fig.~\ref{fig-transition} for two different choices of the initial position, $x_0=11$ and $x_0=30$. Moreover, in each panel, we show the splitting probability for three different system sizes, $N=200$ (blue), $N=300$ (red) and $N=400$ (green) for a fixed $x_0$. We observe the data collapse across all panels, indicating that the splitting probability is essentially independent of $N$ once the system is sufficiently large. For smaller $N$, we do not get the data collapse, see Fig.~\ref{fig-transition-2}. For $N \gg 1$, $P_L(x_0)$, as shown in Figs.~\ref{fig-transition}(a) and \ref{fig-transition}(b), remain essentially equal to $1/2$, independent of the sampling time till a critical sampling time $\gamma \tau_c $, shown by the dashed vertical line in Fig.~\ref{fig-transition}. In this regime, the splitting probability is also independent of the initial position, as also seen before in Fig.~\ref{fig-tau-N15}(b). Beyond this critical sampling time, $\tau > \tau _c$, we see a sharp change in the splitting dynamics, with a fluctuating pattern of pronounced peaks and dips emerging in the probability as a function of $\gamma \tau$. In this regime, $P_L(x_0)$ deviates from its flat value of $1/2$, and develops a clear dependence on the initial position. Since $P_L(x_0)+P_R(x_0)=1$ [see Eq.~\eqref{aux-eq-5}], the same behavior can also be found for $P_R(x_0)$.
%\bluew{The transition is continuous but non-analytic, since the first derivatives of $P_L(x_0)$ and $P_R(x_0)$ change abruptly from zero to oscillatory at $\tau = \tau_c$, even though the probabilities themselves remain continuous. \textbf{PS: This is correct?}}

The flat-to-fluctuating transition emerges in the limit of large system size. We later demonstrate in Fig.~\ref{fig-Ptau} that the splitting probabilities for small system size $(N=5)$ do not exhibit the transition. However, somewhat surprisingly, one does not need to go to very large systems: even for moderate sizes such as $N=15,20,25,30$, the transition is already clearly visible, as shown in Fig.~\ref{fig-transition-2}. We already mentioned that the critical sampling time in Eq.~\eqref{ajbidhiguqok} is related to the width of the energy band and also emerges in single-target problems \cite{QW6,QW12,Yi2011}. We next provide an understanding of the transition based on the poles associated with the interference term in Eq.~\eqref{aux-eq-2}. {Such pole analysis has been instrumental in the context of other first hitting time problems \cite{QW6, QW11, QW13,QW12, ExptQC-3}. Here, we leverage it to gain some understanding on the flat-to-fluctuating transition, including the location of the transition point given before in Eq.~\eqref{ajbidhiguqok}.}

%To explain this transition, we now look at the poles of the probabilities.

\subsection{Pole analysis}
From Eq.~\eqref{aux-eq-5}, it is clear that the deviation from the flat profile originates from the term $\zeta_{+-}(x_0)$, which itself arises from the interference effect between the two auxiliary single-target detection problems with target states $\ket{d_{\pm}}$ (see Eq.~\eqref{aux-eq-2}). We now demonstrate that the behavior of this interference term changes qualitatively at $\tau_c$. For $\tau<\tau_c$, $\zeta_{+-}(x_0)$ remains negligible in the large-system limit, whereas for $\tau>\tau_c$, it generally becomes appreciable, leading to deviations of the splitting probabilities from $1/2$ via Eq.~\eqref{aux-eq-5}.

%We now demonstrate that for $\tau < \tau _c$, the interference effect is destructive, \textit{i.e.,} $\zeta_{+-}(x_0)=0$, while for $\tau \geq \tau _c$, it is constructive, $\zeta_{+-}(x_0)\neq 0$. This constructive interference, in turn, leads to deviations from $1/2$ via Eq.~\eqref{aux-eq-5}.

Recall that in Eq.~\eqref{pole-eq-1}, $\zeta_{+-}(x_0)$ is given by the real part of a contour integral of a complex function $\mathcal{I}(x_0, z)$ in Eq.~\eqref{pole-eq-2}. It is therefore natural to expect that the analytic structure of $\mathcal{I}(x_0,z)$ controls the behavior of $\zeta_{+-}(x_0)$. We therefore use Eq.~\eqref{pole-eq-2} to numerically find the singularities of $\mathcal{I}(x_0, z)$ in the complex plane. Our analysis shows that the only singularities present are simple poles, whose positions depend on the system parameters. Let $z_j^*$ denote these poles, with $j$ indexing them. 
According to the residue theorem, only the poles enclosed within the unit circle $|z|=1$ contribute to the contour integral in Eq.~\eqref{pole-eq-1}, yielding
\begin{align}
\oint_{|z|=1} \mathcal{I}(x_0, z)\,dz
= 2\pi i 
\sum_{ |z_j^*|<1} 
\operatorname{Residue}\!\left[\mathcal{I}(x_0, z);\, z_j^*\right].
\label{pole-eq-3}
\end{align}
Here, $\operatorname{Residue}\!\left[\mathcal{I}(x_0, z);\, z_j^*\right]$ denotes the residue of the function $\mathcal{I}(x_0, z)$ at the pole $z=z_j^*$. 
The summation in Eq.~\eqref{pole-eq-3} runs over all poles $z_j^*$ that satisfy the condition $|z_j^*|<1$, i.e., all singularities of $\mathcal{I}(x_0, z)$ enclosed within the unit circle.

\begin{figure}[t]
	\centering
	\includegraphics[scale=0.18]{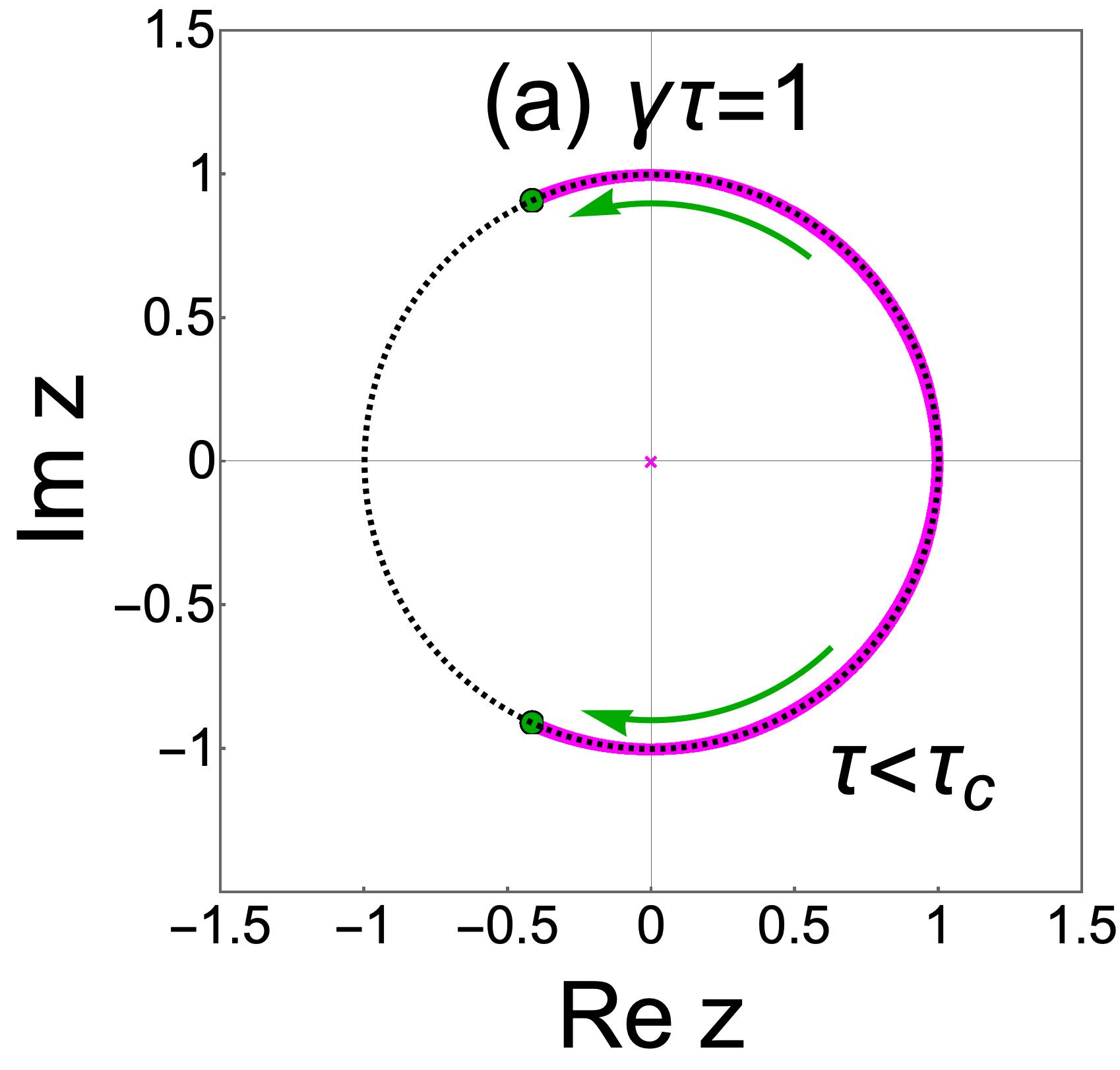}
	\includegraphics[scale=0.18]{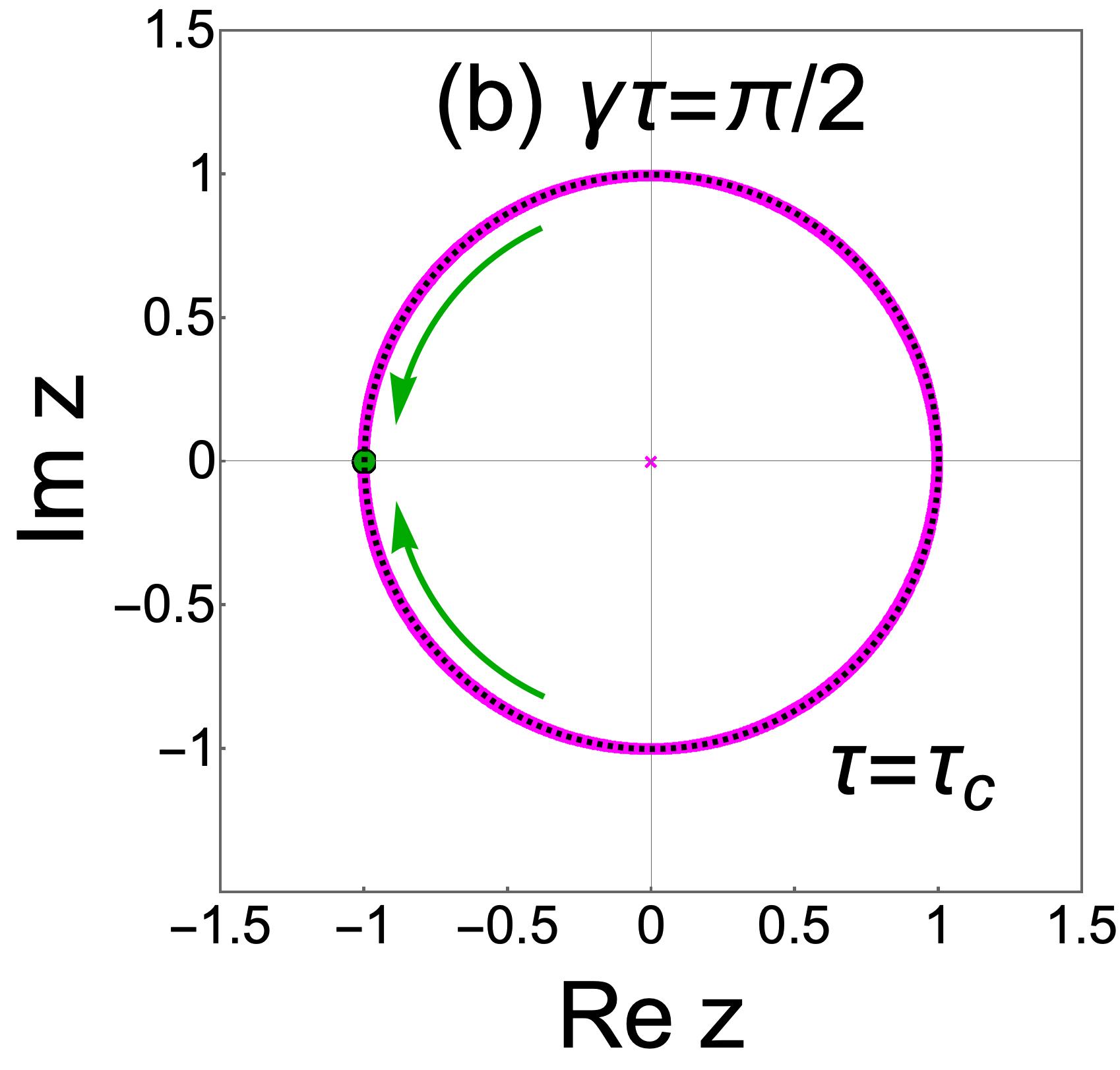}
	\includegraphics[scale=0.18]{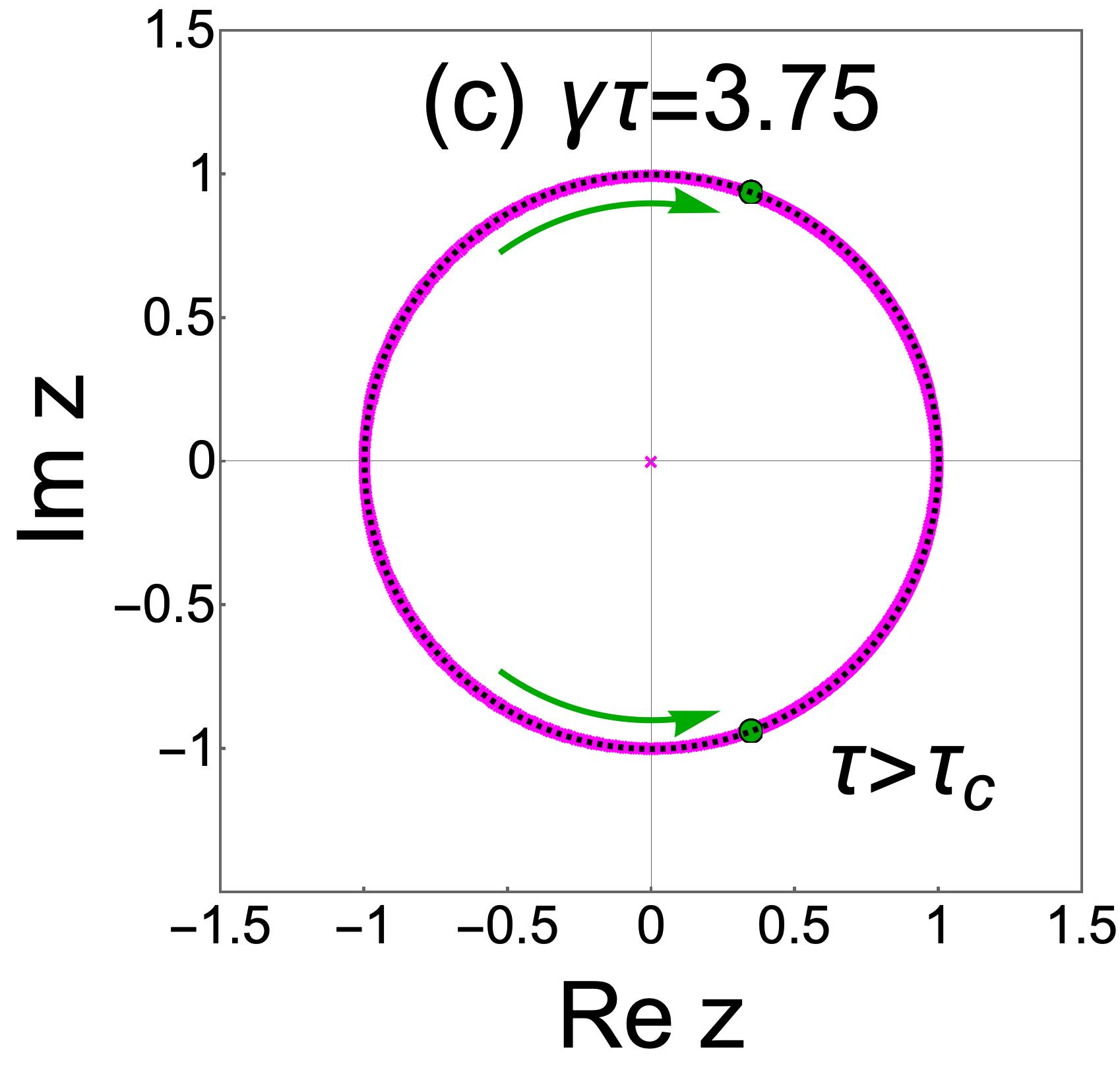}
	\includegraphics[scale=0.18]{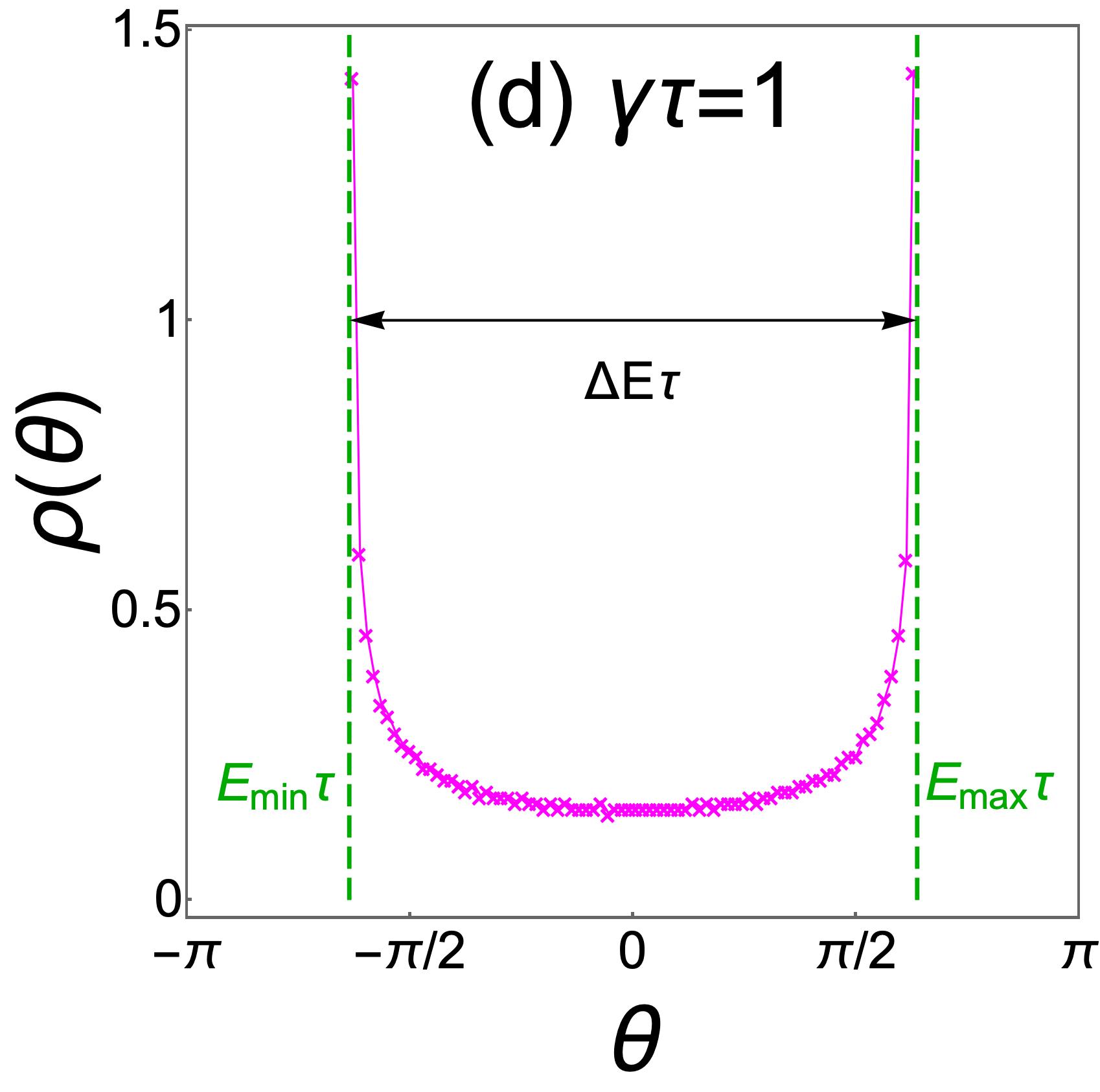}
	\includegraphics[scale=0.18]{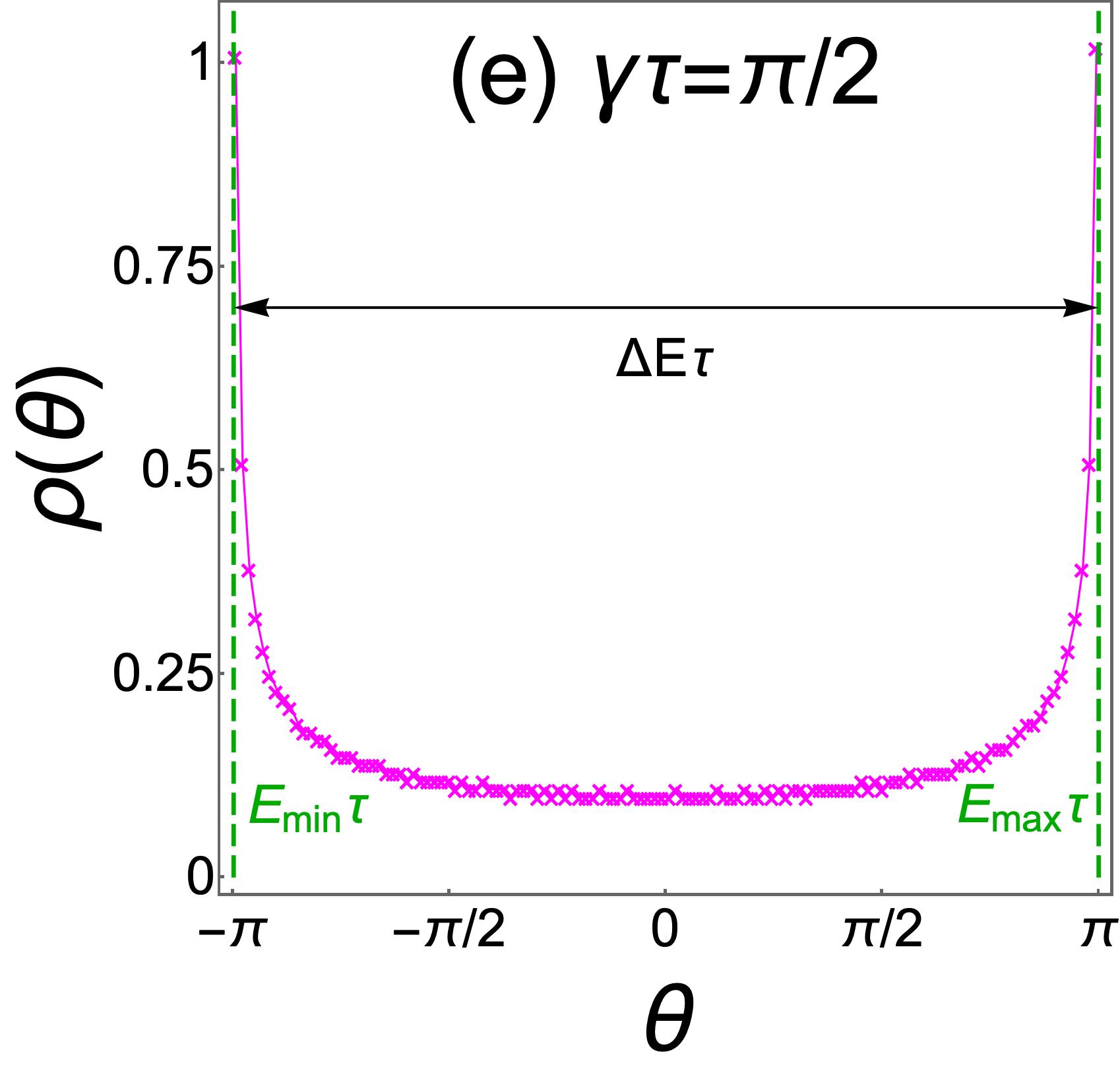}
	\includegraphics[scale=0.18]{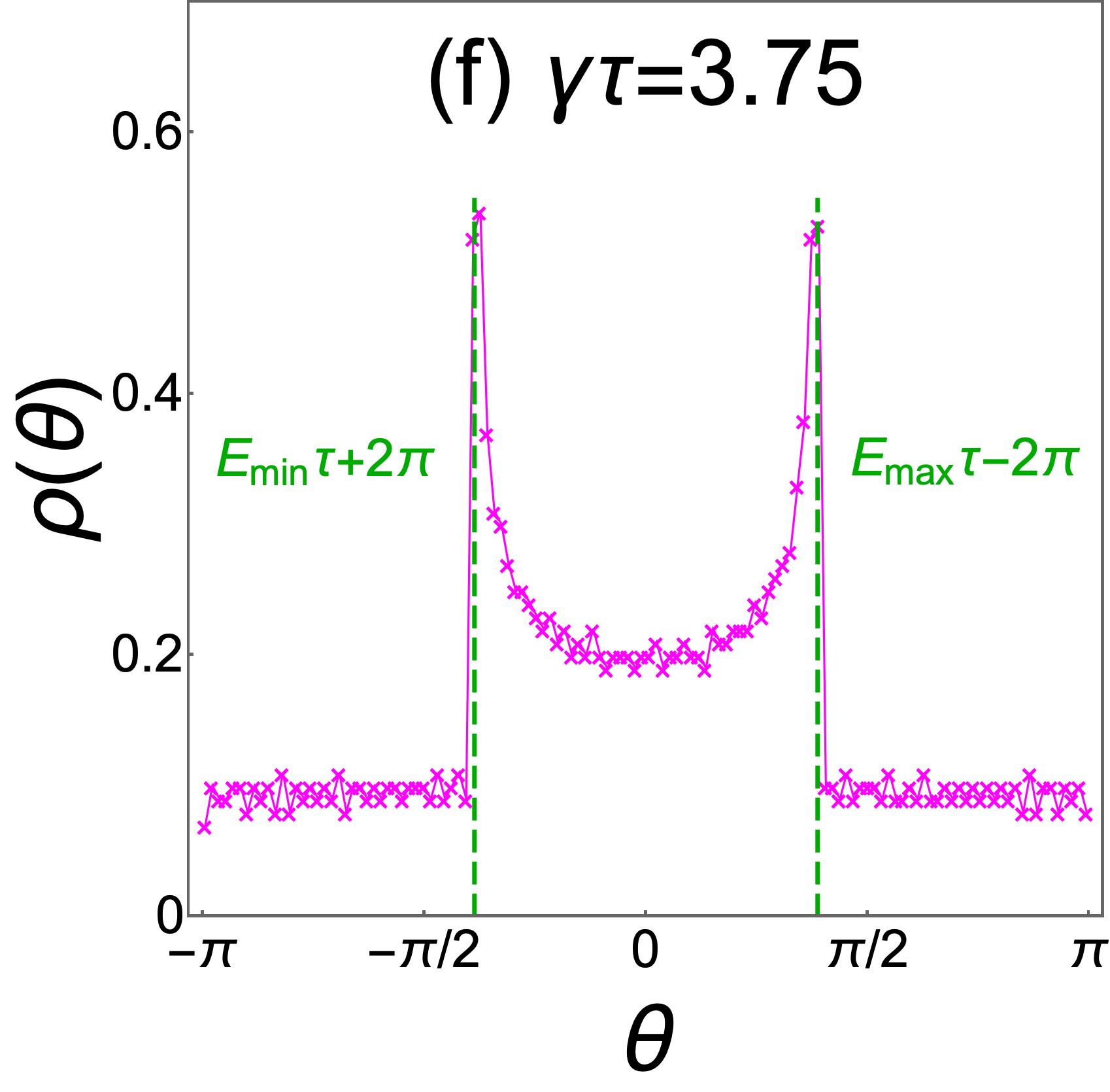}
	\includegraphics[scale=0.13]{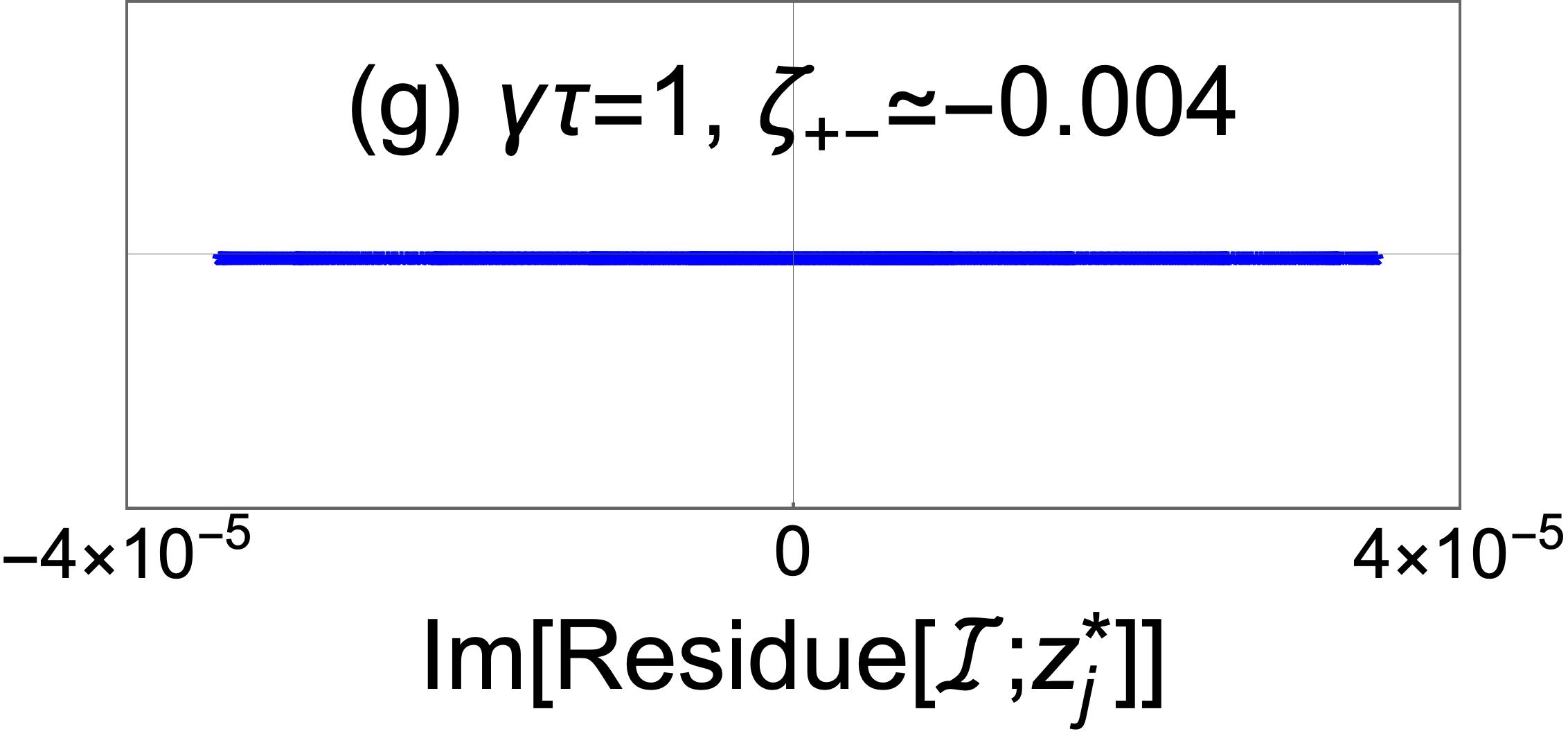} \hspace{0.55cm}
	\includegraphics[scale=0.13]{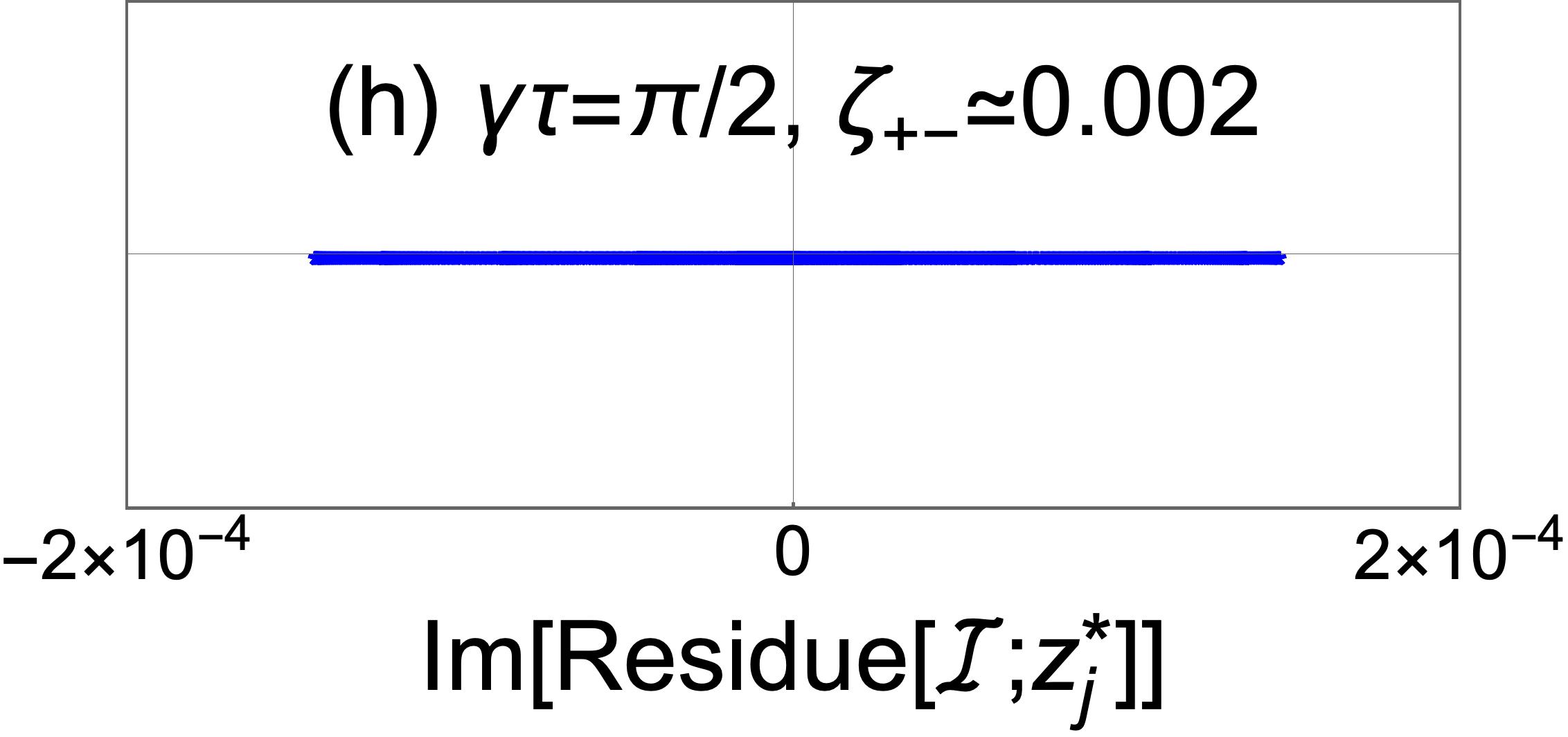} \hspace{0.25cm}
     \includegraphics[scale=0.13]{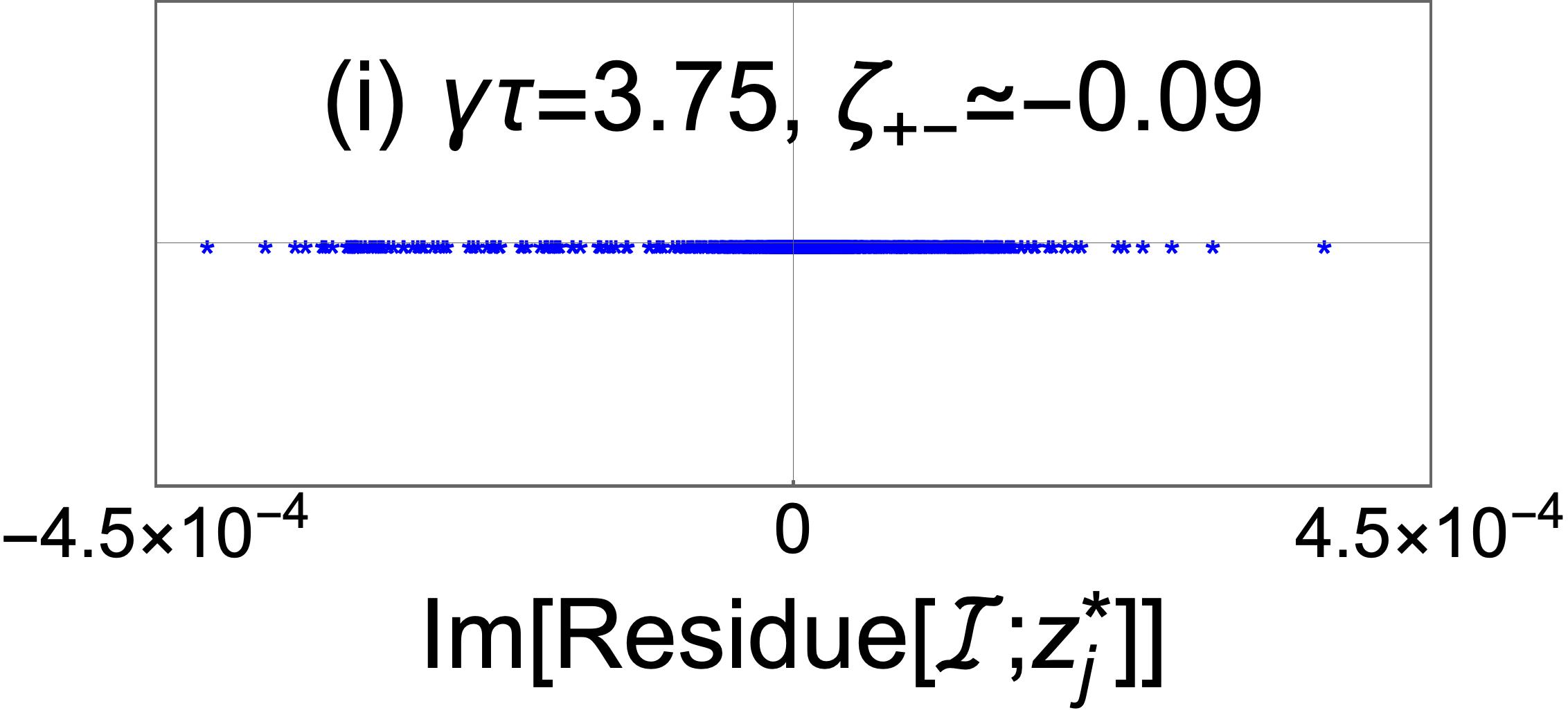}
	\caption{Distribution of poles $(z_j^*)$ and corresponding residues of the integrand $\mathcal{I}(x_0, z)$, defined in Eqs.~\eqref{pole-eq-2}, in the complex $z$-plane, which determines the behavior of the interference term $\zeta_{+-}(x_0)$ in Eq.~\eqref{pole-eq-1}. We show the results for three different sampling times with $\ket{\psi_0}=\ket{x_0}=\ket{11}$ and $N=4000$. 
Panels~(a-c) in the top show the poles (in magenta) for (a) $\tau < \tau _c$, (b) $ \tau = \tau _c$, and (c) for $\tau > \tau _c$. Only poles located inside or on the unit circle will contribute to Eq.~\eqref{pole-eq-1} and are thus shown. For $\gamma \tau = 1$ in (a), all poles, except the one at the origin, lie close to the unit circle. They are distributed along a single arc between the points $a_{*}=\exp \left( -i E_{\rm max} \tau \right)$ and $b_{*}=\exp \left(- i E_{\rm min} \tau \right)$, shown by green dots (the green arrows indicate the direction of the movement of these points as $\tau$ is increased). At the critical point $\gamma \tau_c = \pi/2$ in (b), the two arc endpoints coalesce at $z=-1$, signaling the onset of the transition. For $\tau > \tau _c $, the points $a_{*}$ and $b_{*}$ move past each other, and the distribution of the poles change. To characterize this change, we plot the density of poles, $\rho(\theta)$, along the angular coordinate $\theta \in [-\pi, \pi]$ in the complex plane in panels~(d-f). For $\tau<\tau_c$, $\rho(\theta)$, shown in magenta, exhibits a U-shaped profile, bounded between $E_{\rm min}\tau$ and $E_{\rm max}\tau$ depicted by green lines. At $\tau=\tau_c$, the profile expands over the full range of $\theta$, with $\Delta E \tau=2 \pi$. For $\tau>\tau_c$, the profile changes qualitatively and develops discontinuities at $\theta=E_{\rm max}\tau-2\pi$ and $\theta=E_{\rm min}\tau+2\pi$. The corresponding residues of $\mathcal{I}(x_0, z)$ are symmetrically distributed about the origin for $\tau \leq \tau _c$, but are asymmetrically distributed for $\tau > \tau _c$, see (g-i) for the imaginary part of the residues (in blue) of $\mathcal{I}(x_0, z)$. Due to the asymmetric distribution, the sum of the residues, in general, do not cancel for $\tau > \tau _c$, leading to an appreciable interference contribution $\zeta_{+-}(x_0)$ via Eq.~\eqref{pole-eq-4}. Using this result in Eq.~\eqref{aux-eq-5}, we obtain that the splitting probabilities are essentially equal to $1/2$ for $\tau \leq \tau_c$ and deviate from $1/2$ for $\tau > \tau _c$.
}
	\label{fig-poles}
\end{figure}

Combining Eqs.~\eqref{pole-eq-1} and \eqref{pole-eq-3}
\begin{align}
\zeta _{+-}(x_0) = -2 \pi~ \text{Im} \left[  \sum_{ |z_j^*|<1} 
\operatorname{Residue}\!\left[\mathcal{I}(x_0, z);\, z_j^*\right] \right]. \label{pole-eq-4}
\end{align} 
In Figs.~\ref{fig-poles}(a)-\ref{fig-poles}(c), we employ Eq.~\eqref{pole-eq-2} to numerically obtain the poles of $\mathcal{I}(x_0, z)$ for three different sampling times, (a) $\gamma\tau = 1$, (b) $\gamma\tau = \pi/2$, and (c) $\gamma\tau = 3.75$, corresponding to $\tau < \tau_c$ in Fig.~\ref{fig-poles}(a), $\tau = \tau_c$ in Fig.~\ref{fig-poles}(b), and $\tau > \tau_c$ in Fig.~\ref{fig-poles}(c). We choose $N=4000$ and $x_0=11$ for all panels. Since only poles lying inside or on the unit circle contribute to the residue in Eq.~\eqref{pole-eq-4}, only these poles are displayed. For $\gamma\tau=1$ in Fig.~\ref{fig-poles}(a), all poles except the one at the origin lie close to the unit circle and form a single arc between the points $a_{*}(\tau)$ and $b_{*}(\tau)$, marked by green dots. By varying $\tau$, we identify these endpoints as $a_{*}(\tau) = \exp \left( -i E_{\rm max} \tau \right)$ and $b_{*}(\tau)=\exp \left(- i E_{\rm min} \tau \right)$ ($E_{\rm max} \approx 2 \gamma $ and $E_{\rm min} \approx -2 \gamma $ being the maximum and the minimum energies of $H$ in Eq.~\eqref{unitary-eqn-23} for $N \gg 1$). Increasing $\tau$ further keeps the qualitative feature the same: poles continue to remain along the arc between the points $a_{*}(\tau)$ and $b_{*}(\tau)$. The green arrows indicate the direction of the movement of these points as $\tau$ is increased. At the critical sampling time, $\tau _c = 2 \pi / \Delta E$, $a_{*}(\tau) = b_{*}(\tau)=-1$. The two endpoints meet and the arc therefore closes in Fig.~\ref{fig-poles}(b). For $\tau > \tau _c $ in Fig.~\ref{fig-poles}(c), the points $a_{*}(\tau)$ and $b_{*}(\tau)$ move past each other, and the distribution of the poles changes.

To characterize this change, we study the density of poles, $\rho(\theta)$, along the angular coordinate $\theta \in [-\pi, \pi]$ in the complex plane in Figs.~\ref{fig-poles}(d)-\ref{fig-poles}(f). More specifically, $\rho(\theta)d\theta$ gives the fraction of poles whose angular coordinates lie in the interval $[\theta, \theta + d \theta]$. We plot $\rho(\theta)$ vs $\theta$ for the three sampling times mentioned above. For $\tau<\tau_c$, $\rho(\theta)$, shown in Fig.~\ref{fig-poles}(d), exhibits a characteristic U-shaped profile bounded between $E_{\rm min}\tau$ and $E_{\rm max}\tau$, indicated by the green lines. As $\tau$ increases, the two boundaries move apart and the support of $\rho(\theta)$ expands, while its qualitative U-shaped form remains unchanged. At the critical point, $\Delta E\,\tau_c=2\pi$, the density $\rho(\theta)$ spans the entire possible range of $\theta$, while still having the $U$-shape; see Fig.~\ref{fig-poles}(e). This corresponds to the meeting of the two endpoints $a_*(\tau)$ and $b_*(\tau)$ in Fig.~\ref{fig-poles}(b). For $\tau>\tau_c$, the U-shaped profile of $\rho(\theta)$ changes qualitatively, as shown in Fig.~\ref{fig-poles}(f), as it develops discontinuities at $\theta= \left( E_{\rm max}\tau-2\pi \right)$ and $\theta= \left( E_{\rm min}\tau+2\pi \right)$, indicated by the green lines.

According to Eq.~\eqref{pole-eq-4}, the qualitative change in the distribution of poles affects the imaginary part of the residue corresponding to $\mathcal{I}(x_0, z)$ in Eq.~\eqref{pole-eq-2} arising due to the poles. For $\tau \leq \tau_c$, we find in Figs.~\ref{fig-poles}(g)-\ref{fig-poles}(h) that the residues exhibit a symmetric distribution about the origin. As a result, their summed contribution is negligible, this, according to Eq.~\eqref{pole-eq-4}, implies that the resulting $\zeta_{+-}(x_0)$ is negligible. For example, we obtain $\zeta_{+-}(x_0) \approx -0.004$ for $\gamma \tau = 1$ in Fig.~\ref{fig-poles}(g) and $\zeta_{+-}(x_0) \approx 0.002$ for $\gamma \tau = \pi/2$ in Fig.~\ref{fig-poles}(h). Plugging this result in Eq.~\eqref{aux-eq-5} then gives $P_{L}(x_0) \approx P_{R}(x_0) \approx 1/2$. By contrast, the residues for $\tau > \tau_c$ are not symmetrically  distributed; see Fig.~\ref{fig-poles}(i). Summing all the residues using Eq.~\eqref{pole-eq-4}, we find that $\zeta _{+-}(x_0) $ is, in general, not negligible. For instance, for Fig.~\ref{fig-poles}(i), $\gamma \tau = 3.75$, $\zeta _{+-}(x_0) \approx -0.09$. This result signals that the probabilities $P_L(x_0)$ and $P_R(x_0)$ differ from the flat value of $1/2$ in the $\tau > \tau _c$ regime. This, in turn, leads to the flat-to-fluctuating transition in Fig.~\ref{fig-transition} in the large-$N$ limit.

Although the discussion above focuses on the large-$N$ limit, the flat-to-fluctuating transition, as noted before in Fig.~\ref{fig-transition-2}, is already clearly visible for moderate system sizes like $N=15,20,25,30$. For moderate $N$, the poles can also display a radial rearrangement in the complex plane, in addition to the angular rearrangement discussed above \cite{myletter}. In the regime $\tau\leq\tau_c$, the poles remain close to the unit circle, as in Figs.~\ref{fig-poles}(a) and \ref{fig-poles}(b), even for moderate $N$, whereas for $\tau>\tau_c$, some of the poles move appreciably into the interior of the unit disk, giving rise to a irregular radial arrangement. However, as the system size is increased, the radial irregularity becomes progressively less pronounced and the poles again accumulate close to the unit circle, as observed in Fig.~\ref{fig-poles}(c). Consequently, in the $\tau>\tau_c$ regime, the moderate-$N$ splitting probabilities in Fig.~\ref{fig-transition-2} exhibit more pronounced and irregular fluctuations than those observed for large system sizes in Fig.~\ref{fig-transition}.

\section{Resonance conditions and exceptional sampling times}
\label{sec-resonance}
We now turn to the finite-$N$ case and focus on the special sampling times, as was mentioned above in Sec.~\ref{summary-sec}. As shown in Fig.~\ref{fig-Ptau} for $N=5$ and different initial positions, both $P_L(x_0)$ and $P_R(x_0)$ exhibit point-wise discontinuous jumps at these special sampling times, and satisfy $P_L(x_0)+P_R(x_0) < 1$. To account for these jumps, one must use the general expression given in Eq.~\eqref{neww-eg-eq-17}. In this section, we explicitly derive the resonance conditions, which was previously pointed out in Eq.~\eqref{main-eqq-3}.

%Most of our analysis so far is based on Eq.~\eqref{aux-eq-5}, which was derived under the assumption that all quasi-energies are distinct. This expression captures only the smooth component of the splitting probabilities. However, as shown in Fig.~\ref{fig-Ptau} and was mentioned before in Sec.~\ref{summary-sec}, for certain exceptional measurement times, $P_L(x_0)$ and $P_R(x_0)$ exhibit point-wise discontinuous jumps that are not described by Eq.~\eqref{aux-eq-5}. We saw an example of this jump for $\tau = 0$. Here, we show it for other non-zero values of $\tau$. To account for these resonances, one must use the general expression given in Eq.~\eqref{neww-eg-eq-17}. In this section, we explicitly derive the resonance conditions, which was previously pointed out in Eq.~\eqref{main-eqq-3}.

%However, for certain exceptional measurement times, as pointed out in Eq.~\eqref{main-eqq-3}, some quasi-energies coincide, giving rise to sharp and discontinuous jumps in $P_L(x_0)$ and $P_R(x_0)$. In this section, we explicitly derive this resonance condition.

%In Eqs.~\eqref{main-eqq-3} and \eqref{main-eqq-4}, we stated the conditions required for observing resonances in the splitting probability. To derive them, 

It is instructive to note that, using Eq.~\eqref{neww-eg-eq-1}, the energy eigenfunction \( \psi_\ell(x)=\braket{x|E_\ell} = \sqrt{\frac{2}{N+1}} \sin\!\left( \frac{\pi \ell x}{N+1} \right)\) (recall $\ell \in \{1,2, \ldots,N \} $) possesses definite parity with respect to reflection about the center of the chain. Under the transformation \(x \to(N+1-x) \), the eigenfunctions transform as 
\begin{align}
\psi_\ell(N+1-x) = (-1)^{\ell+1} \psi_\ell(x),
%\langle N+1-x | E_k \rangle = (-1)^{k+1} \langle x | E_k \rangle,
\end{align}
indicating that eigenfunctions with odd \(\ell\) are symmetric under reflection, 
while those with even \(\ell\) are antisymmetric. 
Substituting this property into Eqs.~\eqref{neww-eg-eq-8} and \eqref{neww-eg-eq-82}, we find that the auxiliary target state \(\ket{d_+}\) involves only symmetric eigenstates, 
whereas \(\ket{d_-}\) involves antisymmetric eigenstates.

This identification allows us to recast the auxiliary probabilities $\rho _{\pm} \left( x _0 \right)$ in Eq.~\eqref{new-pseb-eq-2} in terms of the parity-specific sectors of the spectrum
\begin{align}
& \rho _{+}(x _0) = \sump_{\ell \in ~\rm{odd}} 
      \frac{\big| \sum_{m=1}^{g_{\ell}} 
         \braket{d_{+} \big| E_{\ell,m}} \braket{E_{\ell,m} \big| x _0} \big|^2}
           { \sum_{m=1}^{g_{\ell}} \big| \braket{d_{+} \big| E_{\ell,m}} \big|^2},  \label{aux-eq-6} \\
           & \rho _{-}(x _0) = \sump_{\ell \in ~\rm{even}} 
      \frac{\big| \sum_{m=1}^{g_{\ell}} 
         \braket{d_{-} \big| E_{\ell,m}} \braket{E_{\ell,m} \big| x _0} \big|^2}
           { \sum_{m=1}^{g_{\ell}} \big| \braket{d_{-} \big| E_{\ell,m}} \big|^2}, \label{aux-eq-7}
\end{align}
%Since $\braket{E_k|d_+} = 0$ when $k$ is even, $\rho_{+}(x_0)$ receives contributions only from the odd-$k$ sector of the quasi-energy spectrum. 
%Similarly, $\rho_{-}(x_0)$ involves contributions solely from the even-$k$ sector. 
and similarly for the auxiliary generating functions $\widetilde{\chi}_{\pm} \left( x _0 \right)$ in Eq.~\eqref{aux-eq-1}
\begin{align}
& \widetilde{\chi} _{+}(z) =  \frac{\sum _{\ell \in \text{ odd}} \sum _{m=1}^{g_{\ell}} \bra{d_+} E_{\ell,m} \rangle \langle E_{\ell,m} \ket{x_0} ~\frac{z e^{-i \lambda _{\ell}}}{1-z e^{-i \lambda _{\ell}}} }{\sum _{\ell \in \text{ odd}} \sum _{m=1}^{g_{\ell}} \big| \braket{d_+|E_{\ell,m}} \big|^2 ~\frac{1}{1-z e^{-i \lambda _{\ell}}}}  , \label{aux-eq-8} \\
& \widetilde{\chi} _{-}(z) =  \frac{\sum _{\ell \in \text{ even}} \sum _{m=1}^{g_{\ell}} \bra{d_-} E_{\ell,m} \rangle \langle E_{\ell,m} \ket{x_0} ~\frac{z e^{-i \lambda _{\ell}}}{1-z e^{-i \lambda _{\ell}}} }{\sum _{\ell \in \text{ even}} \sum _{m=1}^{g_{\ell}} \big| \braket{d_-|E_{\ell,m}} \big|^2 ~\frac{1}{1-z e^{-i \lambda _{\ell}}}}  . \label{aux-eq-9}
\end{align}
In all of these expressions, the summation in a given sector runs over the distinct quasienergies $\lambda_\ell = E_\ell \tau$ mod $2 \pi$. As noted previously, for most values of $\tau$, the number of distinct $\lambda _\ell$ is same as the number of $E_\ell$, leading to Eq.~\eqref{aux-eq-5}. For certain exceptional measurement times, however, two or more quasienergies from the same parity-sector can become equal, as shown in Fig.~\ref{fig-phase-merge}(b). Then
\begin{align}
\lambda _\ell = \lambda_{\ell'} \implies e^{-i E_\ell \tau } = e^{-i E_{\ell'} \tau }, \label{resonance-eqn}
\end{align}
where $\ell \neq \ell'$ and $\ell, \ell'$ are both odd or both even, so that the associated eigenstates, $\ket{E_\ell}$ and $\ket{E_{\ell'}}$, have the same parity under reflection. At such exceptional sampling times, the sums appearing in $\rho_{\pm}(x_0)$ and $\widetilde{\chi}_{\pm}(z)$ in Eqs.~\eqref{aux-eq-6}--\eqref{aux-eq-9} are reorganized, because several previously separate terms merge into a single degenerate contribution. This reorganization produces discontinuous changes in the auxiliary quantities, which in turn appear as sharp jumps in the splitting probabilities $P_L(x_0)$ and $P_R(x_0)$. Thus, Eq.~\eqref{resonance-eqn} gives the resonance condition, equivalent to Eq.~\eqref{main-eqq-3} noted before.

By contrast, quasienergy coincidences between different parity sectors (i.e., odd $\ell$, even $\ell'$ or vice versa) do not alter the structure of the summation, and therefore do not generate such resonances. We next turn to a more physical picture that illustrates that the appearance of the resonance is related to the formation of dark states that are orthogonal to both targets, $\ket{D_1} = \ket{1}$ and $\ket{D_2} = \ket{N}$.

Contrast the phase-matching condition associated with the resonances in Eq.~\eqref{resonance-eqn} with that in Eq.~\eqref{SS-pos-eq-11}. Although both conditions correspond to matching of the dynamical phases, they involve different parity sectors and lead to different physical effects. The resonance condition in Eq.~\eqref{resonance-eqn} involves states of the same parity and is associated to the formation of dark states (as shown next). In contrast, Eq.~\eqref{SS-pos-eq-11} involves states of opposite parity and governs the interference between the two parity sectors, which gives rise to the spatial oscillations of the splitting probabilities.

\subsection{A physical picture based on dark-states}
For simplicity, consider that $\ell$ and $\ell'$ are both odd $(\ell \neq \ell')$, and we have the phase-matching condition in Eq.~\eqref{resonance-eqn}. 
%From Eqs.\eqref{neww-eg-eq-20} and \eqref{neww-eg-eq-23}, it is clear that both $k$ and $l$ contribute only through the state $\ket{d_+}$. 
In this case, the following state is orthogonal with respect to the both target states $\ket{d_{\pm}}$ in the auxiliary single-target detection problems
\begin{align}
\ket{\mathbb{D}_{\ell \ell'}^{(+)}} & \propto \left(  \langle d_+ \big| E_{\ell'} \rangle  \ket{E_\ell}-\langle d_+ \big| E_{\ell} \rangle  \ket{E_{\ell'}} \right), \label{neww-eg-eq-28-1} \\ 
& \propto \left[  \sin \left(\frac{\pi \ell' }{N+1} \right) \ket{E_\ell}-\sin \left(\frac{\pi \ell}{N+1} \right) \ket{E_{\ell'}} \right].  \label{neww-eg-eq-28-11} 
\end{align}
Remember that $\langle d_- \big| E_{\ell} \rangle = \langle d_- \big| E_{\ell'} \rangle = 0$ when both $\ell$ and $\ell'$ are odd [see Eq.~\eqref{neww-eg-eq-82}]. This leads to $\langle d_- \big| {\mathbb{D}_{\ell \ell'}^{(+)}}  \rangle = 0$. We also have $\langle d_+ \big| {\mathbb{D}_{\ell \ell'}^{(+)}}  \rangle = 0$. Together, these conditions establish that $\ket{\mathbb{D}_{\ell \ell'}^{(+)}}$ is orthogonal with respect to the target states, $\ket{D_1}$ and $\ket{D_2}$, in the original splitting problem
\begin{align}
\braket{D_1 \big| \mathbb{D}_{\ell \ell'}^{(+)}} =\braket{D_2 \big| \mathbb{D}_{\ell \ell' }^{(+)}} =0. \label{neww-eg-eq-28-10982} 
\end{align}
The phase-matching condition implies that $\ket{\mathbb{D}_{\ell \ell'}^{(+)}}$ is an eigenstate of the unitary operator precisely at the measurement time: $U(\tau)\ket{\mathbb{D}_{\ell  \ell' }^{(+)}} = e^{-i E_\ell \tau} \ket{\mathbb{D}_{\ell \ell' }^{(+)}}$. This result, along with Eq.~\eqref{neww-eg-eq-28-10982}, implies that $\ket{\mathbb{D}_{\ell   \ell'   }^{(+)}}$ is also an eigenstate of the survival operator $\mathcal{S} = \Big(\mathbb{1}-\hat{D} \Big) U(\tau)$ appearing in Eq.~\eqref{unitary-eqn-13}
\begin{align}
\mathcal{S} \ket{\mathbb{D}_{\ell  \ell'}^{(+)}} = e^{-i E_\ell \tau} \ket{\mathbb{D}_{\ell  \ell'}^{(+)}}, \label{neww-eg-eq-28-2}
\end{align}
where $\hat{D}$, defined in Eq.~\eqref{unitary-eqn-8}, is the projection operator onto the detector subspace. This means that $\braket{D_{1}| U(\tau) \mathcal{S}^{n-1}| \mathbb{D}_{\ell   \ell'   }^{(+)}} =\braket{D_{2}| U(\tau) \mathcal{S}^{n-1}| \mathbb{D}_{\ell    \ell'   }^{(+)}} = 0$, implying that $\ket{\mathbb{D}_{\ell  \ell'   }^{(+)}}$ is a dark state with respect to both $\ket{D_1}$ and $\ket{D_2}$ at the measurement times $\tau,~2 \tau,~3 \tau$ and so on. This state is never detected by the target detectors at the moments of measurement, and the corresponding splitting probabilities vanish. 

If the initial state $\ket{x_0}$ of the quantum walker has a finite overlap with this dark state, \textit{i.e.,} $\braket{x_0 \big| \mathbb{D}_{\ell  \ell' }^{(+)}} \neq 0$, then the phase-matching condition implies that this component is completely decoupled from the two detectors $\ket{D_1}$ and $\ket{D_2}$, and a dark state is formed during the measurement attempts. As a result, we get a sudden reduction in the splitting probabilities. This will, however, not happen if $\ket{x_0}$ is orthogonal to $\ket{\mathbb{D}_{\ell  \ell'  }^{(+)}}$.

To summarize, the conditions for observing a resonance in $P_L(x_0)$ and $P_R(x_0)$ arising from the symmetric parity-sector (i.e., both $\ell$ and $ \ell' $ are odd) are
\begin{align}
\boxed{e^{-i E_\ell \tau} = e^{-i E_{\ell'} \tau},~~\text{and }\braket{x_0 \big| \mathbb{D}_{\ell \ell' }^{(+)}} \neq 0,} \label{neww-eg-eq-28-3}
\end{align}
and it related to the formation of a dark state right at the stroboscopic measurement times $\tau,~2 \tau,~3 \tau$ and so on.

Similarly, when $\ell$ and $ \ell' $ are both even, the resonance conditions are
\begin{align}
\boxed{e^{-i E_\ell \tau} = e^{-i E_{\ell'} \tau},~~\text{and }\braket{x_0 \big| \mathbb{D}_{\ell  \ell'  }^{(-)}} \neq 0,} \label{neww-eg-eq-28-4}
\end{align}
where
\begin{align}
\ket{\mathbb{D}_{\ell  \ell'  }^{(-)}} &= \mathbb{N} \left(  \langle d_- \big| E_{\ell'} \rangle  \ket{E_\ell}-\langle d_- \big| E_\ell \rangle  \ket{E_{\ell'}} \right), \label{neww-eg-eq-28-5} \\ 
&= \mathbb{N} \left[  \sin \left(\frac{\pi \ell'  }{N+1} \right) \ket{E_\ell}-\sin \left(\frac{\pi \ell}{N+1} \right) \ket{E_{\ell'}} \right], \label{neww-eg-eq-28-6} 
\end{align}
is a dark-state with respect to $\ket{D_1}$ and $\ket{D_2}$
\begin{align}
\braket{D_1 \big| \ket{\mathbb{D}_{   \ell   \ell'  }^{(-)}}} =\braket{D_2 \big| \ket{\mathbb{D}_{    \ell  \ell'   }^{(-)}}} =0. \label{neww-eg-eq-28-109821} 
\end{align}
Eqs.~\eqref{neww-eg-eq-28-3} and \eqref{neww-eg-eq-28-4} thus specify the special $\tau$ values, characterized by the quasienergy phase matching together with a nonvanishing overlap of the initial state with the corresponding dark subspace. They were stated before in Eq.~\eqref{main-eqq-3}. Furthermore, Eqs.~\eqref{neww-eg-eq-28-10982} and \eqref{neww-eg-eq-28-109821} show that no resonance arises when the initial position coincides with one of the targets, \textit{i.e.}, $\ket{x_0} = \ket{D_1}$ or $\ket{x_0} = \ket{D_2}$.

\section{A case study: $N=5$}
\label{sec-example}
As a concrete example, we apply our framework to the nearest-neighbor tight-binding Hamiltonian in Eq.~\eqref{main-eqq-1} for $N=5$, and compute the splitting probabilities, and the resonance points. Following Eq.~\eqref{neww-eg-eq-1}, the eigenvalues of $H$ in Eq.~\eqref{main-eqq-1} are $\{ -\sqrt{3}\gamma, -\gamma, 0, \gamma, \sqrt{3} \gamma \} $, and its eigenstates are $\ket{E_\ell} =  \left(  \sin \omega _\ell,~ \sin 2\omega _\ell, ~\sin 3 \omega _\ell , ~\sin 4 \omega _\ell ,~\sin 5 \omega _\ell \right)^{T}  /\sqrt{3} $ with $\omega _\ell = \pi \ell /6$ ($T$ is the transpose). Observe that all eigenvalues are distinct. Furthermore, recall that the measurements are made at the boundary sites, $\ket{D_1} =\ket{1}$ and $\ket{D_2} =\ket{5}$. Using Eqs.~\eqref{neww-eg-eq-8}-\eqref{neww-eg-eq-82}, we have
\begin{align}
&\ket{d_+} = \frac{1}{\sqrt{6}} \ket{E_1} + \sqrt{\frac{2}{3}} \ket{E_3} + \frac{1}{\sqrt{6}} \ket{E_5}, \label{neww-eg-eq-34} \\
& \ket{d_-} = \frac{1}{\sqrt{2}} \ket{E_2}+  \frac{1}{\sqrt{2}} \ket{E_4}. \label{neww-eg-eq-35} 
\end{align}
Below, we compute the splitting probabilities for three different localized initial conditions, $\ket{x_0}=\ket{3}, \ket{4}, \ket{5}$. The other two localized initial conditions, $\ket{x_0}=\ket{1}, \ket{2}$, can be treated by exploiting the symmetry
\begin{align} 
\begin{cases}
 & P_R(1) = P_L(5),~~P_L(1) = P_R(5) ,\\& P_R(2) = P_L(4),~~P_L(2) = P_R(4). 
 \end{cases}
 \label{nauq}
\end{align}

\subsection{Example I: $\ket{x_0} = \ket{3}$} 
We choose to first consider the initial condition $\ket{x_0} = \ket{3}$, located at the center of the chain and therefore symmetric relative to the boundaries. Eq.~\eqref{neww-eg-eq-2} gives
\begin{align}
\ket{3} = \frac{1}{\sqrt{3}} \left( \ket{E_1}-\ket{E_3}+\ket{E_5}  \right). \label{neww-eg-eq-36} 
\end{align}
Turning to the splitting probability in Eq.~\eqref{aux-eq-5}, we first consider the non-special $\tau$, for which all quasienergies are distinct. To evaluate the interference term in this expression, we use the auxiliary $\widetilde{\chi} _{-}(z)$ from Eq.~\eqref{aux-eq-9} with $g_{\ell}=1$. Since $\ket{x_0} = \ket{3}$ involves only odd values of $\ell$ in Eq.~\eqref{neww-eg-eq-36}, we find $\widetilde{\chi} _{-}(z)=0$. Hence, the interference term in Eq.~\eqref{aux-eq-5} vanishes, and we get
\begin{align}
\boxed{P_L(3) = P_R(3) = 1/2.} \label{neww-eg-eq-39} 
\end{align}
Given that $\ket{x_0}=\ket{3}$ is exactly in the middle of the chain, equidistant from both boundaries, the two splitting probabilities are equal to each other, as expected from the symmetry.

%Eq.~\eqref{neww-eg-eq-39} is derived assuming all quasi-energy values are non-degenerate. 
We next turn to special values of $\tau$, where two or more quasienergies become degenerate. The first case is $\tau=0$, in which all quasienergies collapse to zero. As discussed in Eq.~\eqref{neww-eg-eq-28}, the walker is completely restricted to the interior of the chain, and is never detected at the boundaries, \textit{i.e.}, $P_L(3) = P_R(3) = 0$.

The second case corresponds to the special $\tau$ where the phases coincide $e^{-iE_1 \tau} = e^{-iE_3 \tau} $. Since $E_3 = 0$ and $E_5 = -E_1 = \sqrt{3}\gamma$, this also implies $e^{-iE_5 \tau} = e^{-iE_3 \tau}$; see Fig.~\ref{fig-phase-merge}(b). The corresponding time interval is $\sqrt{3}\gamma \tau = 2 m \pi$ with $m=1,2,\ldots$. At these times, the initial state $\ket{x_0}=\ket{3}$ is fully revived [see Eq.~\eqref{neww-eg-eq-36}], so the walker is never detected at the boundaries. Consequently, the splitting probabilities vanish, $P_L(3) = P_R(3) = 0$.

The third case arises when $e^{-iE_1 \tau} = e^{-iE_5 \tau}$ but they are not equal to unity. In this situation, the dark state from Eq.~\eqref{neww-eg-eq-28-11} takes the form $\ket{\mathbb{D}_{15}^{(+)}} \sim \left( \ket{E_1} -\ket{E_5} \right) $, and Eq.\eqref{neww-eg-eq-36} gives $\braket{3 | \mathbb{D}_{15}^{(+)}} = 0$. Thus, the overlap condition in Eq.\eqref{neww-eg-eq-28-3} is not satisfied, and no resonance occurs. For the same reason, the case $e^{-iE_2 \tau} = e^{-iE_4 \tau}$ also does not lead to a resonance.

\begin{figure}[t]
	\centering
	\includegraphics[scale=0.23]{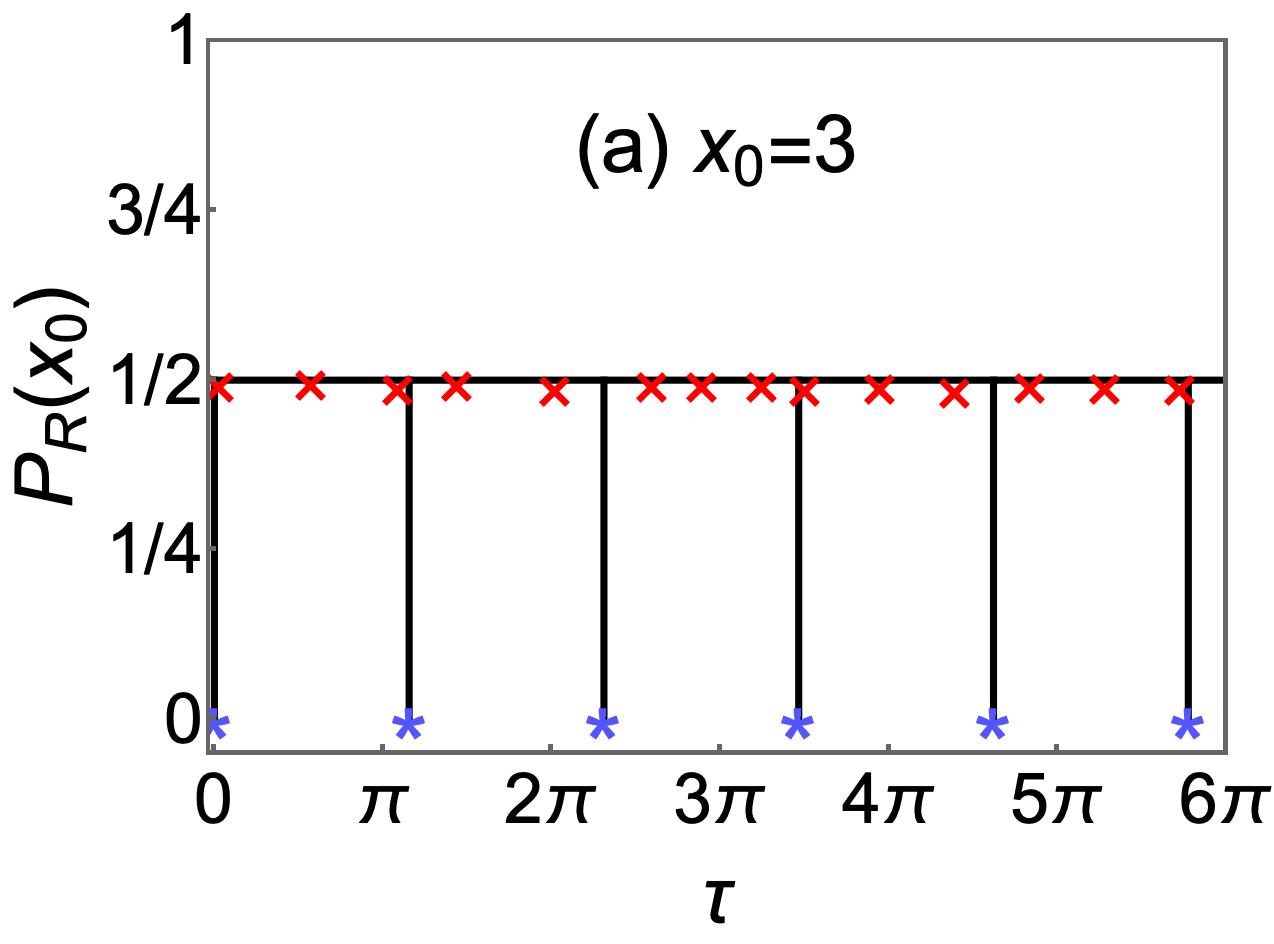}
	\includegraphics[scale=0.22]{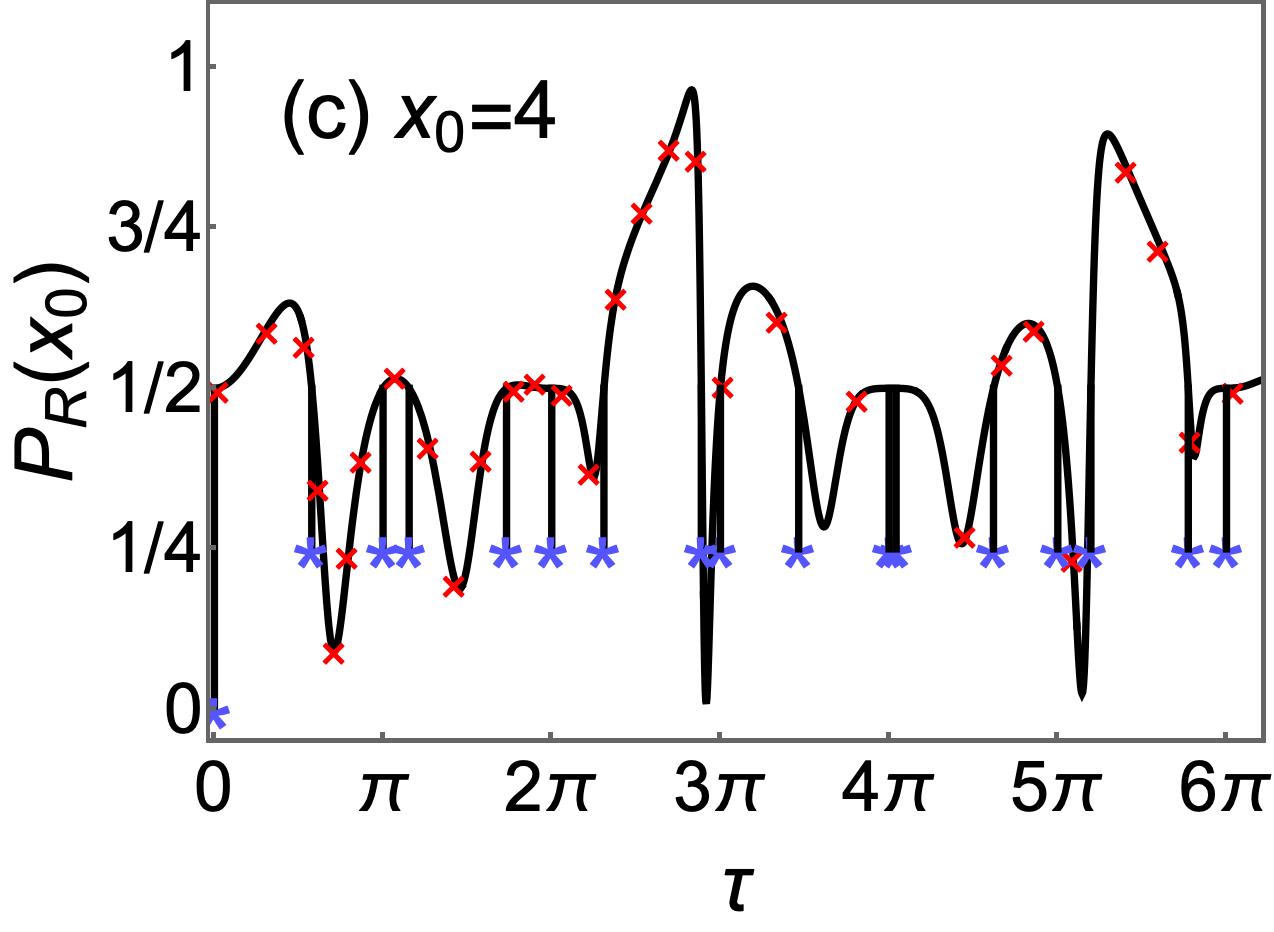}
	\includegraphics[scale=0.22]{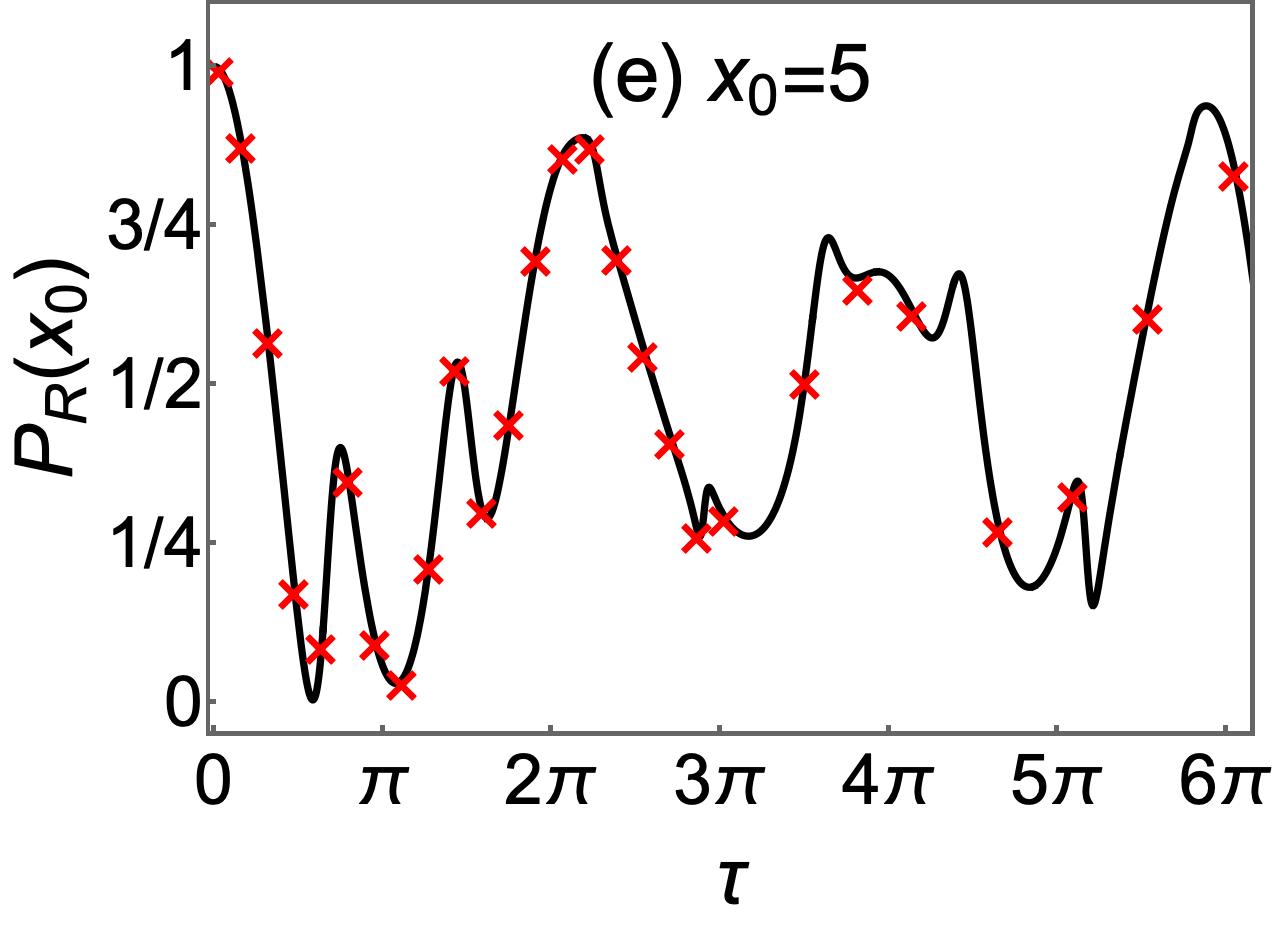}
	\includegraphics[scale=0.23]{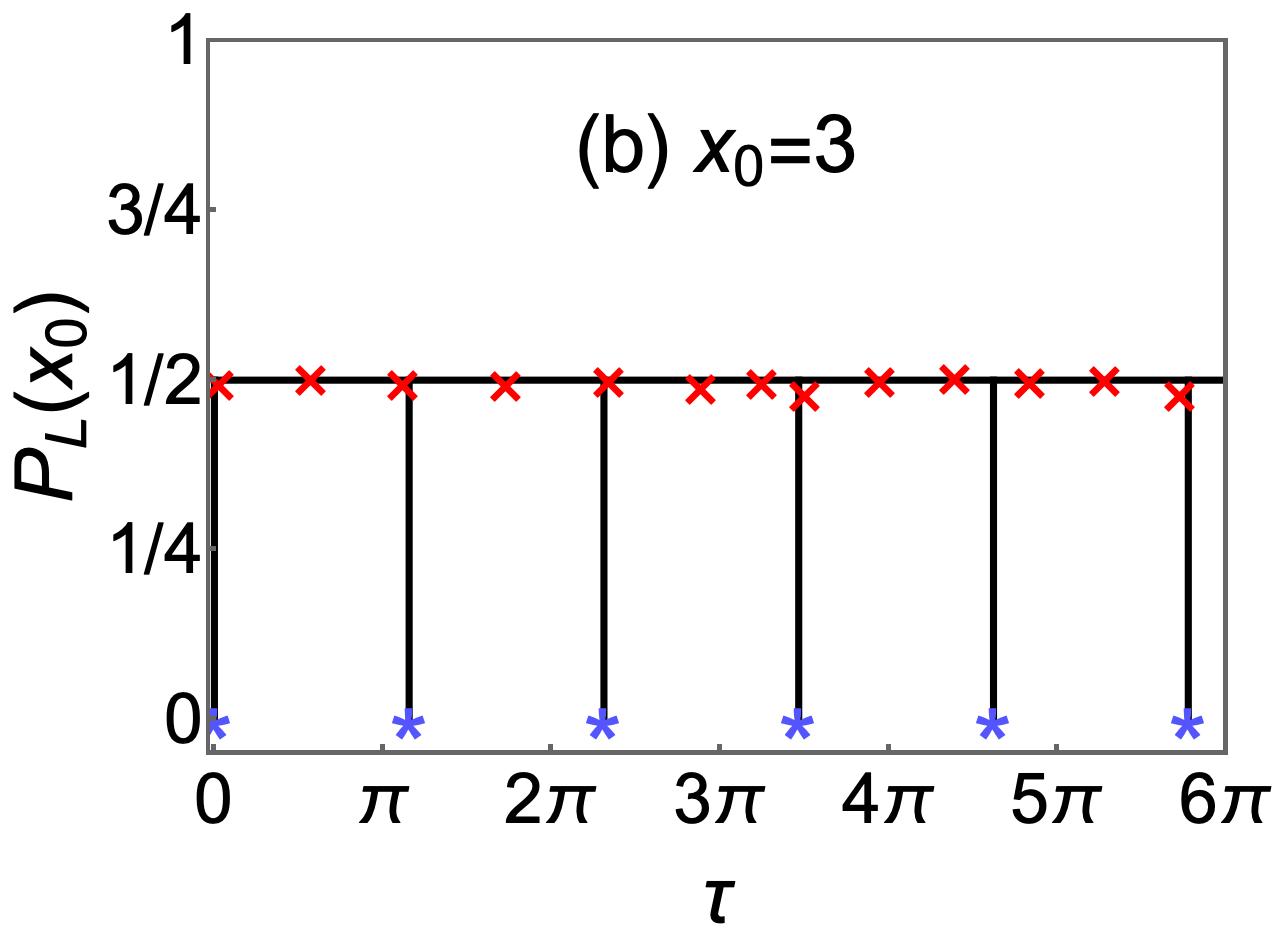}
	\includegraphics[scale=0.22]{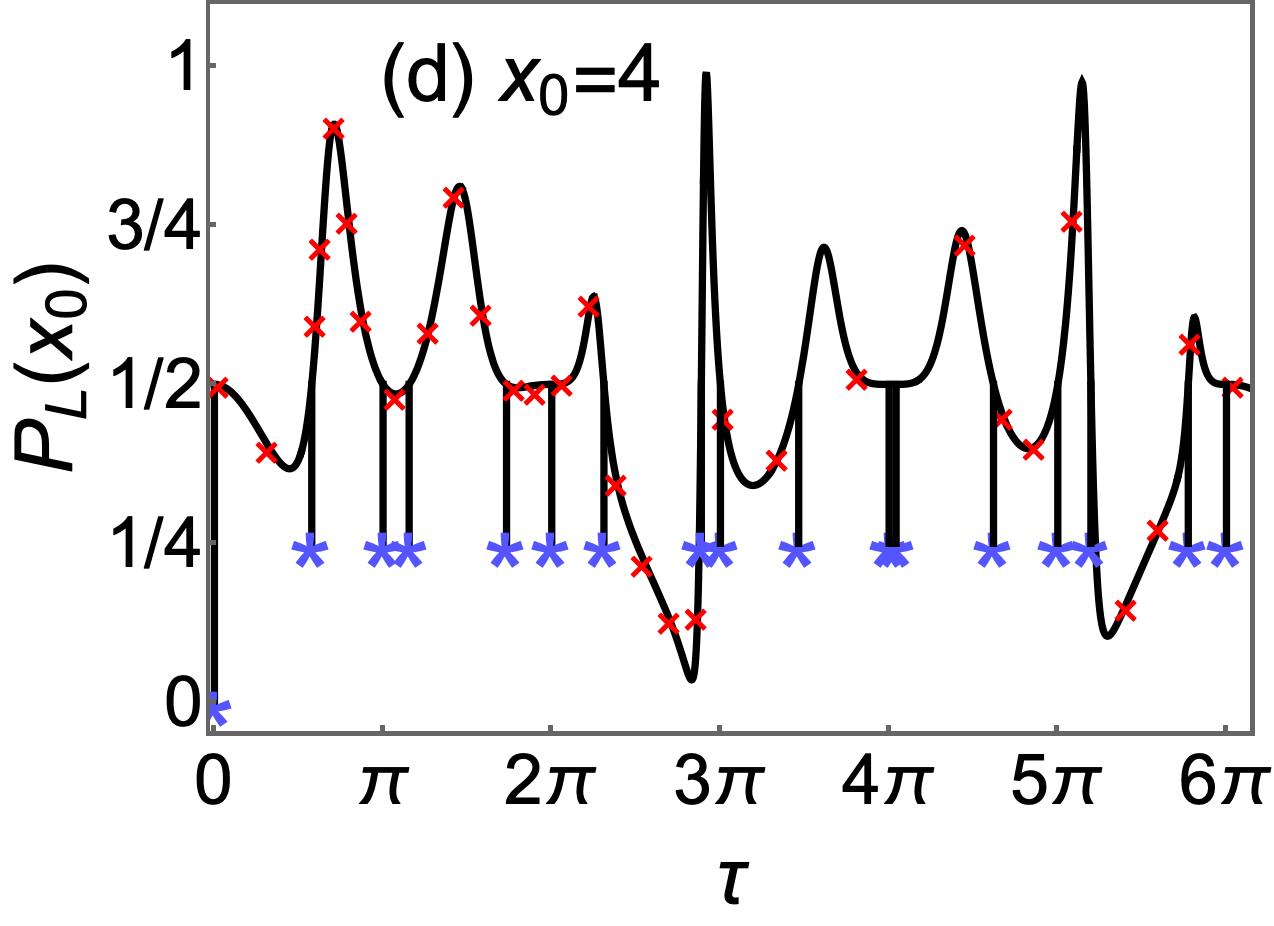}
	\includegraphics[scale=0.22]{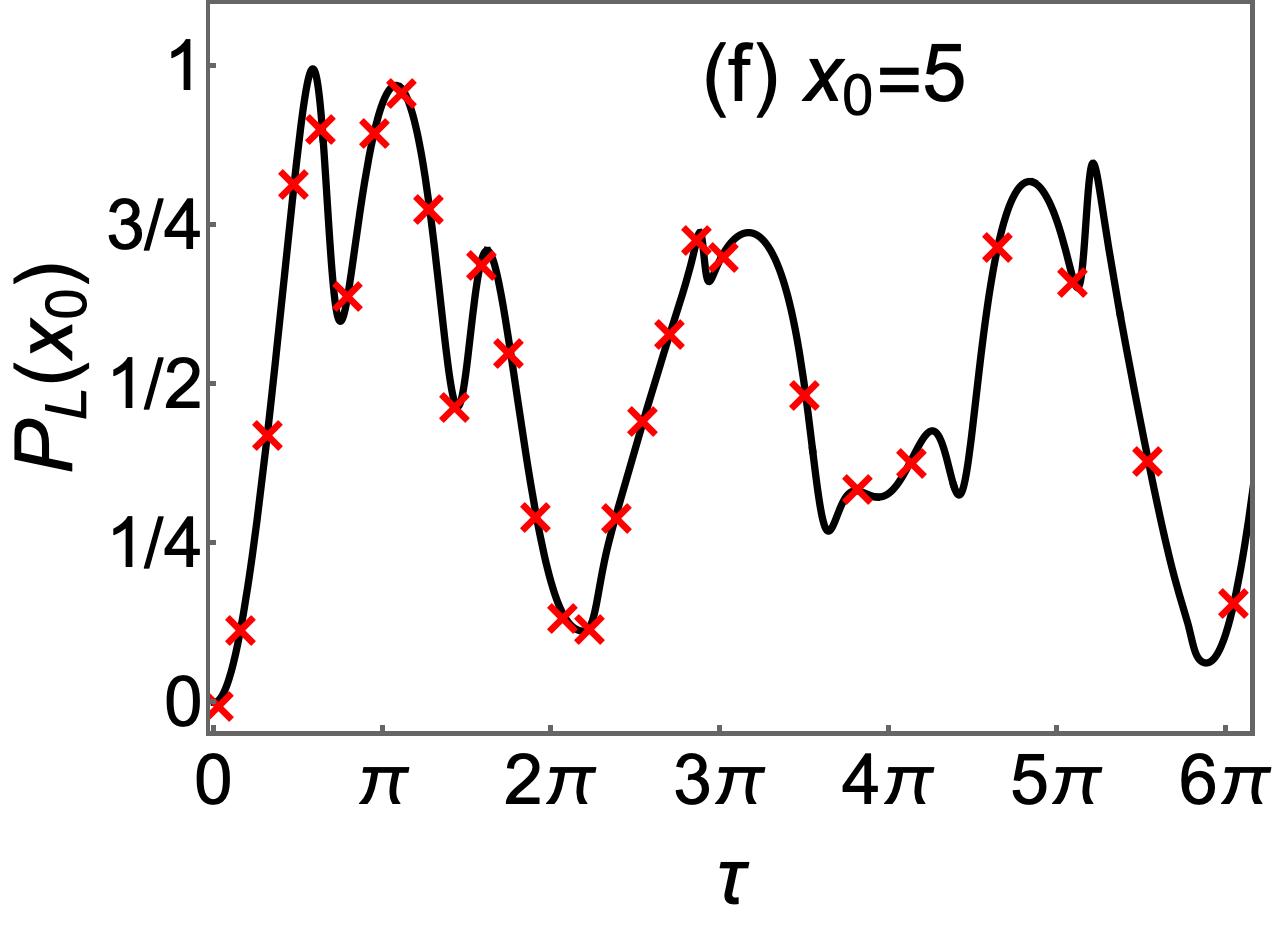}
	\caption{Quantum splitting probabilities $P_R(x_0)$ and $P_L(x_0)$ are shown as functions of the measurement time interval $\tau$ for a one-dimensional chain of size $N=5$, $H$ in  Eq.~\eqref{main-eqq-1}, and three different localized initial conditions: $\ket{x_0}= \ket{3}, \ket{4}, \ket{5}$ (from the symmetry in Eq.~\eqref{nauq}, these also give $\ket{x_0}= \ket{1}, \ket{2}$). In all cases, the detectors are situated at the boundaries, $\ket{D_1} = \ket{1}$ and $\ket{D_2} = \ket{5}$, and the hopping parameter is taken to be $\gamma =1$. The black solid lines correspond to the theoretical predictions, given in Eq.~\eqref{neww-eg-eq-39} for $\ket{x_0}=\ket{3}$, Eqs.~\eqref{unitary-eqn-31}-\eqref{unitary-eqn-313} for $x_0=4$, and  Eq.~\eqref{unitary-eqn-35} for $\ket{x_0}=\ket{5}$. The red and blue symbols denote the data obtained from the numerical simulations, with blue symbols depicting the resonances at special values of $\tau$. Details of the simulations are given in \aref{sec-appen-simulation}. Analytic predictions of the resonances for $\ket{x_0}=\ket{3}$ and $\ket{x_0}=\ket{4}$ are summarized in Table~\ref{table-exceptional}. In contrast, for $\ket{x_0}=\ket{5}$, no resonances occur, since the overlap conditions in Eqs.~\eqref{neww-eg-eq-28-3} and \eqref{neww-eg-eq-28-4} are not satisfied.}
	\label{fig-Ptau}
\end{figure}

Collecting all cases together, these special points and the corresponding values of the probabilities are summarized in Table~\ref{table-exceptional}. In Figs.~\ref{fig-Ptau}(a) and \ref{fig-Ptau}(b), the splitting probabilities are shown as functions of $\tau$. The red and blue symbols correspond to the numerical simulations, with the blue symbols marking the resonances that occur at special values of $\tau$.
The details of the simulations are given in \aref{sec-appen-simulation}. Except at these special values, both $P_L(3)$ and $P_R(3)$ remain equal to $1/2$ irrespective of $\tau$, confirming our theoretical result in Eq.~\eqref{neww-eg-eq-39}. 
%Later we will see that the limiting behavior $P_R(x_0) = P_L(x_0) =1/2$ is, in fact, generic for all initial states $x_0$ located in the bulk in the quasi-continuous monitoring limit ($\tau$ is close to but not strictly $0$).

\subsection{Example II: $\ket{x_0} = \ket{4}$}
We next consider the initial condition $\ket{x_0} = \ket{4}$. Unlike the previous example, this initial condition is no longer symmetric with respect to the two detectors, $\ket{D_1} = \ket{1}$ and $\ket{D_2} = \ket{5}$. The initial position, being closer to the right boundary, would, in a classical random walk, naturally bias the splitting probability toward the right. This is the proximity effect that we explained in the Introduction. Here, we illustrate the breakdown of this effect for the quantum case.

%The central question is whether, and to what extent, such a proximity effect persists in the quantum case.

For $x_0 = 4$, $\widetilde{\chi} _{\pm}(z)$ in Eqs.~\eqref{aux-eq-8} and \eqref{aux-eq-9} take the form
\begin{align}
& \widetilde{\chi} _{+}(z) = - i \sqrt{\frac{3}{2}} ~\frac{(1-z)~ \sin \left( \sqrt{3} \gamma \tau \right) }{\left[  z(1-2z) + z \cos \left( \sqrt{3} \gamma \tau \right)(5-z)-3 \right]}, \\
&\widetilde{\chi} _{-}(z) =  -  \frac{ i \sin \left( \gamma \tau \right)  }{\sqrt{2} \left[ 1-z \cos \left( \gamma \tau \right) \right]}.
\end{align}
Inserting them in Eq.~\eqref{aux-eq-2}, and carrying out the integration over $\omegaa$, we obtain
\begin{align}
\zeta _{+-}(4) = -\frac{\sqrt{3} \sin \left(\frac{\gamma \tau}{2} \right)^2 \sin \left( \gamma \tau \right) \sin \left( \sqrt{3} \gamma \tau \right) }{4+ \cos \left( 2 \gamma \tau \right)  + \cos \left( \gamma \tau \right) \left[  -1+ \cos \left( \sqrt{3 }\gamma \tau \right) \left\{ -5+  \cos \left( \gamma \tau \right) \right\} \right]}. \label{neww-eg-eq-43} 
\end{align}
Finally, the splitting probabilities in Eq.~\eqref{neww-eg-eq-17} are
%\begin{align}
%&P_L(4) = \frac{1}{2} -\frac{\sqrt{3} \sin \left(\frac{\gamma \tau}{2} \right)^2 \sin \left( \gamma \tau \right) \sin \left( \sqrt{3} \gamma \tau \right) }{4+ \cos \left( 2 \gamma \tau \right)  + \cos \left( \gamma \tau \right) \left[  -1+ \cos \left( \sqrt{3 }\gamma \tau \right) \left\{ -5+  \cos \left( \gamma \tau \right) \right\} \right]}, \label{unitary-eqn-31} \\
%& P_R(4) = 1-P_L(4).
% \label{unitary-eqn-313}
%\end{align}
\begin{center}
\fbox{
\begin{minipage}{0.95\textwidth}
\begin{align}
P_L(4)
&=
\frac{1}{2}
-\frac{
\sqrt{3}\sin^2\left(\frac{\gamma\tau}{2}\right)
\sin(\gamma\tau)\sin(\sqrt{3}\gamma\tau)
}{
4+\cos(2\gamma\tau)
+\cos(\gamma\tau)
\left[
-1+\cos(\sqrt{3}\gamma\tau)
\left\{-5+\cos(\gamma\tau)\right\}
\right]
},
\label{unitary-eqn-31}
\\
P_R(4)
&=
1-P_L(4).
\label{unitary-eqn-313}
\end{align}
\end{minipage}
}
\end{center}
Both $P_L(4)$ and $P_R(4)$ show a clear dependence on the measurement time $\tau$.  Nevertheless, they continue to satisfy the normalisation condition, $P_L(4)+P_R(4)=1$, for all values of $\tau$, except at the special cases.
%However, their sum is still equal to unity, $P_L(4) + P_R(4)=1$, for all $\tau$. It holds true for most values of $\tau$, but not at the exceptional times, where different phases coincide. 

These special times, corresponding to phase matching and overlapping conditions in Eqs.~\eqref{neww-eg-eq-28-3} and \eqref{neww-eg-eq-28-4}, are summarized in Table~\ref{table-exceptional}. At these values, the splitting probabilities change abruptly to $P_L(4) = P_R(4) = 1/4$. Since $P_L(4)+P_R(4)=1/2$ in this case, the total detection probability is less than unity. As explained before, this occurs because a part of the wavefunction behaves as a dark state that remains decoupled from the two detectors at the moments of measurement, and therefore does not get detected. This reduces the overall detection probability.

\begin{table}[t]
\centering
\renewcommand{\arraystretch}{2.2} % Increase row height
%\begin{tabular}{p{2.0cm}|p{3.0cm}|p{4.2cm}|p{3.0cm}}
\begin{tabular}{l|l|l|l}
\hline\hline
$~~~~x_0$ & $~~~~~~\tau = 0$ &  $~~~~~~\tau = m \pi / \sqrt{3}\gamma$ & $~~~~\tau = m \pi / \gamma$ \\
\hline
$~~~~3$ & $~~P_L=P_R=0$ & $P_L=P_R=0$ if $m$ is even, No resonance if $m$ is odd & $~~~$No resonance \\
\hline
$~~~~4$ & $~~P_L=P_R=0$ & $~~~~~~P_L=P_R=\tfrac{1}{4}$ & $~~~P_L=P_R=\tfrac{1}{4}$ \\
\hline
$~~~~5$ & $~P_L=0,\; P_R=1$ & $~~~~~~$No resonance & $~~~$No resonance \\
\hline\hline
\end{tabular}
\caption{Splitting detection probabilities $P_L(x_0)$ and $P_R(x_0)$ at the special values of $\tau$ with $m=1,2,\ldots$ for a chain of size $N=5$, and measurements performed at $\ket{D_1} = \ket{1}$ and $\ket{D_2} = \ket{5}$. These resonant points are determined by the phase-matching and overlap conditions given in Eqs.~\eqref{neww-eg-eq-28-3} and \eqref{neww-eg-eq-28-4}. We have shown results for the initial conditions $\ket{x_0}=\ket{3},\ket{4},\ket{5}$. The other two initial conditions, $\ket{x_0}=\ket{1},\ket{2}$, can be analyzed using the symmetry in Eq.~\eqref{nauq}.
}
\label{table-exceptional}
\end{table}

Figs.~\ref{fig-Ptau}(c) and \ref{fig-Ptau}(d) demonstrate a comparison of our theoretical results with the numerical simulations. We see excellent agreement for both $P_R(4)$ and $P_L(4)$, including at the resonance points (shown by the blue symbols). From the figure, we also notice that, unlike in our previous symmetric example with $\ket{x_0} = \ket{3}$, the two splitting probabilities are generally unequal $\left[ P_L(4) \neq P_R(4) \right] $, owing to the initial state, $\ket{x_0} = \ket{4}$, being not symmetric with two detectors.

Let us now examine the proximity effect. For a classical random walk, this can be seen from Eq.~\eqref{unitary-eqn-classical}, which satisfies $P_R^{(\rm cl)}(4)> P_L^{(\rm cl)}(4)$. However, for the quantum walk, Figs.~\ref{fig-Ptau}(c) and \ref{fig-Ptau}(d) reveal that this classical expectation is generally not true, and the opposite behavior $P_R(4) < P_L(4)$ is observed for certain values of $\tau$. For instance, when $\tau = 2$, we obtain $P_R(4) \approx 0.28$ and $P_L(4) \approx 0.72$, clearly departing from the classical proximity effect. This departure arises because, for $\tau =2$, some components of the wave function interfere constructively at the left boundary, while interfering destructively at the right boundary. 
As a result, the probability of detecting the walker on the right boundary becomes lower than the left boundary, leading to the observed reversal of the classical expectation. This demonstrates that, in the quantum case, proximity to a detector does not necessarily guarantee detection at that side, and interference effects can even enhance the probability of detection at the opposite detector.

Finally, in the Zeno limit, where $0<\gamma \tau \ll 1$, one can observe in Figs.~\ref{fig-Ptau}(c) and \ref{fig-Ptau}(d) that both $P_L(4)$ and $P_R(4)$ are close to $1/2$. This can also be validated by expanding Eqs.~\eqref{unitary-eqn-31} and \eqref{unitary-eqn-313} for $\gamma \tau \ll 1$
\begin{align}
P_{R}(4)  \simeq \frac{1}{2} + \frac{\gamma ^2 \tau^2}{8},~~ P_{L}(4)  \simeq \frac{1}{2} -\frac{\gamma ^2 \tau^2}{8}. \label{abcg1}
\end{align}
The leading order term is $1/2$, which is consistent with our previous analysis of the Zeno limit for general $N$ [see Eq.~\eqref{neww-eg-eq-33}]. 
\subsubsection{Loss of the flat-to-fluctuating transition for small $N$}~~\\
\indent 
Beyond the Zeno-limit, we observe in Figs.~\ref{fig-Ptau}(c) and \ref{fig-Ptau}(d) that both $P_L(4)$ and $P_R(4)$ begin to deviate from $1/2$ as the sampling time increases. In particular, we do not observe the flat-to-fluctuating transition discussed earlier in Sec.~\ref{sec-transition} or Fig.~\ref{fig-transition}. This transition, as noted before, occurs only in the limit of large system size. For small system sizes, such as $N=5$, this transition is lost.

%in the small-$\tau$ limit, also see Fig.~\ref{fig-Ptau}(c) and (d). This behavior is reminiscent of our symmetric example $(x_0=3)$ in Eq.~\eqref{unitary-eqn-28}, but only in the regime of small $\tau$. The equality of $P_L(4)$ and $P_R(4)$ arises because, when measurements are very frequent, the quantum walker takes very long time to be detected, unless its initial position coincides with one of the detectors. Any initial asymmetry present in the system is effectively erased at such long detection times. We emphasize that this behavior is intrinsically quantum because for a classical random walk, one has $P_{R}^{\rm (cl)}(4) = 3/4$ and $P_{L}^{\rm (cl)}(4) = 1/4$ when $\tau$ is small.

%Observe that the splitting probabilities, in general, are not equal, $P_L(4) \neq P_R(4) $, but satisfy $P_L(4) + P_R(4)=1 $ for all values of $\tau$, except at the special points discussed before. To sum up, we obtain

\subsection{Example III: $\ket{x_0} = \ket{5}$}
Here, we consider the situation when the initial position, $x_0=5$, coincides with the right boundary
\begin{align}
\ket{5} = \frac{1}{\sqrt{3}} \left[   \frac{1}{2} \ket{E_1} - \frac{\sqrt{3}}{2} \ket{E_2} +\ket{E_3} - \frac{\sqrt{3}}{2} \ket{E_4} +  \frac{1}{2} \ket{E_5}  \right].
\end{align}
For this case, the overlap conditions in Eqs.~\eqref{neww-eg-eq-28-3} and \eqref{neww-eg-eq-28-4} are never satisfied (also see Eqs.~\eqref{neww-eg-eq-28-10982} and \eqref{neww-eg-eq-28-109821}). Consequently, we do not get any resonant points for this case.

For generic $\tau$, treating this initial condition in the same way as the previous two, we find that the splitting probabilities take the form
\begin{align}
\boxed{P_R(5) = \frac{1}{2}+\frac{\mathcal{A}(\gamma \tau)}{\mathcal{B}(\gamma \tau)},~~~P_L(5) = 1-P_R(5),}  \label{unitary-eqn-35} 
\end{align}
with $\mathcal{A}(y)$ and $\mathcal{B}(y)$ defined as
\begin{align}
\mathcal{A}(y) & = 33 \cos (y) - 6 \left( 2+ \cos (2y)\right) + 9 \cos (3y) +2 \cos (2\sqrt{3} y) \left( \cos (y)-5\right) \nonumber \\
&-2 \cos (\sqrt{3}y) \left\{ 7-14 \cos (y) +15 \cos (2y)  \right\}
,  \label{unitary-eqn-36} \\
 \mathcal{B}(y) & = 24 \Big[ 4 +   \cos  (2y)   +\cos  (y)  \Big\{  \cos  (\sqrt{3} y) \left(  \cos  (y)-5 \right)-1  \Big\}  \Big]. \label{unitary-eqn-37} 
\end{align}
Expansion for small $y$ gives $\mathcal{A}(y) \simeq 72 y^2-108 y^4$ and $\mathcal{B}(y) \simeq 144 y^2-72 y^4$. From these, the splitting probabilities simplify to
\begin{align}
P_R(5) \simeq 1-\frac{\gamma ^2 \tau ^2}{2},~~P_L(5) \simeq \frac{\gamma ^2 \tau ^2}{2}, \label{unitary-eqn-38} 
\end{align}	
in the Zeno limit of $\gamma \tau \ll 1$. To the leading order, the walker, as expected, is detected with unit probability at the right detector, since the initial position coincides with this detector. This result is again consistent with our general Zeno-analysis in Eq.~\eqref{neww-eg-eq-33}.
	
In Figs.~\ref{fig-Ptau}(e) and \ref{fig-Ptau}(f), we use Eq.~\eqref{unitary-eqn-35} to plot $P_R(5)$ and $P_L(5)$ as functions of $\tau$, together with the numerical simulations, see \aref{sec-appen-simulation} for the simulation details. Once again, our theoretical predictions show excellent agreement with the simulation data, with no resonances. As seen in the figure, these probabilities display pronounced oscillations, and at certain values of $\tau$, the splitting probability at the right boundary can even fall below the left one. For example, at $\gamma \tau = \pi$, we find $P_R(5) \approx 0.055$ while $P_L(5) \approx 0.945$. The behavior $P_L(5) > P_{R}(5)$ again reflects the loss of the classical proximity effect. It highlights that the quantum interference effect can modify the splitting probabilities in ways that have no classical analog.

\begin{figure}[t]
	\centering
	\includegraphics[scale=0.25]{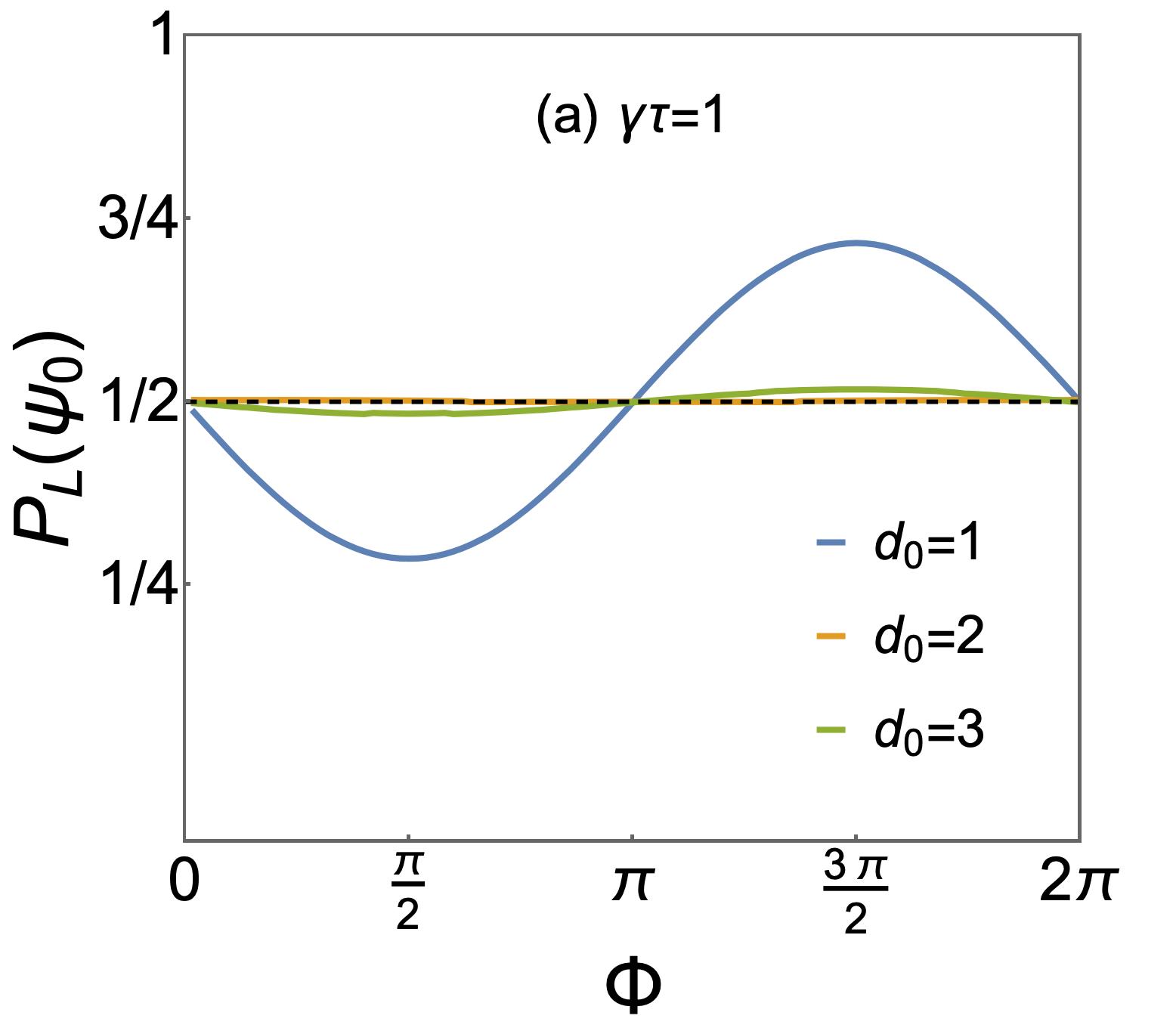}
     \includegraphics[scale=0.25]{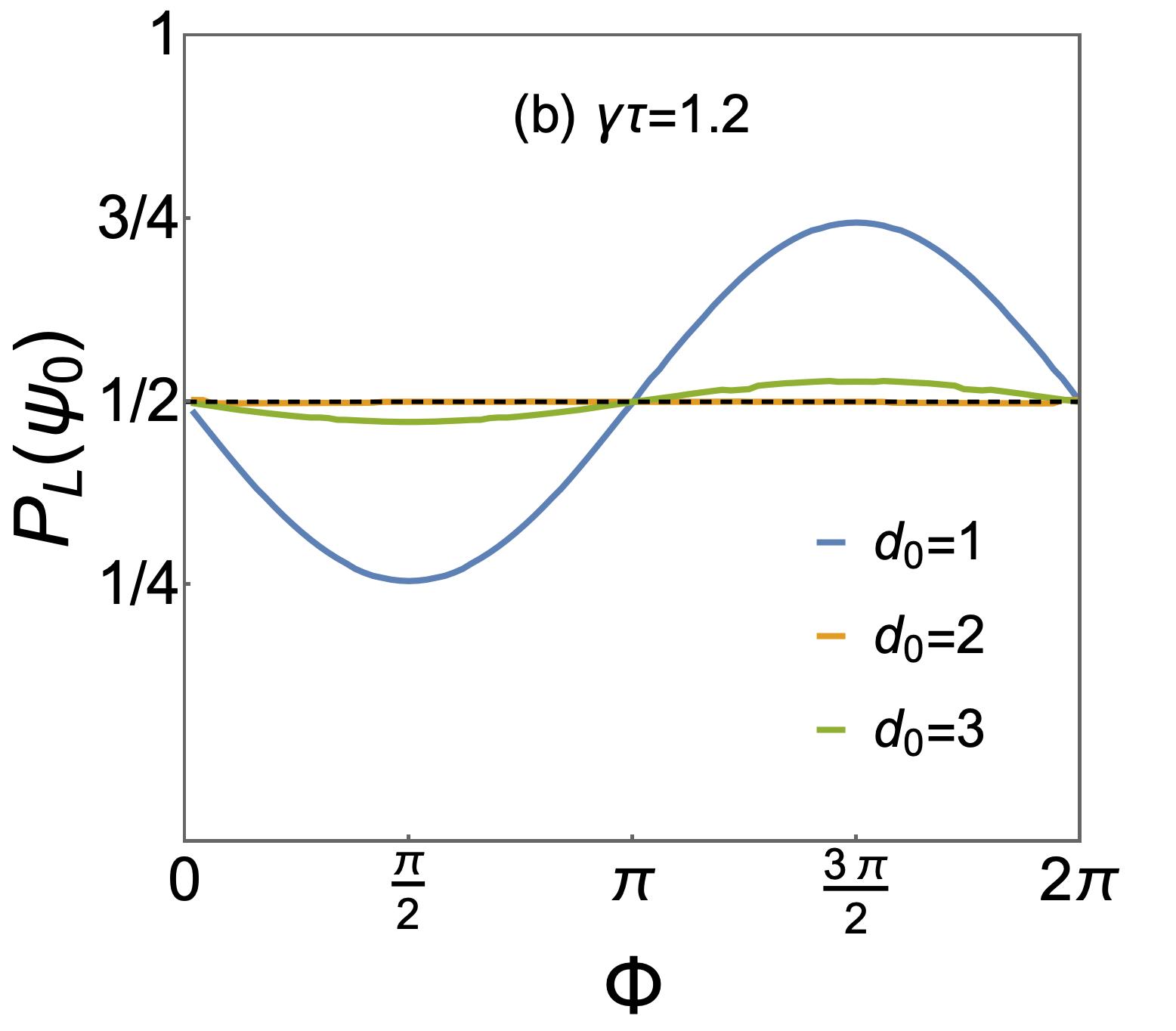}
     \includegraphics[scale=0.25]{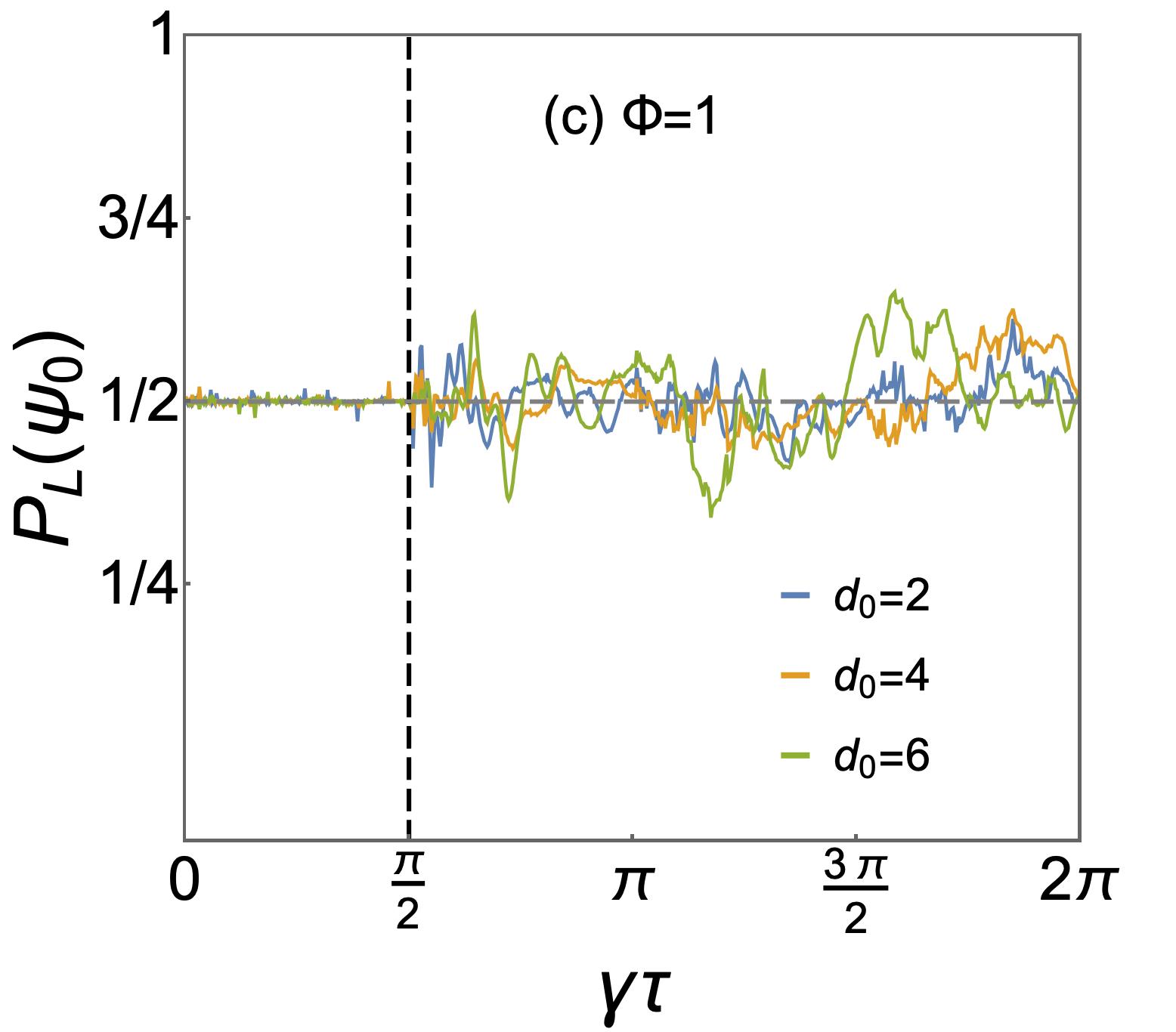}
     \includegraphics[scale=0.25]{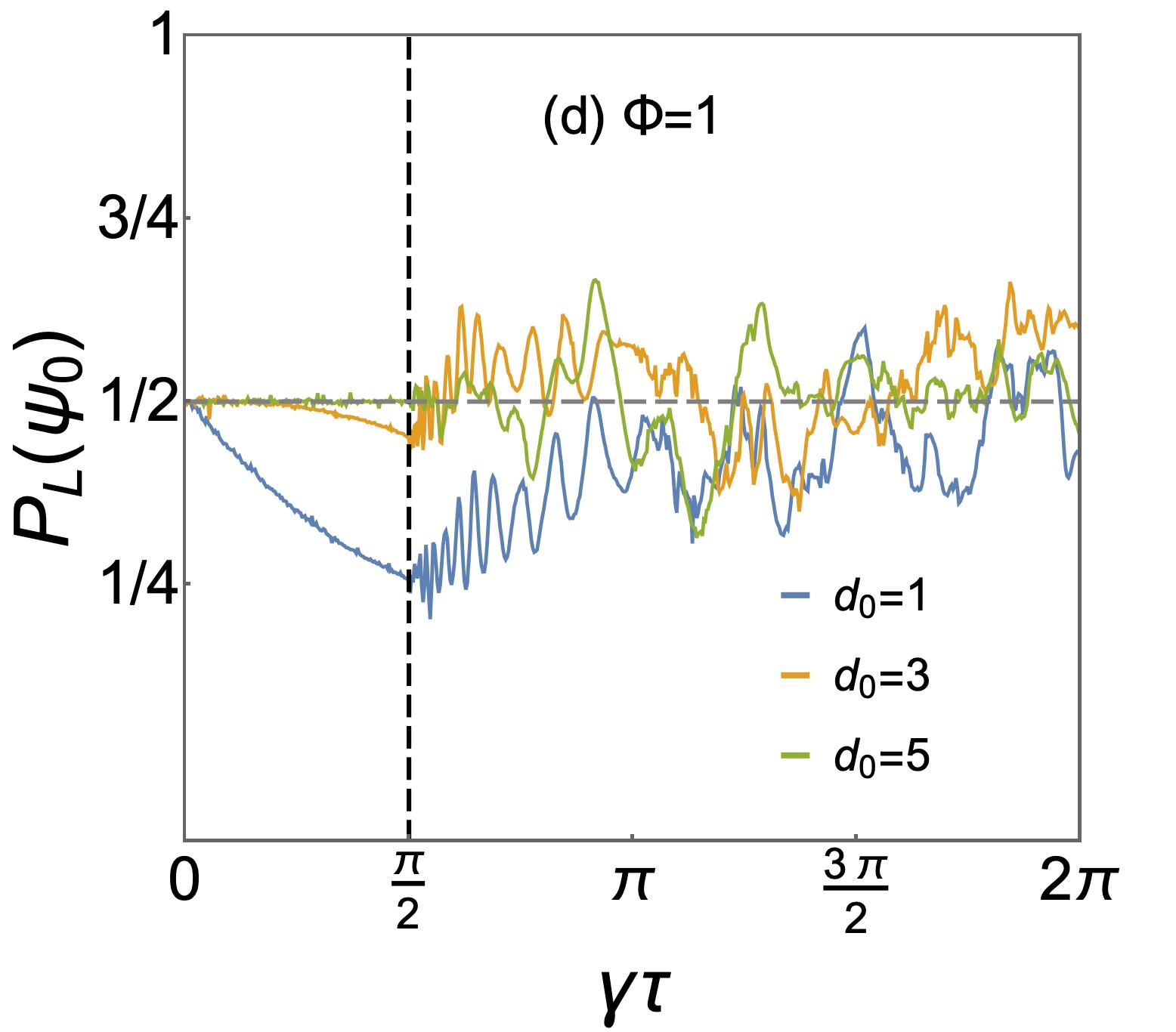}
	\caption{Splitting probability $P_L(\psi_0)$ for the  initial state $\ket{\psi_0}=\left(\ket{x_0}+e^{i\Phi}\ket{x_0+d_0}\right)/\sqrt{2}$, calculated using Eqs.~\eqref{aux-eq-5}--\eqref{pole-eq-2}. We have plotted $P_L(\psi_0)$ as a function of the relative phase $\Phi$ for (a) $N=50$, $\gamma=1$, $\tau=1$, $x_0=10$, and (b) $N=50$, $\gamma=1$, $\tau=1.2$, $x_0=17$. In both panels, we have shown the results for three separations, $d_0=1,2,3$. For odd $d_0$, $P_L(\psi_0)$ in both panels exhibits the $\sin\Phi$ dependence predicted by Eq.~\eqref{OIC-eq-7}. For even $d_0$, by contrast, $P_L(\psi_0) \approx 1/2$, essentially independent of $\Phi$, since the prefactor $\mathcal{K}_{d_0}(x_0,\tau)$ multiplying the $\cos\Phi$ term in Eq.~\eqref{OIC-eq-7} is vanishingly small, as discussed in the text. To see this more explicitly, in panel~(c) we plot $P_L(\psi_0)$ as a function of $\gamma\tau$ for $\Phi=1$, $N=50$, $x_0=10$ and three even separations, $d_0=2,4,6$. In all cases, $P_L(\psi_0)$ remains equal to $1/2$, independent of $\tau$, for $\tau\leq\tau_c$, where $\tau_c$ is given in Eq.~\eqref{ajbidhiguqok}. Thus, the flat behavior below the critical sampling time, previously found for localized initial conditions in Fig.~\ref{fig-transition}, persists also for $\ket{\psi_0}$ with even separation $d_0$. On the other hand, panel~(d) shows the corresponding results for three odd separations, $d_0=1,3,5$. In this case, $P_L(\psi_0)$ generally depends on $\tau$ already for $\tau\leq\tau_c$, with the strongest dependence observed for the smallest separation $d_0=1$. As $d_0$ increases, this dependence becomes markedly weaker, and the flat-to-fluctuating behavior is recovered; see $d_0=5$.}
	\label{fig-OIC}
\end{figure}

\section{Coherent initial states}
\label{sec-OIC}
Having analyzed in detail the case of localized initial states in the tight-binding chain with Hamiltonian in Eq.~\eqref{unitary-eqn-23}, we now investigate more general $\ket{\psi_0}$ in which the walker initially can be in a coherent superposition over several sites. For localized initial conditions, we saw in Fig.~\ref{fig-transition} that the splitting probabilities exhibit flat-to-fluctuating transition. We now investigate it for general $\ket{\psi_0}$.

 Using Eqs.~\eqref{aux-eq-5}--\eqref{pole-eq-2}, we first note that if $\ket{\psi_0}$ has odd parity, then $\braket{E_\ell|\psi_0}=0$ for odd $\ell$, since $\ket{E_\ell}$ then has even parity [see Eq.~\eqref{neww-eg-eq-1}]. Substituting this into Eq.~\eqref{pole-eq-2}, we find $\mathcal{I}(\psi_0,z) = 0$, which according to Eq.~\eqref{pole-eq-1} leads to $\zeta_{+-}(\psi_0) = 0$. Similarly, for $\ket{\psi_0}$ with even parity, one has $\zeta_{+-}(\psi_0) = 0$. As a result,  Eq.~\eqref{aux-eq-5} reduces to
\begin{align}
P_L \left( \psi_0 \right) = P_R \left( \psi_0 \right) = 1/2, \quad \text{for $\ket{\psi_0}$ of definite parity.} \label{OIC-eq-1}
\end{align}
For these initial states, the splitting probabilities therefore are independent of the sampling time $\tau$, and remain equal to $1/2$ for all non-exceptional sampling times. Thus, the flat behavior observed below the critical sampling time for localized initial conditions in Fig.~\ref{fig-transition} extends to the entire non-exceptional values of $\tau$ for $\ket{\psi_0}$ of definite parity.
%However, at special sampling times in Eq.~\eqref{main-eqq-3}, both $P_L \left( \psi_0 \right)$ and $P_R \left( \psi_0 \right)$ deviate from $1/2$, and exhibit discontinuous jumps due to the formation of dark states, as explained before.

We also noted in Eq.~\eqref{neww-eg-eq-33} that, in the Zeno limit, the splitting probabilities are equal to $1/2$ for all initial conditions localized in the interior of the chain. More generally, as indicated in Eq.~\eqref{main-eqq-6}, the Zeno-limit splitting probabilities depend only on the projections of $\ket{\psi_0}$ onto the two boundary sites, and are independent of its detailed structure in the interior of the chain. Therefore, whenever $\ket{\psi_0}$ satisfies $|\braket{1|\psi_0}|^2=|\braket{N|\psi_0}|^2$, Eq.~\eqref{main-eqq-6} implies that both splitting probabilities are equal to $1/2$ in the Zeno limit. Thus, the Zeno-limit result, $P_L (\psi_0) \simeq P_R(\psi_0) \simeq 1/2$, applies to a broad class of initial states beyond the localized ones.

We next turn to the role of coherence in the initial state in shaping the splitting probabilities. As seen above, initial states of definite parity lead to a splitting probability of $1/2$ for either target for generic sampling times. A general coherent superposition, however, need not possess definite parity, and the relative phase between its components can therefore influence the splitting probabilities. To examine this effect, we consider the initial state
\begin{align}
\ket{\psi_0}=\left(\ket{x_0}+e^{i\Phi}\ket{x_0+ d_0}\right)/\sqrt{2}, \label{OIC-eq-2}
\end{align}
where $\Phi\in[0,2\pi]$ is the relative phase between the two components, while recall $1\leq x_0\leq N$ and $1 \leq d_0 \leq (N-x_0)$. Using Eqs.~\eqref{aux-eq-5}-\eqref{pole-eq-2}, we calculate $P_L(\psi_0)$ for this initial condition and plot its dependence on the phase $\Phi$ in Fig.~\ref{fig-OIC}(a). We have used $N=50,~\gamma=1,~\tau=1,~x_0=10$, and chosen three separations between the two components, namely $d_0=1,2,$ and $3$. As shown in Fig.~\ref{fig-OIC}(a), the dependence of $P_L(\psi_0)$ on $\Phi$ is strongest for $d_0=1$ and becomes considerably weaker for $d_0=3$. For the even separation $d_0=2$, the splitting probability is essentially independent of $\Phi$, remaining $1/2$ over the entire range of $\Phi$. This behavior qualitatively persists for other choices of the parameters also. For example, Fig.~\ref{fig-OIC}(b) shows the corresponding results for $\tau=1.2$ and $x_0=17$, again exhibiting an essentially $\Phi$-independent splitting probability for even $d_0$, while for odd $d_0$ the phase dependence of $P_L (\psi_0)$ becomes markedly weaker as $d_0$ increases.

To understand this behavior, let us take the initial state in Eq.~\eqref{OIC-eq-2}, and plug it in Eq.~\eqref{anjiuq1} with $\ket{D_1}=\ket{1}$ and $\ket{D_2}=\ket{N}$. The resulting first-detection amplitude can be expressed as
\begin{align}
\phi_{n}^{(L)}\left( \psi_0 \right) =\frac{1}{\sqrt{2}} \left[ \phi_{n}^{(L)}(x_0) + e^{i \Phi}  \phi_{n}^{(L)}(x_0+d_0) \right].
\label{OIC-eq-3}
\end{align}
The analysis for the right boundary proceeds similarly. From Eq.~\eqref{OIC-eq-3}, the corresponding splitting probability is
\begin{align}
P_L(\psi_0) = \frac{1}{2} \left[ P_L(x_0)  + P_L(x_0+d_0) \right]  +  \text{Re} \left[ \sum_{n=1}^{\infty} \left( \phi_n^{(L)}(x_0) \right)^*   \phi_n^{(L)}(x_0+d_0)  e^{i \Phi} \right], \label{OIC-eq-4}
\end{align}
where recall $P_L(x_0)$ denotes the splitting probability for a walker initially localized at site $x_0$, and similarly for $P_L(x_0+d_0)$. As discussed before in Fig.~\ref{fig-tau-N15}, for large system sizes and sampling times $\tau\leq\tau_c$, where $\tau_c$ is given in Eq.~\eqref{ajbidhiguqok}, the splitting probabilities for localized initial states are equal to $1/2$. Hence, we rewrite Eq.~\eqref{OIC-eq-4} as
\begin{align}
P_L(\psi_0) = \frac{1}{2}  + \text{Re} \left[ \sum_{n=1}^{\infty} \left( \phi_n^{(L)}(x_0) \right)^*   \phi_n^{(L)}(x_0+d_0)  e^{i \Phi} \right], \label{OIC-eq-5}
\end{align}	
where we are looking at $\tau \leq \tau_c$ and $N \gg 1$. For the nearest-neighbor hopping Hamiltonian, we show in Appendix~\ref{appen-real-imag} that
\begin{align}
\left[\phi_n^{(L)}(x_0)\right]^*
=
(-1)^{1+x_0}\phi_n^{(L)}(x_0). \label{OIC-eq-6}
\end{align}	
This means $\phi_n^{(L)}(x_0)$ is real for odd $x_0$, and purely imaginary for even $x_0$. Using this result in Eq.~\eqref{OIC-eq-5}, we find $\left( \phi_n^{(L)}(x_0) \right)^*   \phi_n^{(L)}(x_0+d_0) $ is real for even $d_0$, and purely imaginary for odd $d_0$, independently of $x_0$
	
\begin{align}
\boxed{P_L(\psi_0)
=
\begin{cases}
1/2+\mathcal{K}_{d_0}(x_0,\tau)\sin\Phi,
& d_0~\text{odd},\\[6pt]
1/2  +  \mathcal{K}_{d_0}(x_0,\tau)\cos\Phi,
& d_0~\text{even},
\end{cases}}
\label{OIC-eq-7}
\end{align}	
where
\begin{align}
\mathcal{K}_{d_0}(x_0,\tau)
=
\begin{cases}
i(-1)^{1+x_0}
\displaystyle\sum_{n=1}^{\infty}
\phi_n^{(L)}(x_0)\,
\phi_n^{(L)}(x_0+d_0),
& d_0~\text{odd},
\\[8pt]
(-1)^{1+x_0}
\displaystyle\sum_{n=1}^{\infty}
\phi_n^{(L)}(x_0)\,
\phi_n^{(L)}(x_0+d_0),
& d_0~\text{even}.
\end{cases}
\label{OIC-eq-8}
\end{align}
Eq.~\eqref{OIC-eq-7} was announced before in Eq.~\eqref{OIC-eq-7-sum}. Thus, for odd separations $d_0$, the splitting probability in Eq.~\eqref{OIC-eq-7} exhibits a $\sin \Phi$ dependence on the relative phase $\Phi$ in the initial state $\ket{\psi_0}$ in Eq.~\eqref{OIC-eq-2}. This is in agreement with the behavior observed in Figs.~\ref{fig-OIC}(a) and (b) for $d_0=1$ and $d_0=3$. However, for $d_0=2$, $P_L(\psi_0)$ in these figures is essentially independent of $\Phi$, even though Eq.~\eqref{OIC-eq-7} predicts a $\cos\Phi$ dependence for even $d_0$. This apparent discrepancy arises because the prefactor $\mathcal{K}_{d_0}(x_0,\tau)$ multiplying the $\cos\Phi$ term in Eq.~\eqref{OIC-eq-7} is extremely small for even $d_0$. For example, for the parameters used in Fig.~\ref{fig-OIC}(a), namely $N=50$, $\gamma=1$, $\tau=1$, and $x_0=10$, we find $\mathcal{K}_{d_0}(x_0,\tau)\approx-3\times10^{-4}$ for $d_0=2$. Similarly, for the parameters of Fig.~\ref{fig-OIC}(b), namely $N=50$, $\gamma=1$, $\tau=1.2$, and $x_0=17$, we obtain $\mathcal{K}_{d_0}(x_0,\tau)\approx-1.2\times10^{-4}$ for $d_0=2$. Inserting these values into Eq.~\eqref{OIC-eq-7}, we find $P_L(\psi_0)\approx1/2$ for both cases. Hence for $\ket{\psi_0}$ with even $d_0$, the splitting probability is  $1/2$ in the $\tau\leq\tau_c$ regime. We also verify this behavior in Fig.~\ref{fig-OIC}(c), where we plot $P_L(\psi_0)$ as a function of $\gamma \tau$ for $\Phi=1$ and even separations $d_0=2,4,$ and $6$. In all three cases, $P_L(\psi_0)$ remains essentially equal to $1/2$ for $\tau\leq\tau_c$, while for $\tau>\tau_c$ it develops pronounced fluctuations. Thus, the flat-to-fluctuating transition at $\tau_c$, found previously for localized initial conditions in Fig.~\ref{fig-transition}, persists also for $\ket{\psi_0}$ with even separation $d_0$.

By contrast, $\mathcal{K}_{d_0}(x_0,\tau)$ can be significant for odd values of $d_0$, especially for smaller values of $d_0$. For example, for the parameter set in Fig.~\ref{fig-OIC}(a), we obtain $\mathcal{K}_{d_0}(x_0, \tau) \approx -0.215$ for $d_0=1$ and $\mathcal{K}_{d_0}(x_0, \tau) \approx -0.017$ for $d_0=3$, while for Fig.~\ref{fig-OIC}(b), we have  $\mathcal{K}_{d_0}(x_0, \tau) \approx -0.24$ for $d_0=1$ and  $\mathcal{K}_{d_0}(x_0, \tau) \approx -0.02$ for $d_0=3$. From Eq.~\eqref{OIC-eq-7}, it then follows that $P_L(\psi_0)$ can deviate significantly from $1/2$, particularly for $d_0=1$. We also observed this behavior before in Figs.~\ref{fig-OIC}(a) and (b). This behavior is further illustrated in Fig.~\ref{fig-OIC}(d), where $P_L(\psi_0)$ is plotted as a function of $\gamma \tau$. For $d_0=1$, the splitting probability already exhibits a clear dependence on $\tau$ in the regime $\tau\leq\tau_c$, in contrast to the flat $P_L=1/2$ behavior found above for localized initial conditions. As $d_0$ increases, this dependence is strongly suppressed. For example, for $d_0=5$, $P_L(\psi_0)$ remains essentially equal to $1/2$ below $\tau_c$, and the flat-to-fluctuating transition at $\tau_c$ is recovered.

\section{Finite number of measurements}
\label{sec-finite}
Thus far, we have assumed that it is possible to take an infinite number of measurements. Accordingly, the summations over $n$ appearing in the expressions for the splitting probabilities in Eq.~\eqref{anj8y14} extend to infinity. Yet, as mentioned earlier, only a finite number of measurements can be performed in real experiments \cite{ExptQC-1, ExptQC-2, ExptQC-3}. For instance, starting from $\ket{\psi_0}$, a typical measurement record begins with a sequence of null outcomes, represented as ${\text{no},\text{no},\ldots}$. This sequence is truncated after a finite number of measurements, denoted by $T_{\rm max}$ (`max' for maximum number of measurements that can be performed). If a ``yes" outcome is obtained at either detector on or before the $T_{\rm max}$-th measurement, detection is deemed successful. Otherwise, the record consists entirely of ``no" outcomes, indicating that no detection has occurred within $T_{\rm max}$ number of measurements. The splitting probabilities are obtained by summing the first-detection probabilities, $F^{(L)}_n \left(  \psi _0 \right)$ and $F^{(R)}_n \left(  \psi _0 \right)$, till the $T_{\rm max}$-th measurement,
\begin{align}
P_{L} \left(  \psi _0 \right) = \sum _{n=1}^{T_{\rm max}} F^{(L)}_n \left(  \psi _0 \right) ,~~P_{R} \left(  \psi _0 \right) = \sum _{n=1}^{T_{\rm max}} F^{(R)}_n \left(  \psi _0 \right). \label{vnao01}
\end{align}
Our analysis in previous sections focused on $T_{\rm max} \to \infty$. It is  natural to examine how a finite $T_{\rm max}$ affects the flat-to-fluctuating transition of the splitting probability discussed in Sec.~\ref{sec-transition}. 

\begin{figure}[t]
	\centering
	\includegraphics[scale=0.18]{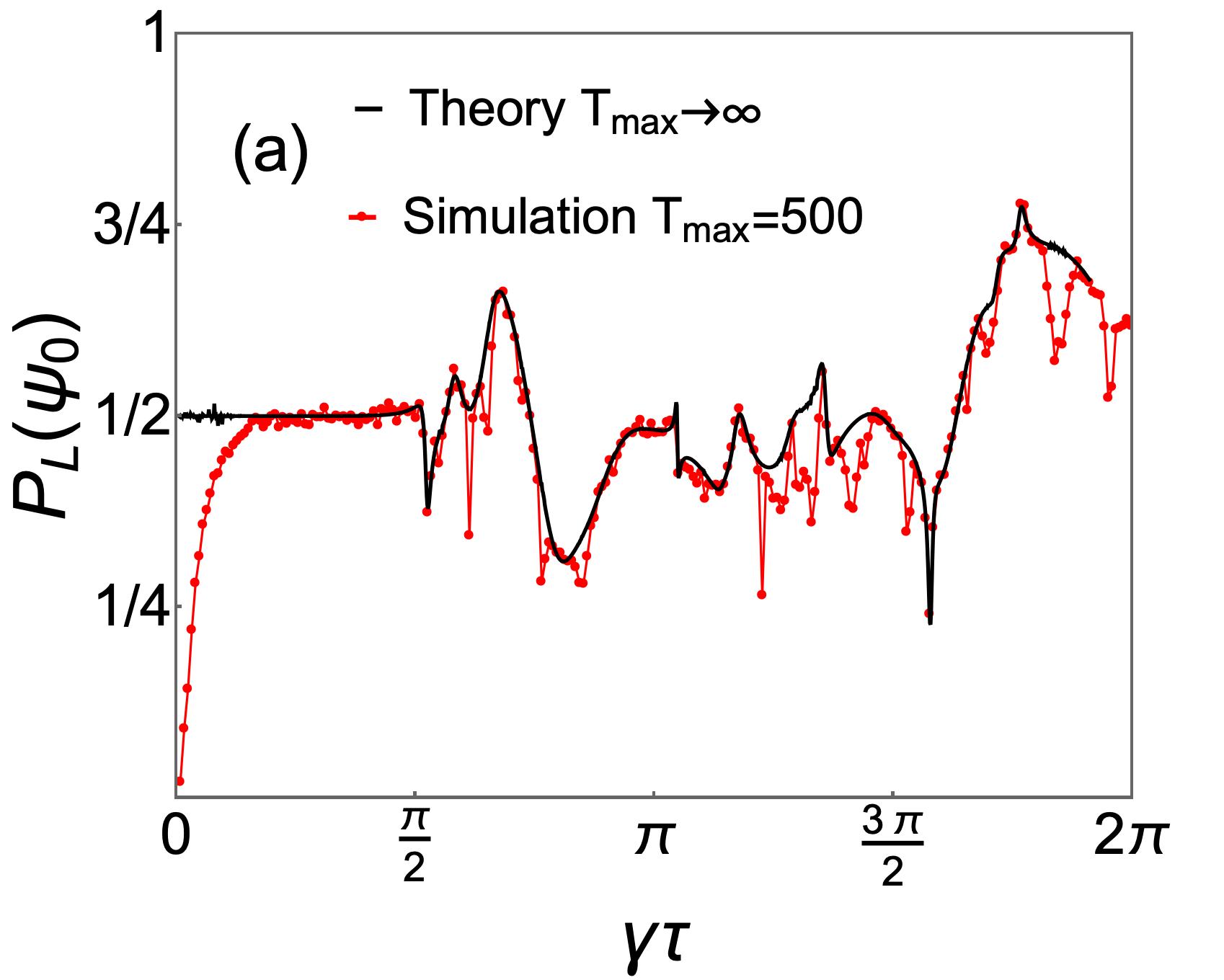}
     \includegraphics[scale=0.18]{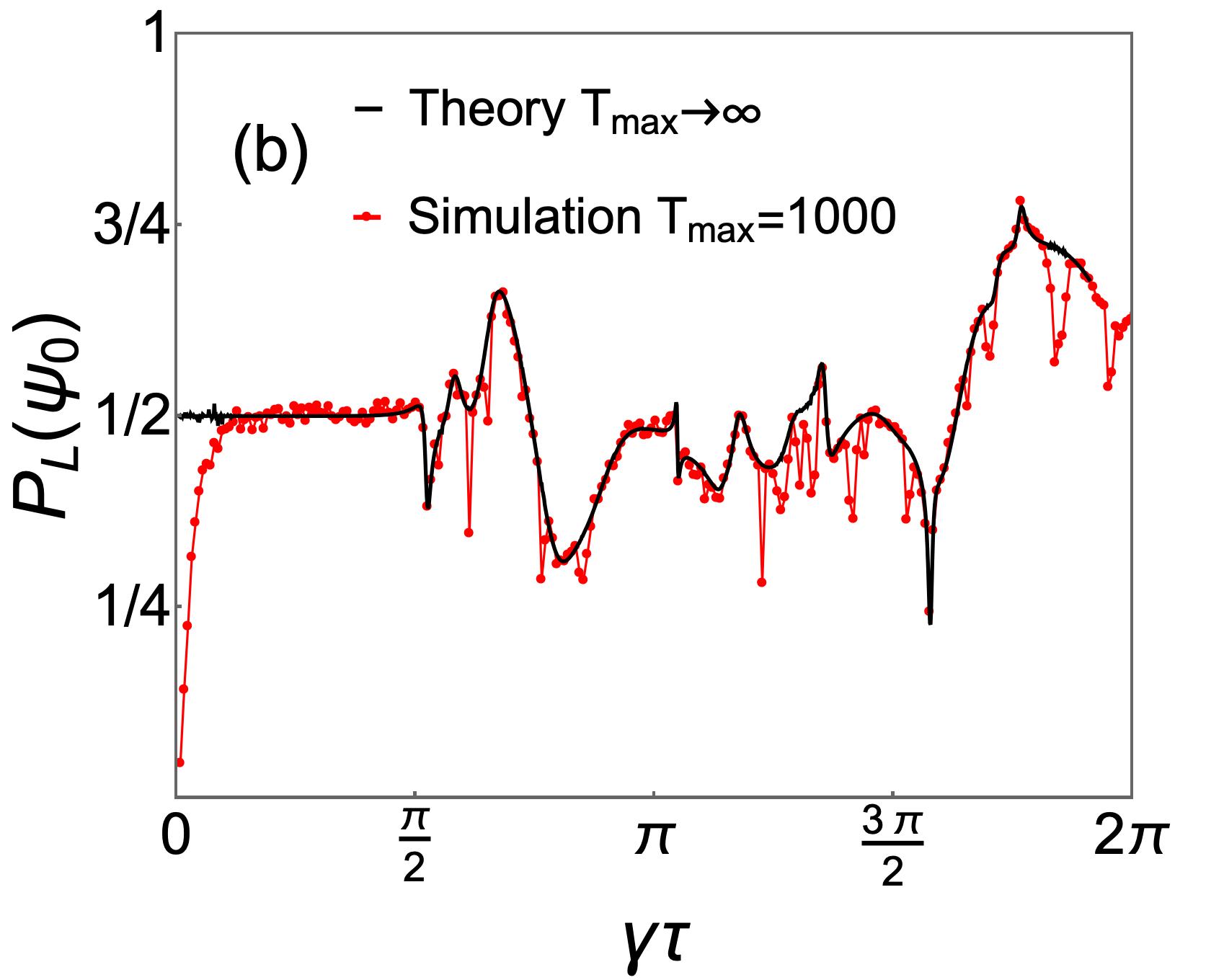}
     \includegraphics[scale=0.18]{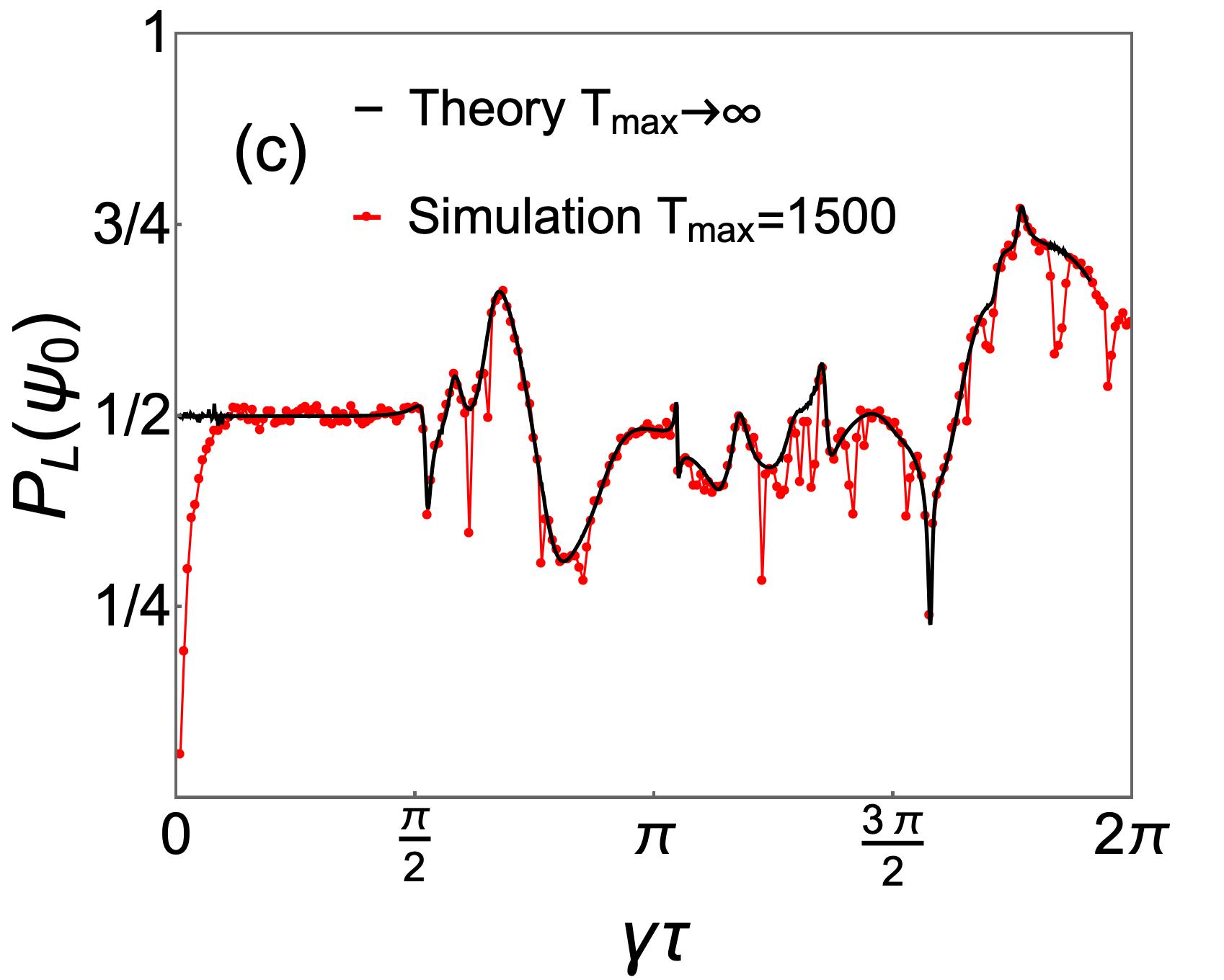}
     \includegraphics[scale=0.18]{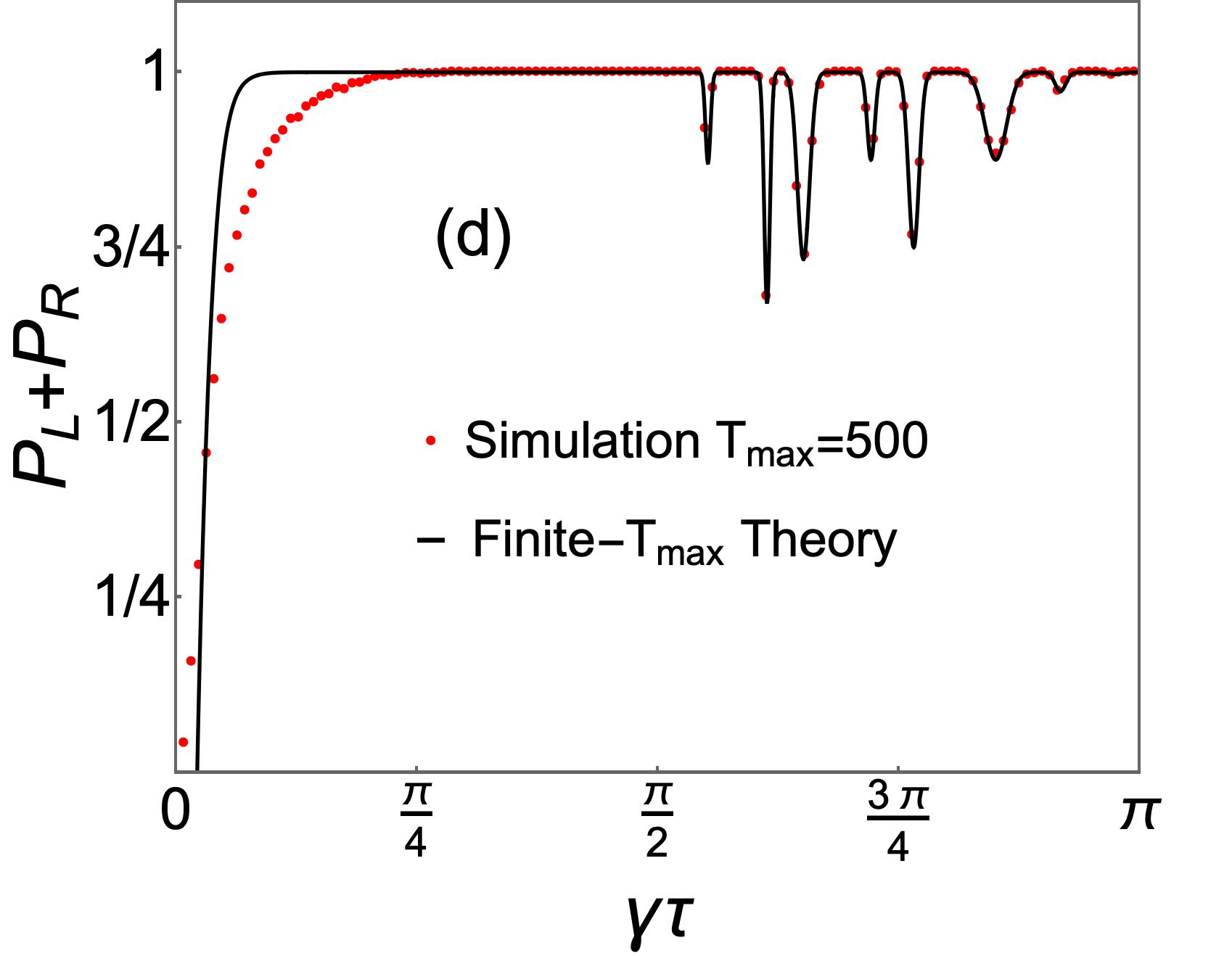}
     \includegraphics[scale=0.18]{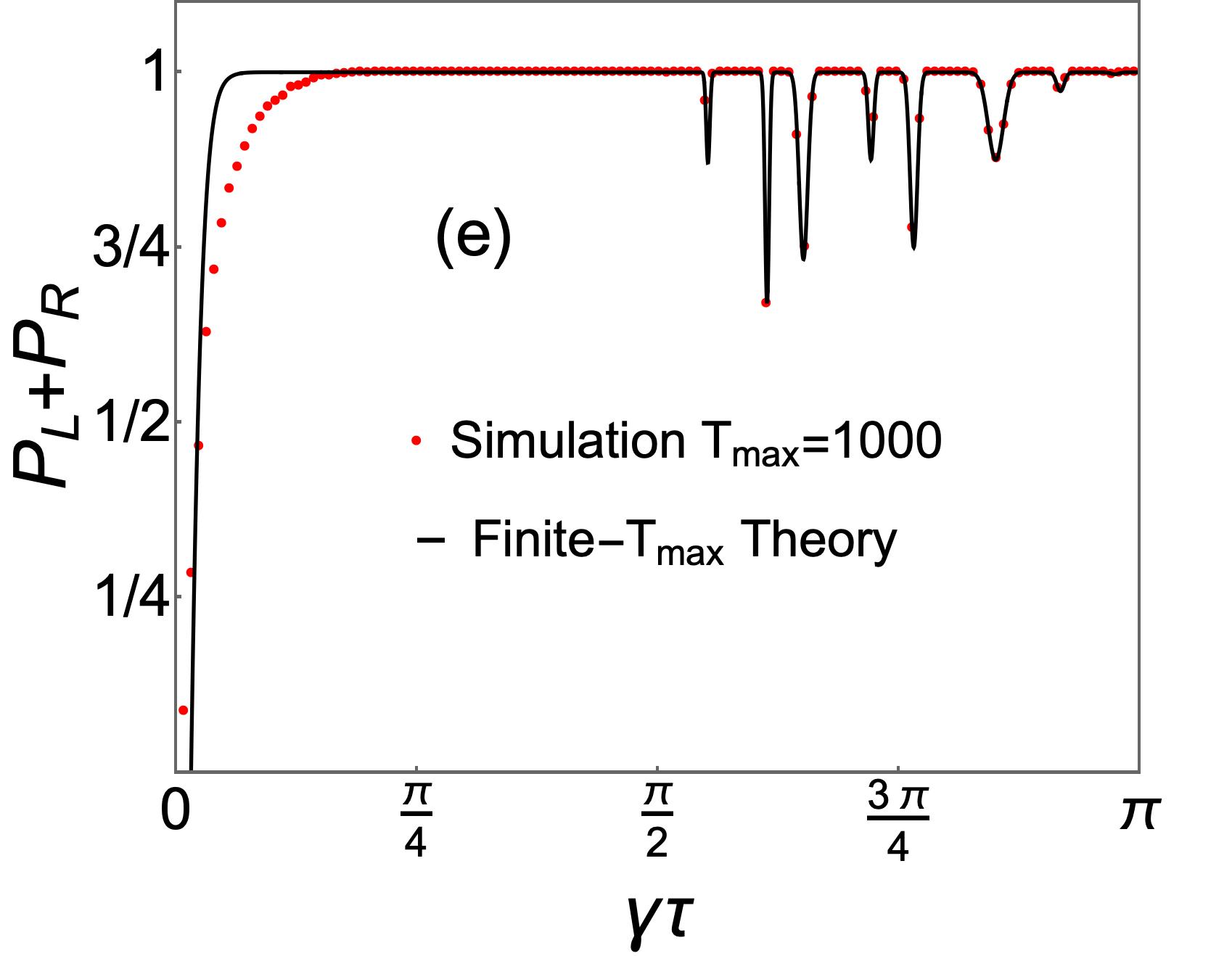}
     \includegraphics[scale=0.18]{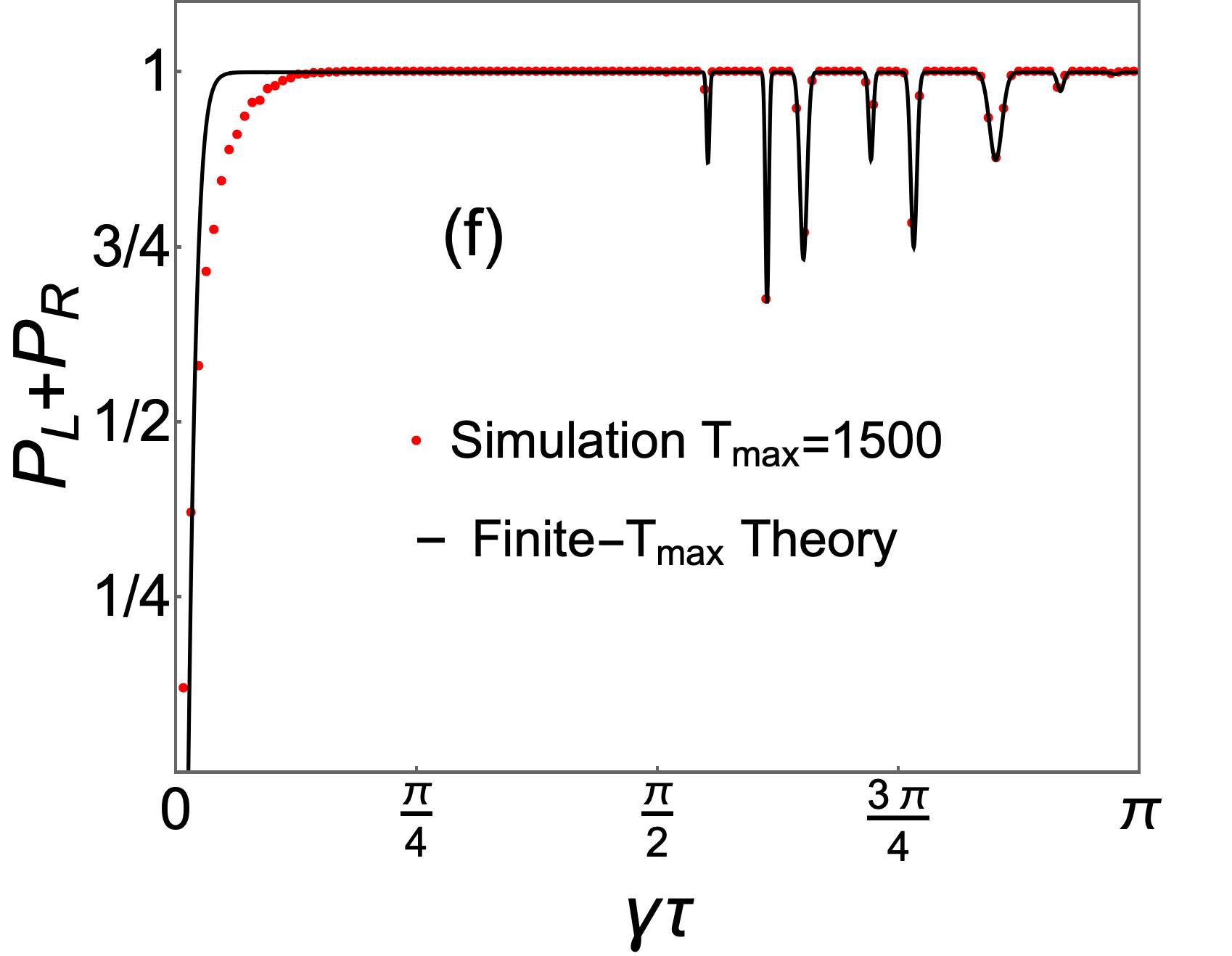}
	\caption{Effect of a finite number of measurements, $T_{\rm max}$, on the splitting probability. Panels~(a-c) show  $P_L(\psi_0)$ as a function of $\gamma\tau$ for $T_{\rm max}=500$, $1000$, and $1500$, respectively. The red dots are numerical simulations based on the nearest-neighbor hopping Hamiltonian in Eq.~\eqref{unitary-eqn-23}, with $N=15$ and $\ket{\psi_0}=\ket{11}$. The black curve is the ideal theory ($T_{\rm max}\to\infty$) for $P_L(\psi_0)$ given by Eq.~\eqref{aux-eq-5}. Finite-$T_{\rm max}$ effects are most pronounced near resonances such as $\tau=0$, where detection typically requires many measurement attempts. Panels~(d-f) show the corresponding normalization, $P_L(\psi_0)+P_R(\psi_0)$, for the same three values of $T_{\rm max}$. The numerical simulations are compared with the finite-$T_{\rm max}$ theory in Eq.~\eqref{app:qndef17}. Recall that in the ideal limit,
$T_{\rm max}\to\infty$, the normalization is mostly equal to unity with pointwise discontinuous jumps at the special sampling times in Eq.~\eqref{main-eqq-3}, where $P_L(\psi_0)+P_R(\psi_0) <1$. For finite $T_{\rm max}$, we see in panels~(d-f) that these discontinuities are broadened into dips of finite width. Except around Zeno limit, our Eq.~\eqref{app:qndef17} agrees very well with the simulation results, and the dips gradually become narrower as $T_{\rm max}$ increases. Deviations from the finite-$T_{\rm max}$ theory appear close to the Zeno limit, where many quasienergies can merge simultaneously. As explained in the paper, Eq.~\eqref{app:qndef17} is no longer valid in this regime.}
	\label{fig-finite}
\end{figure}

In Figs.~\ref{fig-finite}(a)-\ref{fig-finite}(c), we have used numerical simulations to obtain $P_L(\psi_0)$ for the nearest-neighbor hopping Hamiltonian in Eq.~\eqref{unitary-eqn-23}and plot it as a function of $\gamma \tau$ plot for $T_{\rm max}=500, 1000, 1500$. We choose $N=15$ and $\ket{\psi_0}=\ket{11}$. For comparison, we have also shown in black the ideal $(T_{\rm max} \to \infty)$ expression of $P_L(\psi_0)$ in Eq.~\eqref{aux-eq-5}. For $T_{\rm max}=500$, shown in panel~(a), noticeable deviations from the ideal result are observed,  especially at smaller values of the sampling time. However, as $T_{\rm max}$ increases, the simulation data progressively approach the black curve. Close to $\tau=0$, the quantum walker requires a large number of measurement attempts before detection occurs. Imposing a finite $T_{\rm max}$ excludes many long detection trajectories, causing $P_L(\psi_0)$ to deviate from the ideal flat value of $1/2$ in Eq.~\eqref{neww-eg-eq-33} close to $\tau=0$. This deviation is gradually reduced as $T_{\rm max}$ is increased.

Another consequence of a finite $T_{\rm max}$ is seen in the normalization, $P_L(\psi_0)+P_R(\psi_0)$. We noted before that in the ideal
$T_{\rm max}\to\infty$ limit, the normalization is mostly equal to unity as a function of the sampling time, with pointwise discontinuous jumps at the special sampling times given in Eq.~\eqref{main-eqq-3}. We explained that these jumps are related to the formation of the dark states in the system. We now ask how these pointwise discontinuous jumps are modified for a finite $T_{\rm max}$. In Figs.~\ref{fig-finite}(d)-\ref{fig-finite}(f), we use numerical simulations to obtain $P_L(\psi_0)+P_R(\psi_0)$ vs $\gamma \tau$ for $T_{\rm max}=500, 1000, 1500$ and $N=15,~\ket{\psi_0} = \ket{11}$. Interestingly, for finite
$T_{\rm max}$, the pointwise discontinuities are broadened into dips of
finite width. This broadening is especially pronounced for smaller
$T_{\rm max}$, while increasing $T_{\rm max}$ makes the dips 
narrower and drives the normalization toward the ideal result. Broadening effect associated with a finite number of measurements has also been observed in single-target detection problem \cite{ExptQC-3, ExptQC-2}. We now proceed to derive $P_L(\psi_0)+P_R(\psi_0)$ for a finite $T_{\rm max}$.

%We derive here an expression of the total detection probability, $P_L(\psi_0)+P_R(\psi_0)$, for a finite $T_{\rm max}$ number of measurements. To begin with, let us consider the general Hamiltonian $H$ and measurements at two target states $\ket{D_1}$ and $\ket{D_2}$. 
\textit{Normalization for finite $T_{\rm max}$:} Let us consider the one-dimensional lattice chain, with parity symmetric Hamiltonian $H$ and target states at the boundary sites, $\ket{D_1} = \ket{1}$ and $\ket{D_2} = \ket{N}$. The survival probability that the quantum walker has not been detected at either target after $n$ measurements is given by \cite{QW6, QW7}
\begin{equation}
q_n(\psi_0)
=
\bra{\psi_0}(\mathcal{S}^\dagger)^n  \mathcal{S}^n\ket{\psi_0},
\label{app:qndef}
\end{equation}
where remember $\mathcal{S} = \left(\mathbb{1}-\hat{D} \right)U(\tau)$ is the survival operator with $\hat{D}$ given in Eq.~\eqref{unitary-eqn-8} and the unitary operator $U(\tau)$ in Eq.~\eqref{unitary-eqn}. Using Eq.~\eqref{app:qndef}, we get
\begin{align}
q_{n-1} (\psi_0) = \bra{\psi_0}(\mathcal{S}^\dagger)^{n-1}   U^\dagger \hat{D} U \mathcal{S}^{n-1}\ket{\psi_0}+ \bra{\psi_0}(\mathcal{S}^\dagger)^{n-1}   U^\dagger (\mathbb{1}-\hat{D}) U \mathcal{S}^{n-1}\ket{\psi_0}. \label{app:qndef2}
\end{align}
Following the definition of first-detection probabilities from Eq.~\eqref{unitary-eqn-13}, one can recast Eq.~\eqref{app:qndef2} as
\begin{equation}
F_n^{(x_L)} (\psi_0)   +F_n^{(x_R)} (\psi_0)  =q_{n-1} (\psi_0)-q_n (\psi_0). \label{app:qndef3}
\end{equation}
Summing this equation from $n=1$ to $n=T_{\rm max}$ and using Eq.~\eqref{vnao01}, we get
\begin{align}
P_{L}(\psi_0)+P_{R}(\psi_0)=1-q_{T_{\rm max}}(\psi_0).
\end{align}
From Eq.~\eqref{app:qndef}, it then follows
\begin{equation}
P_{L}(\psi_0)+P_{R}(\psi_0) = 1- \bra{\psi_0}(\mathcal{S}^\dagger)^{T_{\rm max}}  \mathcal{S}^{T_{\rm max}}\ket{\psi_0}.
\label{app:Ptot-exact}
\end{equation}
We saw before that for the parity-symmetric Hamiltonian, the dual-target splitting problem decomposes into two parity-specific auxiliary single-target detection problems with target states $\ket{d_{\pm}} = \left( \ket{1} \pm \ket{N} \right)/ \sqrt{2}$. Eq.~\eqref{app:Ptot-exact} can be expressed as
\begin{align}
P_{L}(\psi_0)+P_{R}(\psi_0) = 1 - \bra{\psi_0^{(+)}}(\mathcal{S}_{+}^\dagger)^{T_{\rm max}}  \mathcal{S}_{+}^{T_{\rm max}}\ket{\psi_0^{(+)}}- \bra{\psi_0^{(-)}}(\mathcal{S}_{-}^\dagger)^{T_{\rm max}}  \mathcal{S}_{-}^{T_{\rm max}}\ket{\psi_0^{(-)}}.\label{app:qndef5}
\end{align}
Here $\mathcal{S}_{\pm}= \left(\mathbb{1}-\ket{d_{\pm}} \bra{d_{\pm}} \right)U(\tau)$ are the single-target survival operators and $\ket{\psi_0^{(\pm)}}$ are the even and odd components of the initial state $\ket{\psi_0}$, defined as
\begin{align}
\ket{\psi_0^{(\pm)}} = \left( \frac{\mathbb{1} \pm \mathcal{P}}{2} \right) \ket{\psi_0}, \label{app:qndef6}
\end{align}
with the parity operator $\mathcal{P}$ in Eq.~\eqref{new-map-eqq1}. Introducing the right and the left eigenvectors of $\mathcal{S}_{\pm}$ and its eigenvalues
\begin{equation}
\mathcal{S}_{\pm}\ket{\Lambda_{\pm}^{(\mathrm{right})}}
=
\Lambda_{\pm}\ket{\Lambda_{\pm}^{(\mathrm{right})}},
\qquad
\bra{\Lambda_{\pm}^{(\mathrm{left})}}\mathcal{S}_{\pm}
=
\Lambda_{\pm}\bra{\Lambda_{\pm}^{(\mathrm{left})}},  \label{app:qndef7}
\end{equation}
the survival operators can be spectrally decomposed as
\begin{equation}
\mathcal{S}_{\pm}
=
\sum_{\{ \Lambda_{\pm} \}}
\Lambda_{\pm}
\ket{\Lambda_{\pm}^{(\mathrm{right})}}
\bra{\Lambda_{\pm}^{(\mathrm{left})}},~~~\mathcal{S}_{\pm}^{\dagger}
=
\sum_{\{\Lambda_{\pm}\}}
\Lambda_{\pm}^{*}
\ket{\Lambda_{\pm}^{(\mathrm{left})}}
\bra{\Lambda_{\pm}^{(\mathrm{right})}}.
\end{equation}
Eq.~\eqref{app:qndef5} then becomes
\begin{align}
P_{L}(\psi_0)+P_{R}(\psi_0) &= 1 - 
\sum_{ \{ \Lambda_{+} \}} \sum_{ \{ \widetilde{ \Lambda}_{+} \}} \left( \Lambda_{+} \right)^{T_{\rm max}} \left( \widetilde{ \Lambda}_{+}^* \right)^{T_{\rm max}} \braket{\psi _0^{(+)} |  \widetilde{\Lambda}_+^{(\rm left)} }   \braket{ \widetilde{\Lambda}_+^{(\rm right)}|  \Lambda_{+}^{(\mathrm{right})}     } \braket{\Lambda_{+}^{(\mathrm{left})}| \psi_0^{(+)} } \nonumber \\
&  - \sum_{ \{ \Lambda_{-} \}} \sum_{ \{ \widetilde{ \Lambda}_{-} \}} \left( \Lambda_{-} \right)^{T_{\rm max}} \left( \widetilde{ \Lambda}_{-}^* \right)^{T_{\rm max}} \braket{\psi _0^{(-)} |  \widetilde{\Lambda}_{-}^{(\rm left)} }   \braket{ \widetilde{\Lambda}_{-}^{(\rm right)}|  \Lambda_{-}^{(\mathrm{right})}     } \braket{\Lambda_{-}^{(\mathrm{left})}| \psi_0^{(-)} }
.\label{app:qndef8}
\end{align}
This is an exact result for any parity symmetric Hamiltonian. To gain further insight, we now focus on the vicinity of resonances, where we saw in Figs.~\ref{fig-finite}(d)-\ref{fig-finite}(f) that the total detection probability is less than unity and has a broadening effect.

\textit{Near resonance:} The eigenvalues satisfy $|\Lambda_{\pm}|<1$ in the complex plane and their values approach the unit circle, $|\Lambda_{\pm}| \to 1^-$, when a dark state is formed in the system \cite{QW13}. For simplicity, we now focus on $\mathcal{S}_{+}$, corresponding to the even-parity sector. For this case, resonance occurs when \cite{QW6, QW13}
\begin{align}
\exp(-iE_{\ell} \tau) = \exp(-iE_{\ell'} \tau), \label{app:qndef9}
\end{align}
where the associated energy eigenstates, $\ket{E_{\ell}}$ and $\ket{E_{\ell'}}$, both have even parity. The corresponding dark state is
\begin{align}
& \ket{\mathbb{D}_{\ell \ell'}^{(+)}} = \frac{a_{\ell'} \ket{E_{\ell}} - a_{\ell} \ket{E_{\ell'}}}{\sqrt{p_{\ell}+p_{\ell'}}}, \label{app:qndef10} \\
& \text{with }a_{\ell} = \braket{d_+|E_{\ell}},~~p_{\ell} = | \braket{d_+|E_{\ell}}|^2. \label{app:qndef11}
\end{align}
Close to the resonance condition~\eqref{app:qndef9}, we define
\begin{align}
\Delta_{\ell \ell'} = \operatorname{Arg} \left[ \exp(-i(E_{\ell} - E_{\ell'} )\tau) \right], \label{app:qndef12}
\end{align}
where recall $\operatorname{Arg}(..)$ stands for the argument of the complex number. At exact resonance, $\Delta_{\ell\ell'}=0$, while close to resonance $|\Delta_{\ell\ell'}|\ll 1$.

Following earlier studies on single-target problems \cite{ExptQC-3, ExptQC-2}, we use the following near-resonance approximations
\begin{align}
\ket{\Lambda_+^{(\rm left)}} \approx \ket{\Lambda_+^{(\rm right)}} \approx \ket{\mathbb{D}_{\ell \ell'}^{(+)}},~~|\Lambda_+|^2 \approx 1- \frac{p_{\ell} p_{\ell'}}{(p_{\ell}+p_{\ell'})^3} \Delta_{\ell \ell'}^2, \label{app:qndef13}
\end{align}	
where $\ket{\mathbb{D}_{\ell \ell'}^{(+)}}$ and $p_{\ell}$ are given in Eqs.~\eqref{app:qndef10} and \eqref{app:qndef11}, respectively. Plugging these approximate expressions in Eq.~\eqref{app:qndef8} for $T_{\rm max} \gg 1$, we have
\begin{align}
\sum_{ \{ \Lambda_{+} \}} \sum_{ \{ \widetilde{ \Lambda}_{+} \}} & \left( \Lambda_{+} \right)^{T_{\rm max}} \left( \widetilde{ \Lambda}_{+}^* \right)^{T_{\rm max}}  \braket{\psi _0^{(+)} |  \widetilde{\Lambda}_+^{(\rm left)} }   \braket{ \widetilde{\Lambda}_+^{(\rm right)}|  \Lambda_{+}^{(\mathrm{right})}     } \braket{\Lambda_{+}^{(\mathrm{left})}| \psi_0^{(+)} } \nonumber \\
& \approx \sum_{\ell } \sum_{\ell' < \ell} \left| \braket{\psi_0| \mathbb{D}_{\ell \ell'}^{(+)}} \right|^2 \exp \left[- \frac{p_{\ell} p_{\ell'}}{(p_{\ell}+p_{\ell'})^3} \Delta_{\ell \ell'}^2 T_{\rm max}  \right]  \mathcal{I}_{\ell \ell'}^{\rm (+)}  . \label{app:qndef14}
\end{align}
In the second line, we have introduced the indicator function
$\mathcal{I}_{\ell\ell'}^{(+)}$, which is equal to unity when both
$\ket{E_\ell}$ and $\ket{E_{\ell'}}$ belong to the even-parity sector, and vanishes otherwise.

A similar analysis on the odd-parity sector gives
\begin{align}
\sum_{ \{ \Lambda_{-} \}} \sum_{ \{ \widetilde{ \Lambda}_{-} \}} & \left( \Lambda_{-} \right)^{T_{\rm max}} \left( \widetilde{ \Lambda}_{-}^* \right)^{T_{\rm max}}  \braket{\psi _0^{(-)} |  \widetilde{\Lambda}_-^{(\rm left)} }   \braket{ \widetilde{\Lambda}_-^{(\rm right)}|  \Lambda_{-}^{(\mathrm{right})}     } \braket{\Lambda_{-}^{(\mathrm{left})}| \psi_0^{(-)} } \nonumber \\
& \approx \sum_{\ell } \sum_{\ell' < \ell} \left| \braket{\psi_0| \mathbb{D}_{\ell \ell'}^{(-)}} \right|^2 \exp \left[- \frac{p_{\ell} p_{\ell'}}{(p_{\ell}+p_{\ell'})^3} \Delta_{\ell \ell'}^2 T_{\rm max}  \right]  \mathcal{I}_{\ell \ell'}^{\rm (-)} , \label{app:qndef14b}
\end{align}
where $\mathcal{I}_{\ell\ell'}^{(-)}$ is the indicator function for the odd-parity sector and $\ket{\mathbb{D}_{\ell \ell'}^{(-)}}$ is the dark state corresponding to $\mathcal{S}_-$
\begin{align}
\ket{\mathbb{D}_{\ell \ell'}^{(-)}} = \frac{a_{\ell'} \ket{E_{\ell}} - a_{\ell} \ket{E_{\ell'}}}{\sqrt{p_{\ell}+p_{\ell'}}},~~\text{where both } \ket{E_{\ell}} \text{ and } \ket{E_{\ell'}} \text{ have odd parity}. \label{app:qndef15}
\end{align}
Substituting Eqs.~\eqref{app:qndef14}-\eqref{app:qndef15} in Eq.~\eqref{app:qndef8}, we have
\begin{align}
P_{L}(\psi_0)+P_{R}(\psi_0)&  \approx 1-  \sum_{\ell } \sum_{\ell' < \ell} \left| \braket{\psi_0| \mathbb{D}_{\ell \ell'}^{(+)}} \right|^2 \exp \left[- \frac{p_{\ell} p_{\ell'}}{(p_{\ell}+p_{\ell'})^3} \Delta_{\ell \ell'}^2 T_{\rm max}  \right]  \mathcal{I}_{\ell \ell'}^{\rm (+)} \nonumber \\
&-\sum_{\ell } \sum_{\ell' < \ell} \left| \braket{\psi_0| \mathbb{D}_{\ell \ell'}^{(-)}} \right|^2 \exp \left[- \frac{p_{\ell} p_{\ell'}}{(p_{\ell}+p_{\ell'})^3} \Delta_{\ell \ell'}^2 T_{\rm max}  \right]  \mathcal{I}_{\ell \ell'}^{\rm (-)}. \label{app:qndef16}
\end{align}
We emphasize that this expression is an asymptotic result, derived for large $T_{\rm max}$ and in the vicinity of the finite-$\tau$ resonances that satisfy Eq.~\eqref{app:qndef9}.

\textit{Nearest-neighbor hopping Hamiltonian:} For this case, we use Eq.~\eqref{neww-eg-eq-8} in Eq.~\eqref{app:qndef11} to get $a_{\ell} = 2/\sqrt{N+1} \sin k_{\ell}$ and $p_{\ell} = 4/(N+1) \sin^2 k_{\ell}$ where $k_{\ell} = \pi \ell/(N+1)$. Moreover, we saw before in Eq.~\eqref{neww-eg-eq-1} that the energy eigenstate $\ket{E_{\ell}}$ has even parity for odd $\ell$ and odd parity for even $\ell$. Using these results in Eq.~\eqref{app:qndef16} gives
\begin{align}
\boxed{P_L(\psi_0)+P_R(\psi_0)
\approx
1-
\sum_{\ell=1}^{N}
\sum_{\ell'<\ell}
\frac{1+(-1)^{\ell-\ell'}}{2}
\left|
\braket{\psi_0|\mathbb D_{\ell\ell'}}
\right|^2
\exp\left[
-\frac{N+1}{4}
\frac{
\sin^2 k_\ell\,\sin^2 k_{\ell'}
}{
\left(
\sin^2 k_\ell+\sin^2 k_{\ell'}
\right)^3
}
\Delta_{\ell\ell'}^2 T_{\rm max}
\right],}
\label{app:qndef17}
\end{align}
where the factor $(1+(-1)^{\ell-\ell'})/2$ ensures that only energy eigenstates with the same parity, i.e., with $\ell$ and $\ell'$ both odd or both even, contribute to the sum. We also define $\ket{\mathbb D_{\ell\ell'}}$ as
\begin{align}
\ket{\mathbb D_{\ell\ell'}}
=
\frac{
\sin k_{\ell'}\,\ket{E_\ell}
-
\sin k_\ell\,\ket{E_{\ell'}}
}{
\sqrt{
\sin^2 k_\ell+\sin^2 k_{\ell'}
}
}. \label{app:qndef18}
\end{align}
The theoretical prediction in Eq.~\eqref{app:qndef17} is shown in
Figs.~\ref{fig-finite}(d)-\ref{fig-finite}(f), where we compare it with the numerical
simulations. Away from the Zeno limit, we find excellent agreement between
the analytical expression and the simulation results. Close to the Zeno limit, however, noticeable deviations appear. This is expected, since the derivation of Eq.~\eqref{app:qndef17} relies on the resonance condition in Eq.~\eqref{app:qndef9}, which is not applicable in the Zeno limit. In this limit, all quasienergies become simultaneously degenerate.

However, for the other resonances, Eq.~\eqref{app:qndef9} is satisfied, and consequently there is excellent agreement between Eq.~\eqref{app:qndef17} and the simulations. Close to such a resonance, $|\Delta_{\ell\ell'}|\ll1$ for the corresponding pair of energy levels, so that $\Delta_{\ell\ell'}^2 T_{\rm max}$ is finite inside the exponential factor in Eq.~\eqref{app:qndef17}. Therefore, the terms in the summation in Eq.~\eqref{app:qndef17} are significant. As a result,
$P_L(\psi_0)+P_R(\psi_0)<1$ over a finite range of sampling times close to the resonance, giving rise to the broadening effect, as seen in Figs.~\ref{fig-finite}(d)-\ref{fig-finite}(f). As one moves away from resonance, $\Delta_{\ell\ell'}$ becomes finite, and one gets $\Delta_{\ell\ell'}^2 T_{\rm max} \gg 1$ for sufficiently large $T_{\rm max}$. The corresponding terms in the summation are exponentially suppressed. Consequently, the normalization approaches
$P_L(\psi_0)+P_R(\psi_0)=1$, as observed in Figs.~\ref{fig-finite}(d)-\ref{fig-finite}(f).

\section{Conclusion and Future Directions }
\label{sec-conclusion}
In conclusion, we have constructed a general theoretical framework for the splitting problem of a discrete-state, continuous-time quantum system, and subjected to projective measurements at two distinct, orthogonal target states at regular time intervals $\tau$. We provide explicit tools for analyzing how unitary evolution and repeated measurements combine to shape the competition between the two possible outcomes. Our renewal framework, as exemplified in Eq.~\eqref{main-eqq-00}, expresses the associated first detection amplitudes in terms of the measurement-free unitary evolution. The splitting probability is then given by Eq.~\eqref{anj8y14} or equivalently by Eqs. \eqref{unitary-eqn-19} and \eqref{unitary-eqn-22}. These results are valid for arbitrary time-independent, Hermitian Hamiltonian $H$, target states $\ket{D_{1}}$, $\ket{D_{2}}$, the initial state $\ket{\psi_0}$ and the sampling time $\tau$.

For Hamiltonians possessing reflection symmetry and localized measurements performed at the left and right endpoints of a one-dimensional chain, $\ket{D_1}=\ket{1}$ and $\ket{D_2}=\ket{N}$, the renewal framework reduces to a mapping that relates the splitting probabilities to a pair of single-target first-detection problems. For this class of Hamiltonians, we obtained exact expressions of the splitting probabilities, $P_L(\psi_0)$ and $P_R(\psi_0)$, in Eqs.~\eqref{neww-eg-eq-17}. To get some insights into these expressions, we examined a tight-binding chain with nearest-neighbor hopping Hamiltonian in Eq.~\eqref{main-eqq-1}. An interesting feature emerges in the Zeno limit, where both $P_L(\psi_0)$ and $P_R(\psi_0)$, for any system size $N$, become independent of the detailed structure of the initial state $\ket{\psi_0}$ in the interior of the chain and depend only on the projection of $\ket{\psi_0}$ on the target boundary sites, see Eq.~\eqref{main-eqq-6}.

For localized initial condition, $\ket{\psi_0} = \ket{x_0}$, we found a surprising phase-transition-like behavior in the splitting dynamics at a critical sampling time $\tau_c = \pi/(2\gamma)$ in the limit of large system size for the nearest-neighbor hopping Hamiltonian. This flat-to-fluctuating transition separates two qualitatively distinct regimes: For $0<\tau \leq \tau_c$, the splitting probabilities have a flat value $1/2$ for all $x_0$ and $\tau$, while for $\tau > \tau_c$, both $P_L(x_0)$ and $P_R(x_0)$ differ from $1/2$ and have a fluctuating dependence on $x_0$ and $\tau$. This is illustrated in Fig.~\ref{fig-transition}. Beyond localized initial conditions, we showed that this transition can also persist for initial states involving coherent superpositions over multiple sites. Interestingly, its detailed form can be modified by the spatial separation and relative phase between the components of the initial state. In particular, the splitting probability exhibits an oscillatory dependence on the initial phase, with a parity-dependent phase shift, as shown in Eq.~\eqref{OIC-eq-7} and Fig.~\ref{fig-OIC}.

The fluctuating regime at $\tau>\tau_c$ for localized case also exhibits a characteristic spatial
structure. The splitting probabilities oscillate with the initial position $x_0$, with dominant wavenumbers $Q_{\rm max}(\tau)$ given in Eq.~\eqref{skpssrs}. These dominant modes form the spectral branches seen in the power-spectrum heat map in Fig.~\ref{fig-osc}, with the first branch emerging at $\tau_c$ and additional branches appearing at larger sampling times. The onset of these oscillations can be understood from the inter-parity phase-matching condition in Eq.~\eqref{SS-pos-eq-11}, or equivalently Eq.~\eqref{eq:phase-match-continuous}, between momentum modes. While for $\tau\geq\tau_c$ there exist distinct momentum modes satisfying this phase-matching condition, for $\tau<\tau_c$ no such modes exists. This leads to the spectral interpretation of the oscillations in the splitting probability and the associated flat-to-fluctuating transition. Finally, we derived the resonance conditions in Eq.~\eqref{main-eqq-3} that determine the pointwise discontinuous jumps in the splitting probability at special sampling times. At these sampling times, $P_L(x_0)+P_R(x_0)<1$. For a finite number of measurements, these pointwise discontinuous jumps are broadened into dips of finite width, as described by Eq.~\eqref{app:qndef17} and illustrated in Fig.~\ref{fig-finite}.

For the classical case, these features do not arise, even when the classical random walker is monitored stroboscopically with time period $\tau$ \cite{myletter}. As studied in detail in the Supplementary Material of \cite{myletter}, a continuous-time classical random walk hops on a one-dimensional lattice chain, with its boundaries monitored only at the times $\tau,2\tau,3\tau,\ldots$. If the walker is located at any of the boundaries at the measurement times, it is absorbed and the process is stopped. Otherwise, it continues to evolve until a successful detection is made. Here, $\tau \to 0$ corresponds to the continuous monitoring and leads to the gambler's ruin formula for the splitting probabilities in Eq.~\eqref{unitary-eqn-classical}. By contrast, the $\tau=0$ for quantum case leads to the freezing of dynamics with no detection, unless the starting position coincides with the detectors; see Eq.~\eqref{neww-eg-eq-28}. For arbitrary $\tau$, the classical splitting probabilities vary smoothly with the sampling time, without exhibiting the phase-transition-like behavior found in the quantum case. Moreover, the classical proximity effect is always preserved, with the classical walker having a higher probability of detection at the boundary closer to its initial position, whereas in the quantum case, detection at the more distant boundary can also be favored.

Overall, our work establishes a systematic treatment of splitting probabilities in quantum monitored dynamics with two targets, clarifying the role of quantum measurements and interference effects. Looking ahead, our work points to several future directions. While we concentrated on splitting probabilities between two different targets, in many cases it is useful to characterize the hitting times at various targets in the system and their probabilities \cite{Rednerbook}. Our renewal framework, in principle, can be used to compute these quantities, and are left for the future. Extending our framework for other measurement protocols such as weak measurements \cite{Heine2025QuantumWalksWeak} or random sampling time protocols \cite{QW17, QW26}, and comparing them with our results is a promising future direction. Furthermore, our analysis has focused on local monitoring at fixed target states. It would be interesting to investigate the splitting problem under a tracking protocol, in which the system is monitored over the full set of states rather than only at the designated targets \cite{QW19, ExptQC-2}. An interesting question is whether spatial oscillations also emerge only beyond a critical sampling time under such a protocol. Another direction is to determine the $\tau$-dependence of the amplitudes and phases of the spatial oscillations, beyond the characteristic wavenumbers obtained here for large system sizes. The scaling of the special resonant sampling times with system size is another aspect worth exploring. It would also be interesting to explore the splitting problems for other graphs and  many-body systems \cite{Gambassi2025,Liu2024}. In particular, the transition point $\tau_c$ is dependent only on the width of the energy band via Eq.~\eqref{ajbidhiguqok}, since the discrete energy levels merge into a continuum in the thermodynamic limit. Therefore, the transition between flat to fluctuating regimes is likely to be go beyond the current model.

Finally, the splitting protocol considered in our paper can be implemented in experimental platforms like quantum computer \cite{ExptQC-1, ExptQC-2, ExptQC-3} or optical waveguides \cite{ExptWG-3, ExptWG-2, Chen2026}. In quantum-computing implementations, the quantum-walk dynamics can be mapped to a spin model through the Jordan-Wigner transformation and periodic detection protocol can be implemented by mid-circuit measurements. Similarly, in optical waveguides, quantum walks can be effectively implemented by using electric fields across an array of coupled waveguides, and periodic measurements can be realized using a segmented waveguides. We anticipate that incorporating noise, as encountered in realistic systems, may reveal further insights \cite{ExptQC-3, Ma2026}.

\section*{Acknowledgment}
%The support of Israel Science Foundation's grant $1614/21$ is gratefully acknowledged.
This research was supported by the Israel Science Foundation (grant No. 2311/25). 

\addtocontents{toc}{\protect\setcounter{tocdepth}{-1}}
\appendix

\section{Derivation of the quantum renewal equation}
\label{appen-sec-renewal}
We first recast the first-detection amplitude $\phi _{n}^{\left(D_{\alpha} \right)}\left( \psi _0 \right)$ in Eq.~\eqref{anjiuq1} as
\begin{align}
 \phi _{n}^{\left(D_{\alpha} \right)}\left( \psi _0 \right) = \langle D_{\alpha} \ket{\theta _n}, \label{unitary-eqn-15} 
\end{align}	
where we define
\begin{align}
\ket{\theta _n} =  U(\tau) \Big[ \Big(\mathbb{1}-\hat{D} \Big) U(\tau) \Big]^{n-1}\ket{ \psi _0}. \label{unitary-eqn-14}  
\end{align}
Here $U(\tau) = e^{-i H \tau}$ is the measurement-free unitary operator, and $\hat{D}$ is the projection operator onto the detection subspace, as defined in Eq.~\eqref{unitary-eqn-8}.

Proving the renewal equation \eqref{unitary-eqn-17} is equivalent to showing
\begin{align}
\ket{\theta _n} = U (n \tau ) \ket{\psi _0} - \sum _{l=1}^{(n-1)} U \left( (n-l)\tau \right) \hat{D} \ket{\theta _l}. \label{appen-renewal-eq-1}
\end{align}
Using the expression of $\ket{\theta _n}$ in Eq.~\eqref{unitary-eqn-14}, we get for $n=1$ and $n=2$
\begin{align}
& \ket{\theta _1} = U(\tau) \ket{\psi _0}, \\
& \ket{\theta _2} =  U(2\tau) \ket{\psi _0}- U(\tau) \hat{D} \ket{\theta _1}.
\end{align}
Both of these results are consistent with the recursive relation in Eq.~\eqref{appen-renewal-eq-1}. For general $n$, we will establish this relation by induction. 

Assume that Eq.~\eqref{appen-renewal-eq-1} holds for some arbitrary $n$. Then, using Eq.~\eqref{unitary-eqn-14}, we obtain
\begin{align}
\ket{\theta _{n+1}} & = U(\tau) \ket{\theta _n}-U(\tau)  \hat{D}  \ket{\theta _n}  , \\
& = U \left( (n+1)\tau \right) \ket{\psi _0} - \sum _{l=1}^{n-1} U \left( (n-l+1)\tau \right) \hat{D} \ket{\theta _l} - U(\tau) \hat{D} \ket{\theta _n}, \\
& =U \left( (n+1)\tau \right) \ket{\psi _0} - \sum _{l=1}^{n} U \left( (n+1-l)\tau \right) \hat{D} \ket{\theta _l}.
\end{align} 
This is essentially Eq.~\eqref{appen-renewal-eq-1} evaluated at $(n+1)$. Hence, by induction, this relation holds for arbitrary $n$.

\section{Details of the numerical simulations}
\label{sec-appen-simulation}
To compute the splitting probabilities numerically, we employ a Monte Carlo simulation of the stroboscopic detection protocol for a given Hamiltonian $H$, and the measurement interval $\tau$. 
The procedure can be summarized as follows:

%Starting from the initial state $\ket{x_0}$, the system is evolved with the unitary operator $U(\tau)$ for a duration $\tau$. The probability of finding the system in the first detection state is $ |\langle D_1| U(\tau) \ket{x_0}|^2$. Therefore, we generate a random number $r$ uniformly in the range $[0,1]$. If $r \leq |\langle D_1| U(\tau) \ket{x_0}|^2 $, then the particle is detected at $\ket{D_1}$, and the process is stopped.

\begin{enumerate}
   \item Starting from the initial state $\ket{\psi_0}$, the system evolves under the unitary operator $U(\tau)$ for a duration $\tau$. 
The probability of detecting the system in the first target state $\ket{D_1}$ is given by 
$\omegaa _1(\tau)= |\langle D_1| U(\tau) \ket{\psi_0}|^2$. 
To implement the measurement, we draw a random number, $r_1$, uniformly from the range $[0,1]$. 
If $r_1 \leq \omegaa _1(\tau)$, the system is detected at $\ket{D_1}$, and the process terminates.

\item If this detection attempt is unsuccessful, \textit{i.e.}, $r_1 > \omegaa_1$, 
the state is immediately collapsed to
\begin{align}
\ket{\psi(\tau^+)} = 
\frac{\left( \mathbb{1} - \ket{D_1}\!\bra{D_1} \right) U(\tau) \ket{\psi_0}}
     {\sqrt{1 - \omegaa_1(\tau)}}, \nonumber
\end{align}
where $\tau^+$ recall is the time instant immediately after a failed measurement of $\ket{D_1}$. An immediate measurement is then performed at the second target state $\ket{D_2}$. 
The probability of detection is $\omegaa_2(\tau) = \big|\langle D_2 | \psi(\tau^+) \rangle \big|^2$. 
To employ the measurement, we again generate another random number $r_2$ uniformly in the range $[0,1]$. 
If $r_2 \leq \omegaa_2(\tau)$, the system is detected at $\ket{D_2}$, and the process terminates.

\item On the other hand, if $r_2 > \omegaa_2(\tau)$, this detection attempt also fails. 
The state is then collapsed with respect to $\ket{D_2}$, yielding
\begin{align}
\ket{\psi(\tau^{++})} = 
\frac{\left( \mathbb{1} - \ket{D_2}\!\bra{D_2} \right)}
     {\sqrt{1 - \omegaa_2(\tau)}} \ket{\psi(\tau^+)}. \nonumber
\end{align}
Taking this normalized state in place of the original $\ket{\psi_0}$, the entire procedure is then repeated at the next stroboscopic time step.

\end{enumerate}
This recursive process continues until the system is successfully detected at either $\ket{D_1}$ or $\ket{D_2}$, 
or the maximum $T_{\rm max}$ number of measurement attempts is reached. Subsequently, the simulation is carried out for $R$ independent realizations, from which the splitting probabilities, $P_{D_1}(\psi_0)$ and $P_{D_2}(\psi_0)$, are estimated as ensemble averages. If $R_{\alpha}$ denotes the total number of realizations in which the system is detected at the $\ket{D_{\alpha}}$ state
(with $\alpha \in \{1,2\}$), then the splitting probabilities are 
\begin{align}
P_{D_1}(\psi_0) = \frac{R_{1}}{R}, ~~P_{D_2}(\psi_0) = \frac{R_{2}}{R}.
\end{align}
Note that $R_1+R_2 \leq R$, with strict inequality in the case of  special measurement times. In our simulation, we fix $T_{\rm max}=500,1000, 1500,2000,10^4$ and $R=10^4$.

\section{$\tau = 0$ case}
\label{appen-sec-Zeno}
In this appendix, we will derive the splitting detection probabilities $P_L(\psi_0)$ and $P_R(\psi_0)$ for $\tau =0$, as was quoted in Eq.~\eqref{neww-eg-eq-28kou} in the main text. Our starting point are the auxiliary probabilities $\rho _{\pm}(\psi_0)$ in Eqs.~\eqref{aux-eq-6} and \eqref{aux-eq-7}. Since all quasienergies are identical, we have
\begin{align}
\rho _+(\psi_ 0) =  \frac{\Big|  \sum_{\substack{m=1 \\ m \in  \,\text{odd}}}^{N}  ~ \langle d_+ | E_{m} \rangle \langle E_{m}| \psi_0 \rangle \Big|^2}{   \sum_{\substack{m=1 \\ m \in  \,\text{odd}}}^{N}  \big| \langle d_+| E_{m} \rangle \big|^2}. \label{neww-eg-eq-24}
\end{align}
Using $\ket{d_+}$ from Eq.~\eqref{neww-eg-eq-8}, the denominator in Eq.~\eqref{neww-eg-eq-24} simplifies to
\begin{align}
 \sum_{\substack{m=1 \\ m \in  \,\text{odd}}}^{N} \big| \langle d_+| E_{m} \rangle \big|^2 =  \frac{4}{N+1}\sum_{\substack{m=1 \\ m \in  \,\text{odd}}}^{N} \sin \left( \frac{\pi m}{N+1} \right)^2 = 1.  \label{neww-eg-eq-241}
\end{align}
On the other hand, the numerator depends on the initial state $\ket{\psi_0}$ 
\begin{align}
\sum_{\substack{m=1 \\ m \in  \,\text{odd}}}^{N}  ~ \langle d_+ | E_{m} \rangle \langle E_{m}| \psi_0 \rangle = \braket{d_+| \psi_0}= \frac{1}{\sqrt{2}} \left( \braket{1|\psi_0}+\braket{N|\psi_0} \right).  \label{neww-eg-eq-242}
\end{align}
The simplified forms of the numerator and denominator now yield 
\begin{align}
\rho _+(\psi_0) = \frac{1}{2} \left|  \braket{1|\psi_0}+\braket{N|\psi_0} \right|^2. \label{neww-eg-eq-25}
\end{align}
The same procedure also yields the other auxiliary probability
\begin{align}
\rho _-(\psi_0) = \frac{1}{2} \left|  \braket{1|\psi_0}-\braket{N|\psi_0} \right|^2. \label{neww-eg-eq-26}
\end{align}
Turning to Eq.~\eqref{neww-eg-eq-17}, we also need the interference term $\zeta _{+-}(\psi_0)$ to fully specify the splitting probabilities. This, in turn, requires the auxiliary generating functions $\widetilde{\chi}_{\pm}(z)$ via Eq.~\eqref{aux-eq-2}. For $\tau = 0$, Eq.~\eqref{aux-eq-1} simplifies to  
\begin{align}
\widetilde{\chi} _{\pm}(z) \simeq z \braket{d_{\pm} | \psi_0} \simeq \frac{z}{\sqrt{2}} \left( \braket{1|\psi_0} \pm \braket{N|\psi_0} \right) . 
\end{align}
The interference term in Eq.~\eqref{aux-eq-2} then becomes
\begin{align}
\zeta _{+-}(\psi_0) \simeq \frac{1}{2} \left( |\braket{1| \psi_0}|^2-|\braket{N| \psi_0}|^2 \right).
\end{align}
Substituting this expression into Eq.~\eqref{neww-eg-eq-17}, together with $\rho_{\pm}(\psi_0)$ from Eqs.~\eqref{neww-eg-eq-25} and \eqref{neww-eg-eq-26}, 
we obtain the expressions for the splitting probabilities given in Eq.~\eqref{neww-eg-eq-28kou}.

\section{Dominant wavenumber in the power spectrum of $\zeta_{+-}(x_0)$}
\label{appen-sec-FT}
In this Appendix, we derive the large-$N$ expression of the dominant wavenumber $Q_{\rm max}(\tau)$ in Eq.~\eqref{pos-eq-2}, at which the power spectrum $|\widetilde{\zeta}\left( Q,N \right)|^2$ in Eq.~\eqref{SNRPSS} for the nearest-neighbor Hamiltonian is maximum. Our starting point is Eq.~\eqref{SS-pos-eq-5}, which we rewrite here for simplicity
\begin{align}
\widetilde{\zeta}\left( Q,N \right) & = \sum_{\substack{s=1\\ s\ \mathrm{odd}}}^{N}\sum_{\substack{s'=1\\ s'  \ \mathrm{even}}}^{N}  \frac{\mathcal{C}_{s, s'}(\tau, N) }{4}\Bigg[
e^{-i\frac{(Q-\omega_s+\omega_{s'})(N+1)}{2}}
\frac{\sin\!\left[\frac{N}{2}(Q-\omega_s+\omega_{s'})\right]}
{\sin\!\left[\frac{1}{2}(Q-\omega_s+\omega_{s'})\right]}
+
e^{-i\frac{(Q+\omega_s-\omega_{s'})(N+1)}{2}}
\frac{\sin\!\left[\frac{N}{2}(Q+\omega_s-\omega_{s'})\right]}
{\sin\!\left[\frac{1}{2}(Q+\omega_s-\omega_{s'})\right]} \Bigg. \nonumber \\
&~~~~~~~~~~~~~~~ \Bigg. -
e^{-i\frac{(Q-\omega_s-\omega_{s'})(N+1)}{2}}
\frac{\sin\!\left[\frac{N}{2}(Q-\omega_s-\omega_{s'})\right]}
{\sin\!\left[\frac{1}{2}(Q-\omega_s-\omega_{s'})\right]}
-
e^{-i\frac{(Q+\omega_s+\omega_{s'})(N+1)}{2}}
\frac{\sin\!\left[\frac{N}{2}(Q+\omega_s+\omega_{s'})\right]}
{\sin\!\left[\frac{1}{2}(Q+\omega_s+\omega_{s'})\right]}
\Bigg], \label{appen-FT-eq-5}
\end{align}
where recall $\omega_s = \pi s /(N+1)$ and $\mathcal{C}_{s, s'}(\tau, N) $ is given in Eq.~\eqref{SS-pos-eq-2}. Notice that Eq.~\eqref{appen-FT-eq-5} contains terms of the form
\begin{align}
e^{-i\frac{y(N+1)}{2}} \frac{\sin \left( \frac{N  y}{2} \right)}{\sin \left( \frac{y}{2} \right)} =  \cos \left[ \frac{ y(N+1)}{2} \right]  \frac{\sin \left( \frac{N  y}{2} \right)}{\sin \left( \frac{y}{2} \right)} - i \sin \left[ \frac{ y(N+1)}{2} \right]  \frac{\sin \left( \frac{N  y}{2} \right)}{\sin \left( \frac{y}{2} \right)}, \label{appen-FT-eq-6} 
\end{align}
where $y$ can take possible values $y= 0, \pm \pi/(N+1),\pm 2 \pi / (N+1) \ldots$. Let us look at the real part of Eq.~\eqref{appen-FT-eq-6}. For $y\to0$, the ratio $\sin(Ny/2)/\sin(y/2)$ approaches $N$, while the cosine factor becomes unity, so that the real part scales as $N$. By contrast, for $y=m \pi/(N+1)$ with odd $m$, the real part vanishes since $\cos(m \pi/2)=0$, whereas for nonzero even $m$, the real part is of order unity. Thus, for large $N$, the contribution from $y=0$ dominates the real part of Eq.~\eqref{appen-FT-eq-6}, and we obtain
\begin{align}
\text{Re} \left[ e^{-i\frac{y(N+1)}{2}} \frac{\sin \left( \frac{N  y}{2} \right)}{\sin \left( \frac{y}{2} \right)} \right] \simeq N \delta_{y,0}. \label{appen-FT-eq-7} 
\end{align}
Let us next examine the large-$N$ behavior of the imaginary part of Eq.~\eqref{appen-FT-eq-6}. For $y \to 0$, the imaginary part vanishes, unlike the real part. For $y=m\pi/(N+1)$ with even $m$, the imaginary part again vanishes since $\sin(m\pi/2)=0$. For odd $m$, however, it is nonzero and can be written as $-\cot[m\pi/2(N+1)]$. For large $N$ with fixed $m$, $\cot[m\pi/2(N+1)] \simeq 2(N+1)/(m\pi)$, showing that these contributions are of order $N$. Thus, in the large-$N$ limit, the leading contribution to the imaginary part in Eq.~\eqref{appen-FT-eq-6} arises from odd $m$, while the remaining terms are subleading
\begin{align}
\text{Im} \left[ e^{-i\frac{y(N+1)}{2}} \frac{\sin \left( \frac{N  y}{2} \right)}{\sin \left( \frac{y}{2} \right)} \right] \simeq -\frac{2 N}{ \pi}  \sum_{\substack{m=1\\ m\ \mathrm{odd}}}^{ \infty} \frac{1}{m} \left[ \delta_{y, \frac{m \pi}{N+1}}  -  \delta_{y, -\frac{m \pi}{N+1}} \right]. \label{appen-FT-eq-8} 
\end{align}
Plugging Eqs.~\eqref{appen-FT-eq-7}-\eqref{appen-FT-eq-8} in Eq.~\eqref{appen-FT-eq-6}, we obtain
\begin{align}
e^{-i\frac{y(N+1)}{2}} \frac{\sin \left( \frac{N  y}{2} \right)}{\sin \left( \frac{y}{2} \right)} \simeq  N \delta_{y,0} + \frac{2 i N}{ \pi}  \sum_{\substack{m=1\\ m\ \mathrm{odd}}}^{ \infty} \frac{1}{m} \left[ \delta_{y, -\frac{m \pi}{N+1}}  -  \delta_{y, \frac{m \pi}{N+1}} \right]. \label{appen-FT-eq-9}  
\end{align}
Substituting this form in Eq.~\eqref{appen-FT-eq-5} gives
\begin{align}
\widetilde{\zeta}\left( Q,N \right) & \simeq \frac{N}{4} \sum_{\substack{s=1\\ s\ \mathrm{odd}}}^{N}\sum_{\substack{s'=1\\  s' \ \mathrm{even}}}^{N}  \mathcal{C}_{s, s'}(\tau, N)   \Bigg[ \delta_{Q, \left| \omega_s -\omega_{s'} \right|} - \delta_{Q, \pi- \left|\pi- \omega_s -\omega_{s'} \right|}    + \frac{2i }{\pi}  \sum_{\substack{m=1\\ m\ \mathrm{odd}}}^{ \infty} \frac{1}{m}\Bigg\{           \delta_{Q, \left| \omega_s -\omega_{s'}\right|  -\frac{m \pi}{N+1}         } \Bigg. \Bigg. \nonumber \\
& ~~~~~~~~~~~~  \Bigg. \Bigg. - \delta_{Q, \left| \omega_s -\omega_{s'}\right|  +\frac{m \pi}{N+1}    } - \delta_{Q, \pi-\left| \pi-\omega_s -\omega_{s'}\right|  -\frac{m \pi}{N+1}         }+\delta_{Q, \pi-\left| \pi-\omega_s -\omega_{s'}\right|  +\frac{m \pi}{N+1}         }
\Bigg\}
\Bigg], \label{appen-FT-eq-10}
\end{align}
where the symbol `$\simeq$' indicates that this expression is valid in the large-$N$ limit. Recall that the wavenumber can be written as $Q = \pi j /(N+1)$ where $j  \in \{1,2, \ldots,N \}$. Eq.~\eqref{appen-FT-eq-10} then becomes
\begin{align}
\widetilde{\zeta}\left( Q  = \frac{\pi j }{N+1},N\right) & \simeq \frac{N}{4} \sum_{\substack{s=1\\ s\ \mathrm{odd}}}^{N}\sum_{\substack{s'=1\\  s' \ \mathrm{even}}}^{N}  \mathcal{C}_{s, s'}(\tau, N)   \Bigg[ \delta_{j, \left|s-s' \right|} - \delta_{j, N+1- \left|N+1- s-s' \right|}    + \frac{2i }{\pi}  \sum_{\substack{m=1\\ m\ \mathrm{odd}}}^{ \infty} \frac{1}{m}\Bigg\{           \delta_{j, \left| s-s' \right|  -m      } \Bigg. \Bigg. \nonumber \\
& ~~~~~~~~~~~~  \Bigg. \Bigg. - \delta_{j, \left| s-s' \right|  +m}- \delta_{j, N+1-\left| N+1-s-s'\right|  -m }+  \delta_{j, N+1-\left| N+1-s-s'\right|  +m }
\Bigg\}
\Bigg], \label{appen-FT-eq-11}
\end{align}
where recall $\mathcal{C}_{s, s'}(\tau, N)$ is a real function given in Eq.~\eqref{SS-pos-eq-2}.

\subsection{$Q = \pi j /(N+1)$ with odd $j$}
From Eq.~\eqref{appen-FT-eq-11}, $s$ is odd and $s'$ is even. As a result, both $|s-s'|$ and $N+1-|N+1-s-s'|$ are odd integers, while $|s-s'|\pm m$ and $N+1-|N+1-s-s'|\pm m$ are all even since $m$ is odd. Using this result in Eq.~\eqref{appen-FT-eq-11}, we then see that for odd $j$, only the real part of Eq.~\eqref{appen-FT-eq-11} contributes while for even $j$, only the imaginary part contributes. We first consider $\widetilde{\zeta}(Q,N)$ for $Q=\pi j/(N+1)$ with odd $j$, for which Eq.~\eqref{appen-FT-eq-11} reduces to
\begin{align}
&\widetilde{\zeta}\left( Q,N \right)  \simeq \mathcal{T}_1 \left( Q,N \right) + \mathcal{T}_2 \left( Q,N \right), \label{appen-FT-eq-12}\\
\text{with }& \mathcal{T}_1 \left( Q,N \right) =  \frac{N}{4} \sum_{\substack{s=1\\ s\ \mathrm{odd}}}^{N}\sum_{\substack{s'=1\\ s' \ \mathrm{even}}}^{N}  \mathcal{C}_{s, s'}(\tau, N)    \delta_{j, \left|s-s' \right|}, \label{appen-FT-eq-13}\\
&\mathcal{T}_2 \left( Q,N \right) =  -\frac{N}{4} \sum_{\substack{s=1\\ s\ \mathrm{odd}}}^{N}\sum_{\substack{s'=1\\ s' \ \mathrm{even}}}^{N}  \mathcal{C}_{s, s'}(\tau, N)  \delta_{j, N+1- \left|N+1- s-s' \right|}. \label{appen-FT-eq-14}
\end{align}
We demonstrated in Sec.~\ref{sec-transition} and Fig.~\ref{fig-osc-appen}(a) that $|\widetilde{\zeta}(Q,N)|$ in Eq.~\eqref{appen-FT-eq-12} and $|\mathcal{T}_1(Q,N)|$ in Eq.~\eqref{appen-FT-eq-13} attain their maxima at the same wavenumber $Q_{\rm max}(\tau)$. From Eq.~\eqref{appen-FT-eq-13}, $|\mathcal{T}_1(Q,N)|$ is maximum when $\mathcal{C}_{s,s'}(\tau,N)$ is maximum. As shown in Appendix~\ref{appen-Ckl}, the maximum of $\mathcal{C}_{s,s'}(\tau,N)$ is obtained when $s$ and $s'$ in the large-$N$ limit satisfy 
\begin{align}
  & 2\gamma\tau
\left|
\cos(\omega_s)-\cos(\omega_{s'})
\right| =
2\pi r,
\qquad r=1,2,\ldots, \label{appen-FT-eq-21-n} \\
& \qquad \omega_s+\omega_{s'}=\pi.
\label{appen-FT-eq-24-n}
\end{align}
Using Eq.~\eqref{appen-FT-eq-13}, the corresponding Fourier mode is $Q_{\rm max}(\tau) = \left| \omega_s-\omega_{s'}\right|$. Next using the trigonometric relation
\begin{align}
\left|
\cos(\omega_s)-\cos(\omega_{s'})
\right| = 2 \sin \left( \frac{|\omega_s-\omega_{s'}|}{2} \right) \sin \left( \frac{\omega_s+\omega_{s'}}{2}\right), \label{appen-FT-eq-14-n}
\end{align}
in Eqs.~\eqref{appen-FT-eq-21-n}-\eqref{appen-FT-eq-24-n}, we obtain
\begin{align}
Q_{\rm max} (\tau)= 2 \sin ^{-1} \left(\frac{\pi r}{2 \gamma \tau} \right),~~~r=1,2, \ldots \label{appen-FT-eq-15-n}
\end{align}
Thus, for odd $j$, the dominant wavenumber at which
$|\widetilde{\zeta}(Q,N)|^2$ is maximal is determined by Eq.~\eqref{appen-FT-eq-15-n} in the large-$N$ limit.

\subsection{$Q = \pi j /(N+1)$ with even $j$}

We now consider $\widetilde{\zeta}(Q,N)$ for $Q=\pi j/(N+1)$ with even
$j$. As discussed above, in this case only the imaginary part of
Eq.~\eqref{appen-FT-eq-11} contributes, giving
\begin{align}
\widetilde{\zeta}\left(Q,N\right)
\simeq
\frac{iN}{2\pi}
\sum_{\substack{s=1\\ s\ \mathrm{odd}}}^{N}
\sum_{\substack{s'=1\\ s' \ \mathrm{even}}}^{N}
\mathcal{C}_{s,s'}(\tau,N)
\sum_{\substack{m=1\\ m\ \mathrm{odd}}}^{\infty}
\frac{1}{m}
\Bigg[
&\delta_{j+m,|s-s'|}
-\delta_{j-m,|s-s'|}-\delta_{j+m,N+1-|N+1-s-s'|}
\nonumber\\
&
+\delta_{j-m,N+1-|N+1-s-s'|}
\Bigg].
\label{appen-FT-eq-even-1}
\end{align}
For fixed $s$ and $s'$, the terms in the $m$-summation decay as $1/m$. Hence, the leading contribution is obtained by taking $m=1$
\begin{align}
\left| \widetilde{\zeta}(Q,N) \right|
\simeq
\frac{N}{2\pi}
\left| \sum_{\sigma=\pm1} \sigma \Bigg[
\sum_{\substack{s=1\\ s\ \mathrm{odd}}}^{N}
\sum_{\substack{s'=1\\ s'\ \mathrm{even}}}^{N}
\mathcal{C}_{s,s'}(\tau,N)
\Bigg\{
\delta_{j+\sigma,|s-s'|}
-
\delta_{j+\sigma,N+1-|N+1-s-s'|}
\Bigg\} \Bigg]
\right| .
\label{appen-FT-eq-even-2}
\end{align}
Since $j$ is even and $\sigma=\pm1$, $\left( j+\sigma \right)$ is odd. Therefore, for each $\sigma$, the two terms inside the square brackets in Eq.~\eqref{appen-FT-eq-even-2} have exactly the same structure as the two terms obtained in Eq.~\eqref{appen-FT-eq-12} for the odd-$j$ case. Following the same line of
reasoning, the dominant contribution to $\left| \widetilde{\zeta}(Q,N) \right|$  in Eq.~\eqref{appen-FT-eq-even-2} is then given by $j_{\rm max} + \sigma = |s-s'|$ where the pair $(s, s')$ satisfies Eqs.~\eqref{appen-FT-eq-21-n}-\eqref{appen-FT-eq-24-n}. Translating it to $Q_{\rm max}(\tau) = \pi j_{\rm max}/(N+1)$, we obtain
\begin{align}
Q_{\rm max}(\tau) = 2 \sin ^{-1} \left(\frac{\pi r}{2 \gamma \tau} \right), ~~r=1,2, \ldots, \label{appen-FT-eq-even-3}
\end{align}
in the large-$N$ limit even for the even-$j$ case. Thus, comparing this result with Eq.~\eqref{appen-FT-eq-15-n}, we see that in the large-$N$ limit the dominant wavenumber at which the power spectrum attains its maximum is the same for both odd and even $j$, namely
$Q_{\rm max}(\tau)= 2\sin^{-1}\left(\frac{\pi r}{2\gamma\tau}\right)$.

\subsection{Maximum of $\mathcal{C}_{s,s'}(\tau,N)$}
\label{appen-Ckl}
We now derive the conditions for maximizing
$\mathcal{C}_{s,s'}(\tau,N)$ with respect to $s$ and $s'$, as quoted in Eqs.~\eqref{SS-pos-eq-11}-\eqref{SS-pos-eq-14}. Recall that $\mathcal{C}_{s,s'}(\tau,N)$ is determined by the functions $\mu_s(\eta)$ and $\nu_{s'}(\eta)$ through Eq.~\eqref{SS-pos-eq-2}, with their explicit forms given in Eqs.~\eqref{SS-pos-eq-3} and \eqref{SS-pos-eq-4}. Let us see how these functions depend on $\eta$ for fixed $s$ and $s'$. As illustrated in Fig.~\ref{fig-osc-appen}(b) for $s=19$, $s' = 60$, $N=200$, $\gamma=1$ and $\tau=2$, the absolute values of both  $\mu_s(\eta)$ and $\nu_{s'}(\eta)$ are sharply peaked as functions of $\eta$. We stated before that for large $N$, the peak locations of $\mu_s(\eta)$ and $\nu_{s'}(\eta)$ are given by
\begin{align}
\eta^{{\mu_s} \rm peak}
&\simeq
-2\gamma\tau\cos(\omega_s)
\pmod{2\pi},
\label{appen-FT-eq-15}
\\
\eta^{{\nu_{s'}} \rm peak}
&\simeq
-2\gamma\tau\cos(\omega_{s'})
\pmod{2\pi}.
\label{appen-FT-eq-16}
\end{align}
To prove this, we set $\eta = \left( \eta^{{\mu_s} \rm peak} + \epsilon \right)$ and expand the expression of $\mu_s(\eta)$ in Eq.~\eqref{SS-pos-eq-3} for $\epsilon \to 0$. We obtain
\begin{align}
\mu_s\left( \eta  \right) \simeq \frac{\sin \left( \omega_s \right) e^{2i \gamma \tau \cos(\omega_s)}}{\sin^2 \left( \omega_s \right) -\frac{i \epsilon}{2}  \sum_{\substack{m\neq s\\ m\ \mathrm{odd}}} \sin^2 \left( \omega_m \right)  + \frac{\epsilon}{2}  \sum_{\substack{m\neq s\\ m\ \mathrm{odd}}} \sin^2 \left( \omega_m \right) \cot \left[  \gamma \tau \left\{ \cos (\omega_m)-\cos (\omega_s)\right\} \right]}. \label{appen-FT-eq-17}
\end{align}
We next use
\begin{align}
& \sum_{\substack{m\neq s\\ m\ \mathrm{odd}}} \sin^2 \left( \omega_m \right) = \frac{N+1}{4} - \sin^2 \left( \omega_s \right), \\
& \sum_{\substack{m\neq s\\ m\ \mathrm{odd}}} \sin^2 \left( \omega_m \right) \cot \left[  \gamma \tau \left\{ \cos (\omega_m)-\cos (\omega_s)\right\} \right] \simeq N \alpha_s(\tau) , \label{auiq71aa}
\end{align}
where $\alpha_s(\tau)$ is an order-unity quantity in the large-$N$ limit. Eq.~\eqref{appen-FT-eq-17} now becomes
\begin{align}
\mu_s\left( \eta  \right) \simeq \frac{\sin \left( \omega_s \right) e^{2i \gamma \tau \cos(\omega_s)}}{\sin^2 \left( \omega_s \right) -\frac{i \epsilon N}{8} +\frac{ \epsilon N \alpha_s(\tau)}{2}  }. \label{appen-FT-eq-18}
\end{align}
For $\epsilon N\gg 1$, or equivalently $\epsilon\gg 1/N$, the denominator in Eq.~\eqref{appen-FT-eq-18} becomes large and hence $\mu_s(\eta)$ decays to zero. Thus, $\mu_s(\eta)$ is appreciable only in a narrow region around $\eta= \eta^{{\mu_s} \rm peak}$, whose width shrinks with increasing $N$. This explains the strongly localized structure of $|\mu_s(\eta)|^2$ seen in Fig.~\ref{fig-osc-appen}(b). 

Similarly, expanding $\nu_{s'}(\eta)$ around $\eta =\left( \eta^{\nu_{s'} \rm peak} + \epsilon \right)$ gives
\begin{align}
\nu_{s'}(\eta) \simeq \frac{\sin \left( \omega_{s'} \right) e^{-2i \gamma \tau \cos(\omega_{s'})}}{\sin^2 \left( \omega_{s'} \right) +\frac{i \epsilon N}{8} -\frac{ \epsilon N \beta_{s'}(\tau)}{2}  }, \label{appen-FT-eq-19}
\end{align}
where
\begin{align}
\beta_{s'}(\tau) = \frac{1}{N} \sum_{\substack{m\neq s' \\ m \ \mathrm{even}}} \sin^2 \left( \omega_m \right) \cot \left[  \gamma \tau \left\{ \cos (\omega_{s'})-\cos (\omega_m)\right\} \right]. \label{appen-FT-eq-20}
\end{align}
Following the same argument as for $\mu_s(\eta)$, we find that
$\nu_{s'}(\eta)$ is appreciable only in a narrow region around
$\eta=\eta^{{\nu_{s'}}\rm peak}$, whose width likewise shrinks with
increasing $N$.

We now turn to $\mathcal{C}_{s,s'}(\tau,N)$ in Eq.~\eqref{SS-pos-eq-2}. As we showed, $\mu_s(\eta)$ and $\nu_{s'}(\eta)$ are sharply localized around $\eta^{{\mu_s}\rm peak}$ and $\eta^{{\nu_{s'}}\rm peak}$, respectively, their product in Eq.~\eqref{SS-pos-eq-2} gives an appreciable contribution to the integral only when the two peak locations coincide. If the peaks are separated, their overlap is negligible and $\mathcal{C}_{s,s'}(\tau,N)$ is correspondingly small. Hence, to get pronounced values of $\mathcal{C}_{s,s'}(\tau,N)$ require $\eta^{{\mu_s}\rm peak}
=\eta^{{\nu_{s'}}\rm peak}$. For example, for $N=200$, $\gamma=1$, and $\tau=2$, the pair $(s,s')=(33,150)$ gives $\eta^{\mu_s {\rm peak}}\approx\eta^{\nu_{s'} {\rm peak}}\approx2.8$, and correspondingly $\mathcal{C}_{s,s'}(\tau,N)\sim10^{-3}$. On the other hand, for $(s,s')=(19,60)$, we have $\eta^{\mu_s {\rm peak}}\approx2.4$ and $\eta^{\nu_{s'} {\rm peak}}\approx3.9$. These peak locations are well separated, and the corresponding value is much smaller, $\mathcal{C}_{s,s'}(\tau,N)\sim10^{-5}$. 

Using Eqs.~\eqref{appen-FT-eq-15}--\eqref{appen-FT-eq-17}, we therefore find that pronounced values of $\mathcal{C}_{s,s'}(\tau,N)$ require
\begin{align}
 &  \eta^{{\mu_s}\rm peak}
=\eta^{{\nu_{s'}}\rm peak}, ~~
\implies  2\gamma\tau
\left|
\cos(\omega_s)-\cos(\omega_{s'})
\right| =
2\pi r,
\qquad r=1,2,\ldots,
\label{appen-FT-eq-21}
\end{align}
where recall $s$ is odd and $s'$ is even.
Noting that $E_{s} = -2 \gamma \cos(\omega_s)$, Eq.~\eqref{appen-FT-eq-21} gives the condition for maximizing  $\mathcal{C}_{s,s'}(\tau,N)$, as was quoted earlier in Eq.~\eqref{SS-pos-eq-11}.

The corresponding form of $\mathcal{C}_{s,s'}(\tau,N)$ is obtained by plugging Eqs.~\eqref{appen-FT-eq-18}-\eqref{appen-FT-eq-19} in Eq.~\eqref{SS-pos-eq-2} and changing the integral from $\eta$ to $\epsilon$
\begin{align}
\mathcal{C}_{s,s'}(\tau,N)
\simeq \frac{\sin(\omega_s)\sin(\omega_{s'})}{4\pi}
\operatorname{Re}\Bigg[
\int_{-\infty}^{\infty} d\epsilon\,
\frac{1}{
\left[
\sin^2(\omega_s)
-\frac{i\epsilon N}{8}
+\frac{\epsilon N\alpha_s(\tau)}{2}
\right]
\left[
\sin^2(\omega_{s'})
+\frac{i\epsilon N}{8}
+\frac{\epsilon N\alpha_s(\tau)}{2}
\right]
}
\Bigg].
\label{appen-FT-eq-22}
\end{align}
We have extended the limits of integration to $(-\infty,\infty)$ since, in the large-$N$ limit, the integrand is sharply localized around $\epsilon=0$ and decays rapidly away from this region, as explained before. We have further used $\alpha_s(\tau)\simeq-\beta_{s'}(\tau) $ in the large-$N$ limit. This relation follows directly from Eqs.~\eqref{auiq71aa}-\eqref{appen-FT-eq-20} by replacing the sums with integrals and using Eq.~\eqref{appen-FT-eq-21}.

Performing the integration in Eq.~\eqref{appen-FT-eq-22}, we get
\begin{align}
\mathcal{C}_{s,s'}(\tau,N) \simeq \frac{4}{N}~ \frac{\sin \left( \omega_s \right) \sin \left( \omega_{s'} \right) \left[ \sin^2 \left( \omega_{s} \right)+ \sin ^2\left( \omega_{s'}  \right)\right] }{ \left[ \sin^2 \left( \omega_{s} \right)+ \sin ^2\left( \omega_{s'}  \right)\right]^2 + 16 \alpha_s(\tau)^2 \left[ \sin^2 \left( \omega_{s} \right) - \sin ^2\left( \omega_{s'}  \right)\right]^2 }. \label{appen-FT-eq-23}
\end{align}
The same expression was given earlier in Eq.~\eqref{SS-pos-eq-13}. 

To determine the largest possible value of $\mathcal{C}_{s,s'}(\tau,N)$, we minimize the denominator in Eq.~\eqref{appen-FT-eq-23}. Its second term is nonnegative and disappears when $\sin^2(\omega_s)=\sin^2(\omega_{s'})$. Because $s$ and $s'$ are distinct, $\omega_s\neq\omega_{s'}$, and hence this condition implies
\begin{equation}
\omega_s+\omega_{s'}=\pi,
\label{appen-FT-eq-24}
\end{equation}
as obtained previously in Eq.~\eqref{SS-pos-eq-14}. The corresponding maximum value is $\mathcal{C}_{s,s'}(\tau,N)\simeq 2/N$. Thus, Eqs.~\eqref{appen-FT-eq-22}-\eqref{appen-FT-eq-24} give the conditions for maximizing
$\mathcal{C}_{s,s'}(\tau,N)$ with respect to $s$ and $s'$
in the large-$N$ limit.

\section{Derivation of Eq.~\eqref{OIC-eq-6}}
\label{appen-real-imag}
We stated in Eq.~\eqref{OIC-eq-6} that, for the nearest-neigbor hopping Hamiltonian in Eq.~\eqref{main-eqq-1}, the first-detection amplitude to the left boundary $\phi_n^{(L)}(x_0)$ is real for odd $x_0$ and purely imaginary for even $x_0$ (similarly for $\phi_n^{(R)}(x_0)$). To derive this result, let us introduce the sublattice operator
\begin{align}
\hat{\Gamma}\ket{x}=(-1)^x\ket{x}.
\end{align}
The Hamiltonian in Eq.~\eqref{main-eqq-1} satisfies
\begin{align}
\hat{\Gamma} H \hat{ \Gamma}=-H, \quad \hat{\Gamma} U(\tau) \hat{\Gamma}=e^{iH\tau}=U(\tau)^*
\end{align}
Moreover, $\hat{\Gamma} \ket{x_L} = - \ket{x_L}$ and $\hat{\Gamma} \ket{x_R} = (-1)^N \ket{x_R}$. Therefore, for the survival operator $\mathcal{S} = \Big(\mathbb{1}-\ket{x_L} \bra{x_L}-\ket{x_R} \bra{x_R}\Big) U(\tau)$, we have
\begin{align}
\hat{\Gamma} \mathcal{S} = \mathcal{S}^* \hat{\Gamma}, \implies \hat{\Gamma} \mathcal{S}^{n-1} = (\mathcal{S}^*)^{n-1} \hat{\Gamma}
\end{align}
We then obtain
\begin{align}
\phi_n^{(L)}(x_0) & = \braket{x_L|U(\tau) \mathcal{S}^{n-1}|x_0}, \nonumber \\ 
&= -\braket{x_L|\hat{\Gamma} U(\tau) \mathcal{S}^{n-1}|x_0},
\nonumber\\
& = -\braket{x_L| U^*(\tau) \mathcal{S^*}^{n-1} \hat{\Gamma}|x_0}  \nonumber \\
&=
(-1)^{1+x_0} \left( \phi_n^{(L)}(x_0) \right)^*.
\end{align}

%\section*{References}
%\bibliographystyle{unsrt}
%\bibliographystyle{iopart-num}
\bibliography{Bib_QRW}

\end{document}